\documentclass[twocolumn,
showpacs,
showkeys,
preprintnumbers,
nofootinbib,
superscriptaddress,
amsmath,
amssymb,
floatfix,
secnumarabic,
aps,
pra,
a4paper,
notitlepage,
final,
]{revtex4}%

\usepackage[colorlinks=true,urlcolor=blue]{hyperref}
\usepackage{graphicx}
\usepackage{epsfig}

\newcommand{\beq}{\begin{equation}}
\newcommand{\eeq}{\end{equation}}
\newcommand{\beqar}{\begin{eqnarray}}
\newcommand{\eeqar}{\end{eqnarray}}
\newcommand{\bal}{\begin{aligned}}
\newcommand{\eal}{\end{aligned}}

\def\dalam{\hbox
{\vrule\vbox{\hrule\hbox to 1ex{ \hfill}\kern 1 ex\hrule}\vrule}}
\def\sign{\hbox{sign}}
\def\1/2{\hbox{$ {1 \over 2}$ }}

\def\h{\hbar}
\def\i/h{{i \over \h}}

\def\inf{\infty}
 
\def\v{\vec}

\def\a{\alpha} 
\def\b{\beta} 
 
\def\g{\gamma} \def\G{\Gamma} 
\def\d{\delta} \def\D{\Delta}
\def\l{\lambda} 
\def\e{\epsilon} \def\E{\hbox{$\cal E $}}

\def\ve{\varepsilon}
\def\s{\sigma}
\def\r{\rho}

\def\vf{\varphi}
\def\f{\phi} \def\F{\Phi}
 
\def\p{\psi} \def\P{\Psi}

\def\n{\nu}

  \def\vk{\varkappa}

\def\z{\zeta}

\def\tt{\theta} 
\def\et{\eta}

\def\<{\langle}
\def\>{\rangle}

\def\({\left(}
\def\[{\left[}
\def\){\right)}
\def\]{\right]}

\usepackage{subfigure}

\usepackage[notcite,notref,color]{showkeys}  

\usepackage{tikz}
\usetikzlibrary{decorations.pathreplacing}
\usetikzlibrary{decorations.pathmorphing}    
\usepackage{multirow}
\usepackage{dcolumn}		
\newcolumntype{.}{D{.}{.}{-1}}
\newcolumntype{i}[1]{D{.}{.}{#1}}

\newcommand{\myfrac}[2]{{\ifmmode{}^{#1}\!/_{\!#2}\else${}^{#1}\!/_{\!#2}$\fi}}

\begin{document}
\sloppy

\title{Casimir (vacuum) energy in planar QED with strong coupling}

\author{Yu.~Voronina}
\email{voroninayu@physics.msu.ru} \affiliation{Department of Physics and
Institute of Theoretical Problems of MicroWorld, Moscow State
University, 119991, Leninsky Gory, Moscow, Russia}

\author{K.~Sveshnikov}
\email{costa@bog.msu.ru} \affiliation{Department of Physics and
Institute of Theoretical Problems of MicroWorld, Moscow State
University, 119991, Leninsky Gory, Moscow, Russia}

\author{P.~Grashin}
\email{grashin.petr@physics.msu.ru} \affiliation{Department of Physics and
Institute of Theoretical Problems of MicroWorld, Moscow State
University, 119991, Leninsky Gory, Moscow, Russia}

\author{A.~Davydov}
\email{davydov.andrey@physics.msu.ru} \affiliation{Department of Physics and
Institute of Theoretical Problems of MicroWorld, Moscow State
University, 119991, Leninsky Gory, Moscow, Russia}

\date{\today}


\begin{abstract}

The essentially non-perturbative vacuum polarization effects, caused by an extended external supercritical Coulomb source, are explored for  a planar  Dirac-Coulomb (DC) system with strong coupling (similar to graphene and graphene-based heterostructures).  Taking account of results, obtained in \cite{partI2018} for the induced charge density $\r_{VP}(\v r)$, in  the present paper the evaluation of the Casimir (vacuum) energy $\E_{VP}$ is presented.  The main result  is that for a wide range of the system parameters in the overcritical region $\E_{VP}$  turns out  to be a rapidly decreasing negative function  $\sim -  Z^3/R_0\, $ with $Z\, , R_0$ being the charge and the size of the external source. By an explicit calculation the possibility for complete screening of the  electrostatic reflection self-energy of the external source by such  polarization effects for  $Z \gg Z_{cr,1}$ is demonstrated. The dependence of the  Casimir energy on the screening of the Coulomb asymptotics of the external source at some $R_1>R_0$ is also explored in detail, and some peculiar effects in the partial channels with the lowest rotational numbers $m_j=\pm 1/2\, , \pm3/2$ in the screened case are also discussed.
\end{abstract}


\maketitle

\section{Introduction}\label{sec:intro}

This work continues the study of  essentially nonperturbative vacuum effects for a model of an extended charged impurity with non-zero size $R_0$ in the graphene-like system on a substrate, initiated in ~\cite{partI2018},   with the main attention paid to the   vacuum polarization (Casimir) energy $\E_{VP}$. In such systems due to  the large value of $ \a_g \sim 1$  it is much easier to observe many non-trivial QED-effects experimentally. In particular,  the critical charges of atomic collapse in graphene are subject of condition $Z \a_g > 1/2$ \cite{Shytov2009},\cite{Shytov2007}, the observation of the Klein paradox requires electric fields $\sim 10^5\,\, \text{V/cm}$ (eleven orders of magnitude less than the fields necessary for the observation of
the Klein paradox for elementary particles) \cite{Katsnelson2006a}, the quantum Hall effect  can be observed for much higher temperatures and lower magnetic fields than in the conventional semiconductors \cite{Gusynin2005, Giesbers2008, Cobaleda2011}. Some effects turn out to be strong enough to affect the transport properties of graphene \cite{Shytov2009}.  The main feature inherent in all these effects is that they are essentially non-perturbative due to the large value of $ \a_g$ and therefore cannot be described within the perturbation theory (PT).

In this work we explore another essentially non-perturbative effect in the two-dimensional Dirac-Coulomb (DC) system with application  to graphene-like planar structures with strong coupling, namely, the vacuum polarization, caused by diving of discrete levels into the lower continuum in the  supercritical static or adiabatically slowly varying Coulomb fields, which are created by localized extended sources with $Z> Z_{cr}$.  Such effects have  attracted a considerable amount of theoretical and experimental activity in  3+1 D heavy ions collisions, where for $Z > Z_{cr,1} \simeq 170$ a non-perturbative reconstruction of the vacuum state is predicted, which should be accompanied by a number of nontrivial effects including the vacuum positron emission (\cite{Greiner1985a, Plunien1986, Ruffini2010, Greiner2012, Rafelski2016} and refs. therein).

Similar phenomena could occur in graphene with the charge impurities acting as atomic nuclei, while the graphene itself -- as the QED vacuum and its relativistic electrons and holes  --- as the virtual particles which populate the vacuum. A remarkable circumstance here is that due to the large value of the effective fine-structure constant these effects should take place for relatively small impurity charges $Z \simeq 1-10$. Since for these effects the charge carriers in graphene play the role of the virtual QED-particles, the induced charge density can be measured directly. In Ref.~\cite{Wang2013}, the five-dimer cluster consisting of Ca-atoms was used as a charge impurity and the induced density was measured via STM. Polarization effects in graphene, caused by charged impurities, have also been considered by many authors  (\cite{Katsnelson2006b, Biswas2007, Pereira2007, Kotov2008, Terekhov2008, Nishida2014, Khalilov2017} and refs. therein). Here it should be noted that in most cases the impurity is  modeled  as a point-like charge, what causes some problems in the supercritical case. Our work is aimed mainly at the study of vacuum polarization effects, caused by extended supercritical Coulomb sources with non-zero size $R_0$, which provide a physically clear and unambiguous problem statement like in Refs.~\cite{Shytov2007, Fogler2007, Milstein2010}, where the charge is assumed to be displaced away or smeared over a finite region of the graphene plane.

Taking account of results, obtained in \cite{partI2018} for the induced charge density $\r_{VP}(\v r)$, in  the present paper the evaluation of the Casimir energy $\E_{VP}$ is considered with emphasis on the   renormalization and convergence of the partial expansion for $\E_{VP}$,  matching $\E_{VP}$ with the  reflection self-energy of the external source and dependence on screening of  the external potential at some $R_1 > R_0$.
Here it is worth to note that although the most of works cited above considers  $\r_{VP}(\v r)$ as the main polarization observable,  $\E_{VP}$  turns out to be not less informative and in many respects complementary to $\r_{VP}(\v r)$. Moreover, compared to $\r_{VP}(\v r)$, the main non-perturbative effects, which appear in the vacuum polarization for $Z > Z_{cr,1}$ due to levels diving into the lower continuum, show up in the behavior of Casimir energy even more clear, demonstrating explicitly their possible role in the  overcritical region. The evaluation of $\E_{VP}$ is performed by means of the original method, which recently has  been successfully used  in solving similar problems for the one-dimensional H-like atom~\cite{Davydov2017, Sveshnikov2017, Voronina2017}.

The external Coulomb field $A^{ext}_{0}(\v r)$ is chosen in the form of a projection onto a plane of the potential of the uniformly charged sphere with the radius  $R_0$ and a cutoff of the Coulomb asymptotics at some $R_1>R_0$ in the form
\begin{multline}
\label{1.0}
A^{ext}_{0}(\v r)=  Z |e| \[{1 \over R_0}\tt\(R_0-r\)+ \right. \\ \left. +{1/r-1/R_1 \over 1-R_0/R_1} \tt \(R_0 \leq r \leq R_1\) \] \ ,
\end{multline}
which differs from the one used in \cite{partI2018} by absence of the discontinuity at $r=R_1$. The last circumstance is necessary for convergence of the one-loop vacuum polarization energy, which plays an important role in the calculation of $\E_{VP}$.

The radius of the source is taken as $R_0=a$, where $a\simeq1.42\, A$ is the C-C distance
in the  graphene lattice. Such cutoff of the Coulomb potential at small distances  has been used  in \cite{Pereira2008}. The cutoffs $R_0=a/2$ and $R_0=2a$ are also considered. The external cutoff $R_1$ will be taken as $R_1=2 R_0\, , 5R_0\, , 10 R_0$ for the study of screening effects, and as $R_1=20 R_0\, , 50 R_0\, , 150 R_0$ to establish a smooth transition into the unscreened case $R_1 \to \inf$, which  will  be also considered in detail.

The effective fine-structure constant is defined as
\beq
\label{1.2}
\a=e^2/(\hbar v_F \ve_{eff}) \ , \quad \ve_{eff}=(\ve +1)/2 \ ,
\eeq
with $\ve$ being  the substrate dielectric constant and $v_F = 3ta/2 \hbar$ -- the Fermi velocity in graphene. In its turn, $t$ is the hopping amplitude, while  $\l_c = {\hbar / m v_F}$  is the effective Compton length \cite{Goerbig2011}. Here  $m$ denotes the effective fermion mass, which is related to the local energy mismatch in the tight-binding formulation through the relation $\D=2mv_F^2$. These definitions lead to the relation $\l_c/a \simeq 3t/\D$. In this paper we consider $\a=0.4$ (which corresponds to  graphene on the SiC substrate \cite{Pereira2008}) and $\a=0.8$ (graphene on the h-BN substrate \cite{Goerbig2011, Sadeghi2015}).

Henceforth the system of units in which $\hbar=v_F=m=1$ is used, and so the distances are measured in units of $\l_c$, while the energy --- in units of $mv_F^2$.   For $\a=0.4$ the local energy mismatch is $\D=0.26$ eV and therefore for $R_0=a/2,a,2a$ one obtains  $R_0=1/60,1/30,1/15$ in the  units chosen, while for  $\a=0.8$ one has $\D=0.056$ eV and so $R_0=1/350,1/175,2/175$.

The paper  is arranged as follows. First in Sect. 2 the general approach to essentially non-perturbative evaluation of the Casimir energy for such DC systems is presented, in Sect.3 we consider the unscreened case $R_1 \to \inf$, and thereon in Sect. 4-5 explore the changes, caused by finite $R_1$. To conclusion (Sect.6) the main reasons and consequences of the Casimir energy decline in the overcritical region are discussed.

\section{The general approach to evaluation of the Casimir energy for an extended Coulomb source}\label{sec:EVP}

As it was shown in~\cite{Davydov2017, Sveshnikov2017, Voronina2017, Davydov2018a, partI2018}, the formation of localized vacuum shells, caused by  diving of discrete levels into the lower continuum, significantly affects $\r_{VP}^{ren}(\v r)$. This effect also yields a substantial and essentially nonlinear  contribution to $\E_{VP}^{ren}$ in the overcritical region.  The corresponding changes in $\E_{VP}^{ren}$, caused by  formation of localized vacuum shells with increasing $Z$, depend  strongly on the number of spatial dimensions. In case of 1+1 QED the growth rate of the shells total number is quite moderate, that's why  the non-renormalized $\E_{VP}$ in the overcritical region behaves as  $\sim Z^\n$, $1<\n<2$. Therefore in this case the dominant contribution to the renormalized Casimir energy comes from the renormalization term  ~\cite{Davydov2017, Sveshnikov2017, Voronina2017}. In 2+1 QED the picture changes more significantly, as it is shown in ~\cite{Davydov2018b} and in this paper below for a strongly coupled planar DC system.

First let us  consider such approach to calculation of  $ \E_{VP} $, which takes  account for the non-perturbative effects in the overcritical region from the very beginning. The starting point for this approach is the following expression for the vacuum energy ~\cite{Plunien1986, Greiner2012}
\beq
\label{f39}
\E_{VP}=\<H_D\>_{vac}=\1/2 \(\sum\limits_{\e_n<\e_F} \e_n-\sum\limits_{\e_n \geqslant \e_F}\e_n\) \ ,
\eeq
which follows from the Dirac Hamiltonian, written in the invariant under charge conjugation form, and is defined up to the constant, which depends on the choice of the energy origin. As in \cite{partI2018},  the Fermi level $\e_F$ is chosen at the threshold of the lower continuum ($\e_F=-1$). There follows from (\ref{f39}) that even in the absence of external fields $ A_{ext} = 0 $ the vacuum energy is negative and divergent. Since the induced density (see~\cite{partI2018}, eq.(12)) is defined in such a way that for $A_{ext} = 0 $ it vanishes exactly, it is natural to normalize  $ \E_{VP} $ in the same way. Another point is that in the external Coulomb potentials like (\ref{1.0}) there exists a certain number of bound states (infinite without screening). To keep the interaction effects only, the quantity $mv_F^2$ (which is equal to 1 in the units chosen), corresponding to the electron rest mass in the ,,normal''  QED, should be subtracted from the energy of each bound state. So in the physically well-motivated form  the initial expression for the vacuum energy should be represented as
\beq\label{f40}\bal
\E_{VP}=&\1/2 \(\sum\limits_{\e_n<\e_F} \e_n-\sum\limits_{\e_n \geqslant \e_F} \e_n  + \sum\limits_{-1\leqslant \e_n<1} \! 1
\)_A \ -  \\ & - \1/2 \(\sum\limits_{\e_n<0} \e_n-\sum\limits_{\e_n
	>0} \e_n \)_0 \ ,
\eal\eeq
where the label $A$ denotes the  nonzero external field, while $0$ stands for the free case. The vacuum energy, defined in such way,  vanishes in absence of the external field, while in  presence of the latter it contains only the interaction effects  starting from $ O(Z^2) $.

In the next step let us divide (\ref{f40}) into separate contributions from discrete and continuous spectra, applying to the difference of integrals over the continua $ (\int dk\sqrt{k^2+1})_A-(\int dk\sqrt{k^2+1})_0 $ the well-known techniques, which represent this difference in the form of an integral from the elastic scattering phase $ \d(k) $ (\cite{Rajaraman1982, Sveshnikov1991} and refs. therein). After a number of almost obvious intermediate steps~\cite{Sveshnikov2017} one obtains the final answer for $\E_{VP} $ in the form
\begin{multline}\label{f41a}
\E_{VP}=2\sum\limits_{m_j=1/2,3/2,..} \E_{VP,|m_j|}= \\  =2\sum\limits_{m_j=1/2,3/2,..} \({1 \over 2\pi} \int\limits_0^\inf \!   \  \frac{k \, dk }{\sqrt{k^2+1}} \ \d_{tot,|m_j|}(k) + \right. \\  \left.  + {1 \over 2} \sum\limits_{-2 \leqslant \e_{n,|m_j|}<2} \(2-\epsilon_{n,|m_j|}\)\) \ ,
\end{multline}
where $ \d_{tot,|m_j|}(k) $ is the partial total phase shift for the given $|m_j|$, which includes contributions  from  scattering states in both continua with $\pm m_j$, while $ (2-\e_{n,|m_j|}) $ is the total bound energy of two discrete levels with the same $\pm m_j$ in the radial DC spectral problem for the  external field (\ref{1.0})
\beq
\label{3.7}
\left\lbrace\bal
&\frac{d}{d r}\p_1(r)+\frac{1/2-m_j}{r}\,\p_1(r)=(\e-V(r)+1)\p_2(r) \ ,\\
&\frac{d}{d r}\p_2(r)+\frac{1/2+m_j}{r}\,\p_2(r)=-(\e-V(r)-1)\p_1(r) \ ,
\eal\right.
\eeq
where
\begin{multline}
\label{1.1}
V(r)= - Q \[{1 \over R_0}\tt\(R_0-r\)+ \right. \\ \left.+{1/r-1/R_1 \over 1-R_0/R_1} \tt \(R_0 \leq r \leq R_1\)\] \ ,
\end{multline}
and
\beq
\label{1.11}
Q=Z\a  \ .
\eeq
Here and in what follows we take into account that in 2+1 QED the Dirac matrices  can be chosen either in two- or four-dimensional representations. In the first case there are two inequivalent possible choices of the matrix signature ~\cite{Hosotani1993, Khalilov2016}, while in the latter one the DC spectral problem for the external source (\ref{1.0}) splits into two independent subsystems, which are related  by $m_j \to -m_j$. Therefore the degeneracy factor of the energy eigenstates with the fixed $m_j$ equals to 2 and in what follows this factor will be shown explicitly in all the expressions for $\r_{VP}(\v r)$ and $\E_{VP}$, while the DC spectral problem without any loss of generality will  be considered in the two-dimensional representation with  $\a_i=\s_i$,  $\b=\s_3$.

 Such approach to calculation of $ \E_{VP} $ turns out to be quite effective. As it will be shown by direct calculation below, the total partial phase shift $ \d_{tot,|m_j|}(k) $ is finite for $ k \to 0 $ and behaves like $ O(1/k^3) $ for $ k \to \inf $. Thus, each partial phase integral in (\ref{f41a}) is always convergent. The total partial bound energy of discrete levels is also finite, since $(2-\e_{n,|m_j|})$ behaves like $ O(1/n^2) $ for $ n \to \inf $. So  there is  no special need in any additional regularization of the Coulomb asymptotics of the external potential for $ r \to \inf $ even in the unscreened case ($R_1 \to \inf$).

As a result, for the external potentials like (\ref{1.0}) each separate term in the partial expansion  for $ \E_{VP} $ (\ref{f41a}) turns out to be finite without any special UV-renormalization.
However, there remains a natural question concerning the convergence of this series. For these purposes let us explore the asymptotical behavior of  separate terms in (\ref{f41a}) for $|m_j|\to\inf$ by means of the WKB-approximation for the total partial phase $\d_{tot,|m_j|}(k)$
\beq \bal
\label{int_WKB}
&\d_{WKB,|m_j|}(k) = 2\int \!   dr \ \(\sqrt{\(\e(k)+V(r)\)^2-1-{m_j^2 \over r^2}} \ + \right. \\ & \left. + \ \sqrt{\(\e(k)-V(r)\)^2-1-{m_j^2 \over r^2}}-2 \sqrt{k^2-{m_j^2 \over r^2}}\) \ ,
\eal \eeq
where $\e(k)=\sqrt{k^2+1}$, while the integration is carried out over the regions  where the expressions under the square root are positive. Omitting certain cumbersome calculations connected with the evaluation of corresponding integrals for the WKB-phase and for the phase integral in (\ref{f41a}), as well as of the bound energy of discrete levels, all of which can be performed analytically by means of the computer algebra tools (for more details of calculation see~\cite{Davydov2018b}),    let us give  the final answer for partial $\E_{VP,|m_j|}$ in the limit $|m_j| \to \inf$:
\beq
\label{term_m}
\E_{VP,|m_j|}=\frac{1}{\pi} \int\limits_0^{\infty} \! dr \  V^2(r) + O\({1 \over |m_j|^3}\) \ , \quad |m_j|\to \inf \ .
\eeq
Here it is worth-while to note that  the following circumstance  should be taken into account by calculating  the sum of discrete levels in (\ref{f41a}). Namely, by definition  $ (2-\e_{n,|m_j|}) $ is the sum of bound energies  of two discrete levels of the system  (\ref{3.7}), corresponding to  $\pm m_j$ with the same radial number $n$. However, it can be easily verified that in the system (\ref{3.7})  the lowest level with $n=0$ exists for $m_j>0$ only, whereas for $m_j<0$ the discrete spectrum starts from $n=1$. At the same time, for the mirror-symmetrical system with the opposite signature of the two-dimensional Dirac matrices or, equivalently, for another subsystem in the 4-dimensional representation, which is connected with (\ref{3.7}) via the replacement $m_j\to -m_j$, the same lowest level with $n=0$ appears for $m_j<0$ only. This effect is quite similar to the DC problem in 3+1 D, when in the relativistic H-like atom the levels $nj$ with $j=n-1/2$  turn out to be degenerate twice less than the others \cite{Itzykson1980}.

Taking account for this difference in degeneracies in (\ref{f41a}) the contribution of discrete levels to $\E_{VP, |m_j|}$ should be written more thoroughly, namely
\begin{multline}\label{sum_dm}
2 \sum\limits_{-2 \leqslant \e_{n,|m_j|}<2} \(2-\e_{n, |m_j|}\)= \\ = \(2-\e_{0, |m_j|}\)+ 2 \sum\limits_{n=1} \(2-\e_{n, |m_j|}\) \ ,
\end{multline}
where the common degeneracy factor 2, which in (\ref{f41a}) is taken out of the series in $m_j$, is in accordance with the multiplier 2 before the sums  in (\ref{sum_dm}). To the contrary,  in the contribution from the lowest level such a factor is absent, since $\e_{0, |m_j|}$ contains  the sum of the lowest levels from both subsystems (signatures) simultaneously.  This circumstance underlines once more that in  $\E_{VP}$ and  $\r_{VP}$ both subsystems (signatures) should be considered at the same footing, and hence, the degeneracy of levels except for the lowest one should be indeed 2.

The asymptotics (\ref{term_m}) shows that the partial series in $m_j$ for $\E_{VP}$  diverges linearly, whence it follows the necessity of its regularization and subsequent renormalization. At the same time, each partial channel in (\ref{f41a}) in itself is finite without any additional doings. It should be specially noted that  the degree of divergence of the partial series (\ref{f41a}) for $ \E_{VP} $ is formally  the same (linear), as  within PT in 2+1 QED without virtual photons for the  unique divergent  fermionic loop with two external lines. The latter circumstance shows that by calculation of $ \E_{VP} $ via the principally different non-perturbative approach, which doesn't reveal any connection with PT, we nevertheless meet actually the same  divergence of the theory, as in PT. Therefore in the present approach the cancelation of divergent terms should follow the same rules as in PT, based on the regularization of the fermionic loop with two external lines, which preserves the physical essence of the whole renormalization procedure and simultaneously provides the mutual agreement between perturbative and non-perturbative approaches to the calculation of $ \E_{VP} $.  This conclusion is in complete agreement with results obtained in Ref.~\cite{Gyulassy1975}.

The need in the renormalization via  fermionic loop follows also from the analysis of the properties of $\r_{VP}(\v r)$, which shows that without such UV-renormalization the integral induced charge doesn't acquire the expected integer value in units of $(-2 |e|)$. In fact, the properties of $\r_{VP}(\v r)$ play here the role of a controller, which provides the implementation of the required physical conditions for a correct description of the induced polarization effects beyond the scope of PT, that cannot be tracked via evaluation of $\E_{VP}$ by means of the initial relations (\ref{f40}),(\ref{f41a}).

Another motivation to such renormalization of the vacuum energy is that for $Z \to 0$ it should coincide with  $\E^{(1)}_{VP}$, obtained   to the lowest order of PT via
\beq
\label{2.1}
\E^{(1)}_{VP}=\frac{1}{2} \int d^2r\, \r^{(1)}_{VP}(\vec{r})A_{0}^{ext}(\vec{r}) \ ,
\eeq
where $\r^{(1)}_{VP}(\vec{r})$ is the corresponding lowest-order perturbative induced  density, considered in \cite{partI2018}. In the axially-symmetric case the general expression for $\E^{(1)}_{VP}$, which follows from the initial formulae for $\r^{(1)}_{VP}(\vec{r})$, reads
\begin{multline}
\label{2.2}
\E^{(1)}_{VP}=\frac{\a}{8}\int\limits_{0}^{\infty} q^2dq\,\[\frac{2 }{q} + \(1-\frac{4}{q^2}\) \arctan \( \frac{q}{2}\)\] \times  \\  \times \(\int_0^{\inf} r\, dr J_0(qr) A^{ext}_{0}(r)\)^2 \ .
\end{multline}
Inserting into (\ref{2.2}) the  unscreened $A^{ext}_{0}(r)$ ($R_1 \to \inf$), one finds
\begin{multline}
\label{2.6}
\E^{(1)}_{VP}=\frac{Q^2}{32}\int\limits_{0}^{\infty}dq\,\[\frac{2 }{q} + \(1-\frac{4}{q^2}\) \arctan \( \frac{q}{2}\)\]\times \\
\times \(2\[1+J_1(q R_0)-q R_0 J_0(q R_0)\]  + \right. \\ \left. +\pi q R_0\[ J_0(q R_0) \mathbf{H}_1(q R_0)- J_1(q R_0) \mathbf{H}_0(q R_0)\]\)^2 \ .
\end{multline}
It is easy to verify that  the integrand in (\ref{2.6}) behaves for $q \to \inf$ as $\sim 1/q^3$, and so in the unscreened case $\E^{(1)}_{VP}$ is a well-defined quantity.
Moreover, due to the axial symmetry of the external field both the perturbative density $\r^{(1)}_{VP}$ and the perturbative energy  $\E^{(1)}_{VP}$ correspond to the partial channel with $ |m_j|=1/2 $.  However,  the non-renormalized  $ \E_{VP, 1/2}$ does not reproduce the perturbative answer for $Z \to 0$ even in the unscreened case. Namely, in this case the  direct calculation shows that the analytic answers   for $ \r_{VP}^{(1)}(r) $ (\cite{partI2018}, eq.(10)) and for $ \r_{VP,1/2}(r)$,  found from the first Born approximation for the Green function (\cite{partI2018}, eqs.(37,38)), and hence for $ \E_{VP}^{(1)} $ and $ \E_{VP, 1/2}$,  turn out to be substantially different for $Z \to 0$.

Proceeding further this way,  we  pass from $\E_{VP}$ to $\E_{VP}^{ren}$ by means of the following prescription
\beq \bal
\label{f42}
&\E^{ren}_{VP}(Z) = 2\sum\limits_{m_j=1/2,3/2,..} \E^{ren}_{VP,|m_j|}(Z), \\ &\E^{ren}_{VP,|m_j|}(Z)=\E_{VP,|m_j|}(Z)+ \eta_{|m_j|} Z^2 \,,
\eal\eeq
where the renormalization coefficients $\eta_{|m_j|}$ are defined as
\beq
\label{f43}
\eta_{|m_j|} = \lim\limits_{Z_0 \to 0}  \[{\E_{VP}^{(1)}(Z_0)\d_{|m_j|,1/2}-\E_{VP,|m_j|}(Z_0) \over Z_0^2}\]
\ , \eeq
and depend solely on the profile of the external potential~\cite{Davydov2018b}.

The key-point of (\ref{f42}) is that now the quadratic in  $Z$ components are extracted from  the  initial expressions for non-renormalized partial terms  $ \E_{VP,|m_j|}(Z) $ in (\ref{f41a})  and replaced further by the renormalized  $\E^{(1)}_{VP}\d_{|m_j|,1/2}$, found within PT. This procedure is in complete agreement with the renormalization of $\r_{VP}$ with the only difference, that in the latter case the same procedure is applied to the linear in $Z$ components. Another argument in favor of such renormalization follows from the  well-known Schwinger relation between $ \E_{VP} $ and $ \r_{VP} $ \cite{Plunien1986}
\beq\label{Schwinger}
\d \E_{VP} = \int \r_{VP} \d A_0^{ext} + \d \E_N \ .
\eeq
It could be easily verified that the normalization of $ \E_{VP} $ on the free case and subtraction of the quantity $mv_F^2$ from the energy of each bound state don't change this relation \cite{Sveshnikov2017}.
At the same time,  the replacement $ \r_{VP}\to\r^{ren}_{VP} $ in (\ref{Schwinger}) implies  the corresponding replacement $ \E_{VP}\to\E^{ren}_{VP} $ with $\E^{ren}_{VP} $ defined as in (\ref{f42}).

Moreover, such a renormalization provides simultaneously the convergence of the partial series for  $\E_{VP}^{ren}$, since the divergent terms in the sum (\ref{f41a}), according to (\ref{term_m}), are proportional to $(Z\a)^2$. So the renormalization via fermionic loop turns out to be the universal method, which removes the divergence of the theory both in purely perturbative and essentially non-perturbative approaches to the vacuum polarization.

  Finally, the renormalized expression for the partial terms of the sum over $m_j$ for $ \E_{VP}^{ren} $ takes the form
\beq\bal
\label{f44a}
\begin{aligned}
&\E^{ren}_{VP,|m_j|} = {1 \over 2\pi} \int\limits_0^\inf \!   \  \frac{k \, dk }{\sqrt{k^2+1}} \ \d_{tot,|m_j|}(k) + \\ & + {1 \over 2} \sum\limits_{-2 \leqslant \e_{n, |m_j|}<2}  \(2-\e_{n,|m_j|}\) +\eta_{|m_j|}Z^2 \ .
\end{aligned}
\eal\eeq

It would be worth noticing that actually each partial channel (\ref{f44a}) reproduces by its structure almost exactly the renormalized  $\E_{VP}^{ren}$ in the one-dimensional case \cite{Davydov2017, Sveshnikov2017, Voronina2017}. However, in the one-dimensional case $\eta(R)$ is  a nontrivial  sign-alternating function of the radius $R$ of the Coulomb source ~\cite{Davydov2017,Voronina2017}, whereas in  2+1 D  all the  $ \eta_{|m_j|}$'s, including $ \eta_{1/2} $, turn out to be always strictly negative. For the unscreened case it is shown in~\cite{Davydov2018b}, while for the screened one in Section 5 of the present paper.

\section{Explicit evaluation of  the Casimir energy  in the unscreened case $R_1 \to \inf$}\label{sec:evaluationofEVP}

Now --- having dealt with the general approach to evaluation of the Casimir energy  this way --- let us  consider the calculation of $\E_{VP}^{ren}$  for the external source (\ref{1.0}), first for the unscreened case $R_1 \to \inf$.

 For $0 < r\leqslant R_0$ the solutions of the system  (\ref{3.7}) up to a common normalization factor take the form
\beq \bal \label{f450}
&\psi^{int}_{1, m_j}(r,\e )=(-i)^{(m_j-1/2)\theta(1-|\epsilon+V_0|)}\, \sqrt{|\epsilon+V_0+1|}\, \times \\& \times J_{m_j-1/2}(\z r),\\
&\psi^{int}_{2, m_j}(r, \e )=(-i)^{(m_j+1/2)\theta(1-|\epsilon+V_0|)}(-1)^{\theta(\epsilon +V_0 - 1)} \times
\\& \times \sqrt{|\epsilon+V_0-1|}\,\, J_{m_j+1/2}(\z r) \ ,
\eal\eeq
where $J_{\nu}(z)$ are the Bessel functions,
\beq\label{f4501}
V_0=Q/R_0 \ , \quad \z=\sqrt{(\e+V_0)^2-1} \ .
\eeq
In (\ref{f450}) the phase factors $(-i)^{(m_j \mp 1/2)\theta(1-|\e+V_0|)}$ are inserted in order to ensure the solutions (\ref{f450}) being purely real in the region $|\epsilon+V_0|<1$, where the Bessel functions are replaced by the corresponding Infeld ones.

For $ r>R_0 $ the most convenient form for solutions of (\ref{3.7}) is given by means of the Kummer $\F(b,c,z)$ and Tricomi $\P(b,c,z)$ functions ~\cite{Bateman1953}. Let us consider first the continuum spectra with $|\e| \geq 1$. For $|m_j| > Q$ the corresponding solutions take the following form
\beq
\label{f45}
\begin{aligned}
	&\p_{1,\, m_j}^{ext}(r,\e)= \sqrt{|\e + 1|}\,\, r^{\vk-1/2}\, \times \\ & \times \( \mathrm{Re}\[\mathrm{e}^{i\f_+}\mathrm{e}^{i k r} \F_r\]+B_{m_j}(\e)\, \mathrm{Re}\[i \mathrm{e}^{-i\pi\vk}\mathrm{e}^{i\f_-}\mathrm{e}^{i k r} \tilde \F_{r}\] \),
	\\
	&\p_{2,\, m_j}^{ext}(r,\e)= -\sign(\e)\sqrt{|\e - 1|}\,\, r^{\vk-1/2}\, \times \\& \times  \( \mathrm{Im}\[\mathrm{e}^{i\f_+}\mathrm{e}^{i k r} \F_r\]+B_{m_j}(\e)\, \mathrm{Im}\[i \mathrm{e}^{-i\pi\vk}\mathrm{e}^{i\f_-}\mathrm{e}^{i k r} \tilde \F_{r}\] \) \ ,
\end{aligned}
\eeq
where $\e = \pm\sqrt{k^2 + 1}$ for the upper and lower continua, correspondingly,
\beq\label{f45-1}\bal
& \vk=\sqrt{m_j^2-Q^2} \ , \quad b=\vk - i\e Q / k \ , \quad c=1+2\vk \ , \\
& \f_+=\1/2 \mathrm{Arg}\[{m_j+iQ/k\over b}\], \quad \f_-=\1/2 \mathrm{Arg}\[{b \over m_j-iQ/k}\] \ , \\
&\F_r=\F\(b,c,-2ikr\) \ ,  \\& \tilde \F_r=(-2ikr)^{1-c}\F\(1+b-c,2-c,-2ikr\) \ ,
\eal
\eeq
while the coefficients $B_{m_j}(\e)$ are derived from the matching condition,
imposed on internal and external solutions at $r=R_0$
\begin{widetext}
\beq\label{f45-2}
	B_{m_j}(\e)=-{C_{1,\, m_j}(\e)\, \mathrm{Im}\[\mathrm{e}^{i\f_+}\mathrm{e}^{i k R_0} \F_{R_0}\] - C_{2,\, m_j}(\e)\, \mathrm{Re}\[\mathrm{e}^{i\f_+}\mathrm{e}^{i k R_0} \F_{R_0}\] \over C_{1,\, m_j}(\e)\, \mathrm{Im}\[i \mathrm{e}^{-i\pi\vk}\mathrm{e}^{i\f_-}\mathrm{e}^{i k R_0} \tilde \F_{R_0}\] - C_{2,\, m_j}(\e)\, \mathrm{Re}\[i \mathrm{e}^{-i\pi\vk}\mathrm{e}^{i\f_-}\mathrm{e}^{i k R_0} \tilde \F_{R_0}\]} \ ,
\eeq\end{widetext}
where
\beq\label{f45-3}\bal
&C_{1,\,m_j}(\e)=-\sign(\e)\sqrt{|\e - 1|}\,\, \psi^{int}_{1, m_j}(R_0,\e ) \ , \\
&C_{2,\,m_j}(\e)=\sqrt{|\e + 1|}\,\, \psi^{int}_{2, m_j}(R_0,\e ) \ .\\
\eal\eeq

For $ r>R_0 $ and $|m_j| < Q$ the corresponding solutions of (\ref{3.7}) should be written as
\beq\label{f45-5}\bal
	&\p_{1,\, m_j}^{ext}(r,\e)=\sqrt{|\e + 1|}\, \times \\ &  \mathrm{Re}\[\mathrm{e}^{i\l_{m_j}(\e)} \mathrm{e}^{ikr}(2  k r)^{i|\vk|-\1/2}\(b \F_r(b+)+(m_j+i Q/k)\F_r\)\] \ ,
	\\
	&\p_{2,\, m_j}^{ext}(r,\e)=-\sign(\e)\sqrt{|\e - 1|}\, \times \\  & \mathrm{Re}\[i \,\mathrm{e}^{i\l_{m_j}(\e)} \mathrm{e}^{ikr}(2  k r)^{i|\vk|-\1/2}\(b \F_r(b+)-(m_j+i Q/k)\F_r\)\] \ ,
\eal\eeq
where
\beq\label{f45-6}\bal
	& |\vk|=\sqrt{Q^2-m_j^2} \ ,\quad b=i\(|\vk| - {\e Q \over k}\) \ ,\quad c=1+2i|\vk|  \ , \\
	& \F_r=\F\(b,c,-2ikr\) \ , \quad\F_r(b+)=\F\(b+1,c,-2ikr\) \ ,
\eal\eeq
\beq \label{f45-7}\bal
&\lambda_{m_j}(\e)= -\mathrm{Arg}\left[i \mathrm{e}^{i k R_0}(2 k R_0)^{i |\vk|}\(\(C_{2,\, m_j}+i C_{1,\, m_j}\) \times \right. \right.\\ & \left. \left. \times (m_j+iQ/k)\F_{R_0}+(C_{2,\, m_j}-i C_{1,\, m_j})b\F_{R_0}(b+) \)\right] \ .
\eal\eeq

The discrete levels with $-1 \leqslant \e <1$ are determined from the conditions of vanishing solutions at the spatial infinity combined with their matching at the point $r=R_0$. The internal  solutions  of (\ref{3.7}) remain the same as in (\ref{f450}).  For $|m_j|>Q$ the external solutions of (\ref{3.7})  are represented now in the form
\begin{multline}\nonumber
	\p_{1,\, m_j}^{ext}(r,\e)=\sqrt{1 + \e} \ \mathrm{e}^{-\g r} r^{-1/2 + \vk} \times \\ \times \[(Q/\g-m_j)\P(b+,c,z)+\P(b,c,z)\] \ ,
\end{multline}
\begin{multline}\label{f45-71}	
	\p_{2,\, m_j}^{ext}(r,\e)=\sqrt{1 - \e} \ \mathrm{e}^{-\g r} r^{-1/2 + \vk} \times \\ \times \[ (Q/\g-m_j)\P(b+,c,z)-\P(b,c,z)\] \ ,
\end{multline}
where $\g=\sqrt{1-\e^2}$ \ ,
\beq\label{f45-72}
z=2\g r \ , \quad b=\vk-\e Q/\g \ , \quad c=1+2\vk \ ,
\eeq
while the equation for the energy eigenvalues takes the form
\begin{widetext}\begin{multline}
	\label{f45-8}
	\sqrt{(\e+V_0+1)(1-\e)}\,J_{m_j-1/2}(\z R_0)\,\[-\P(b,c,2 \g R_0)+(Q/\g-m_j)\,\P(b+1,c,2 \g R_0)\] + \\
	+\sqrt{(\e+V_0-1)(1+\e)}\,J_{m_j+1/2}(\z R_0)\,\[\P(b,c,2 \g R_0)+(Q/\g-m_j)\,\P(b+1,c,2 \g R_0)\]=0 \ .
\end{multline}\end{widetext}

For $|m_j|< Q$ the external solutions of (\ref{3.7})  should be written as
\begin{multline}\nonumber
	\p_{1,\, m_j}^{ext}(r,\e)=\sqrt{1 + \e} \ \mathrm{e}^{-\g r}\,
\mathrm{Re}\[\mathrm{e}^{i\l} (2 \g r)^{i|\vk|-\1/2} \times \right.\\  \left. \times \((m_j+ Q/\g)\F(b,c,z)+b \F(b+1,c,z)\)\] \ ,
\end{multline}	
\begin{multline}\label{f45-81}
	\p_{2,\, m_j}^{ext}(r,\e)=\sqrt{1 - \e} \ \mathrm{e}^{-\g r}\,
\mathrm{Re}\[\mathrm{e}^{i\l} (2 \g r)^{i|\vk|-\1/2} \times \right. \\ \left. \times \(-(m_j+ Q/\g)\F(b,c,z)+b \F(b+1,c,z)\)\] \ ,
\end{multline}	
where $b=i|\vk|-\e Q/\g$, $c=1+2i|\vk|$. The phase $\l$ is determined via matching the internal and external solutions. The equation for discrete levels follows from the condition of vanishing for  $r \to \infty$ and reads

\begin{widetext}\begin{multline}
\label{f45-9}
\mathrm{Im}\Big[(2\g R_0)^{i|\vk|}\, \G(c^*)\G(b)\\
\times\Big(\sqrt{(\e+V_0+1)(1-\e)}\,J_{m_j-1/2}(\z R_0)\,\(-(Q/\g+m_j)\F(b,c,2 \g R_0)+b\, \F(b+1,c,2 \g R_0)\)\\
+\sqrt{(\e+V_0-1)(1+\e)}\,J_{m_j+1/2}(\z R_0)\,\((Q/\g+m_j)\F(b,c,2 \g R_0)+b\, \F(b+1,c,2 \g R_0)\)\Big)\Big]=0 \ .
\end{multline}\end{widetext}

For the given $|m_j|$ the total phase shift $\delta_{tot,|m_j|}(k)$, which includes the contributions from both continua and  $\pm m_j$, is determined via
\beq
\label{f45-10}
 \d_{tot,|m_j|}(k)=\sum \d^{\pm}_{\pm |m_j|}(k) \ .
\eeq
 Separate phase shifts are found from the asymptotics of solutions (\ref{f45}) or (\ref{f45-5}) for $r\to\infty$ and contain the Coulomb logarithms $\pm Q\, (|\e|/k)\, \ln (2 k r)$, which cancel mutually in the total phase  (\ref{f45-10}) and henceforth will be omitted in the expressions for separate shifts (\ref{f45-11}),(\ref{f45-12}).

For $|m_j|>Q$ the  separate  shifts turn out to be
 \beq\label{f45-11}\bal
&	\d_{m_j}(k)=\mathrm{Arg}\[\mathrm{e}^{(\pi i/2)|m_j|} \times \right. \\ & \times \left. \( {\mathrm{e}^{i\f_+}\mathrm{e}^{-i \pi\vk/2} \over \G(1+b^*)} + i B_{m_j}(\e) {\G(2-c)\over \G(c)} {\mathrm{e}^{i\f_-}\mathrm{e}^{i \pi\vk/2} \over  \G(1-b)}\)\] \ ,
\eal\eeq
while for $|m_j|<Q$
\beq
\label{f45-12}
\begin{aligned}
&	\delta_{m_j}(k)=\mathrm{Arg}\[\mathrm{e}^{(\pi i/2)|m_j|}  \times \right. \\ & \times \left. \( {(m_j+i Q/k)\Gamma(c) \over \Gamma(c-b)}\,\mathrm{e}^{i\lambda_{m_j}(\e)}\mathrm{e}^{\pi|\varkappa|} +  {\Gamma(c^*)\over \Gamma(b^*)} \,\mathrm{e}^{-i\lambda_{m_j}(\e)}\)\] \ .\\
\end{aligned}
\eeq
Besides  Coulomb logarithms, the separate shifts (\ref{f45-11}),(\ref{f45-12}) contain still singular terms both in IR and UV-limits. However, in the total partial phase all these singularities disappear. In particular, the infrared asymptotics of separate shifts contains the singular terms $\pm Q/k(1-\ln(Q/k))$, which cancel each other in the total phase. Therefore, the total phase is finite for $k \to 0$. The exact infrared asymptotics of $\d_{tot,|m_j|}(k)$ for $|m_j|< Q$ reads
\begin{widetext}\beq
\label{f45-13}
\d_{tot,|m_j|}(k\to 0)=\mathrm{Arg}\[-\(\mathrm{e}^{\pi|\vk|}\mathrm{e}^{i\vf^+_{|m_j|}}
-\mathrm{e}^{-\pi|\vk|}\mathrm{e}^{-i\vf^+_{|m_j|}}\)\(\mathrm{e}^{\pi|\vk|}\mathrm{e}^{i\vf^+_{-|m_j|}}
-\mathrm{e}^{-\pi|\vk|}\mathrm{e}^{-i\vf^+_{-|m_j|}}\)\sin(\vf^-_{|m_j|})\sin(\vf^-_{-|m_j|})
\] \ ,
\eeq
where
\beq\label{f45-14}\bal
&	\vf^+_{\pm|m_j|}=-\mathrm{Arg}\[\pm\sqrt{2QR_0}J_{1+2 i|\vk|}(\sqrt{8QR_0})J_{|m_j|\pm1/2}(R_0\sqrt{V_0(V_0+2)})+J_{2 i|\vk|}(\sqrt{8QR_0})w^+_{\pm|m_j|}\] \ , \\
&	\vf^-_{\pm|m_j|}=-\mathrm{Arg}\Big[(-i)^{(|m_j|-1/2)\tt(2-V_0)}\Big(\sqrt{-2QR_0}J_{1+2 i|\vk|}(\sqrt{-8QR_0})J_{|m_j|\mp1/2}(R_0\sqrt{V_0(V_0-2)})\mp J_{2 i|\varkappa|}(\sqrt{-8QR_0})w^-_{\pm|m_j|}\Big)\Big] \ ,
\eal\eeq
with the coefficients
\beq\label{f45-15}\bal
	&w^+_{\pm|m_j|}=Q\sqrt{V_0+2\over V_0}J_{|m_j|\mp1/2}(R_0\sqrt{V_0(V_0+2)})-(|m_j|\pm i|\vk|)J_{|m_j|\pm1/2}(R_0\sqrt{V_0(V_0+2)}) \ ,\\
	&w^-_{\pm|m_j|}=Q\sqrt{V_0-2\over V_0}J_{|m_j|\pm1/2}(R_0\sqrt{V_0(V_0-2)})-(|m_j|\mp i|\vk|)J_{|m_j|\mp1/2}(R_0\sqrt{V_0(V_0-2)}) \ .
\eal\eeq
For $|m_j|> Q$ the exact infrared asymptotics of $\d_{tot,|m_j|}$ takes the form
\beq\label{f45-16}
\d_{tot,|m_j|}(k\to 0)=\mathrm{Arg}\[-\mathrm{e}^{-2i\pi\vk}v_{1,+} v_{1,-} v_{2,+} v_{2,-}\] \ ,
\eeq
where
\beq
\label{f45-17}
\bal
& v_{1,\pm}=J_{|m_j|\mp1/2}(R_0\sqrt{V_0(V_0+2)})\,\(\mp J_{-2\vk}(\sqrt{8 Q R_0})\pm \mathrm{e}^{2i\pi\vk}J_{2\vk}(\sqrt{8 Q R_0})\)\sqrt{(V_0+2)/ V_0} \ + \\ & + J_{|m_j|\pm1/2}(R_0\sqrt{V_0(V_0+2)})\Big[\(\sqrt{2 Q R_0}J_{1+2\vk}(\sqrt{8 Q R_0})+(\mp|m_j|-\vk)J_{2\vk}(\sqrt{8 Q R_0})\)\mathrm{e}^{2i\pi\vk} \  - \\ &  -\(\sqrt{2 Q R_0}J_{1-2\vk}(\sqrt{8 Q R_0})+(\mp|m_j|+\vk)J_{-2\vk}(\sqrt{8 Q R_0})\)\Big]/Q \ ,\\
& v_{2,\pm}=\mathrm{Im}\[(-i)^{(|m_j|\mp1/2)\theta(2-V_0)}J_{|m_j|\mp1/2}(R_0\sqrt{V_0(V_0-2)})\(J_{-2\vk}
 (\sqrt{-8 Q R_0})\mathrm{e}^{-i\pi\vk}-J_{2\vk}(\sqrt{-8 Q R_0})\mathrm{e}^{i\pi\vk}\)\] \ .
\eal\eeq
For $k\to\infty$ the asymptotics of separate  shifts (\ref{f45-11}),(\ref{f45-12}) contains additional logarithms $\mp Q(|\e|/k)\ln(2 k R_0)$, which in the total phase also cancel each other. As a result, the final UV-asymptotics of $\d_{tot,|m_j|}(k)$ turns out to be a decreasing one and equals to
\beq
\label{f45-18}
\d_{tot,|m_j|}(k\to \infty)=
{Q\over R^3_0 k^3}\({4 Q\over 3}(m_j^2-3R^2_0)-|m_j|\cos(2Q+\pi |m_j|)\sin(2 k R_0)\) + O(1/k^4) \ .
\eeq
In addition, the asymptotics (\ref{f45-18}) indicates  that for the  point-like Coulomb source the  method of calculating the vacuum energy, based on transformation of the contribution from the continua into the phase integral, is not valid (at least in the present form), since for $R_0 \to 0$ the evaluation of the latter becomes ambiguous. For more details concerning this circumstance see Ref.~\cite{Davydov2018b}.
\end{widetext}

\begin{figure*}[ht!]
\subfigure[]{\label{pic:1}
		\includegraphics[width=\columnwidth]{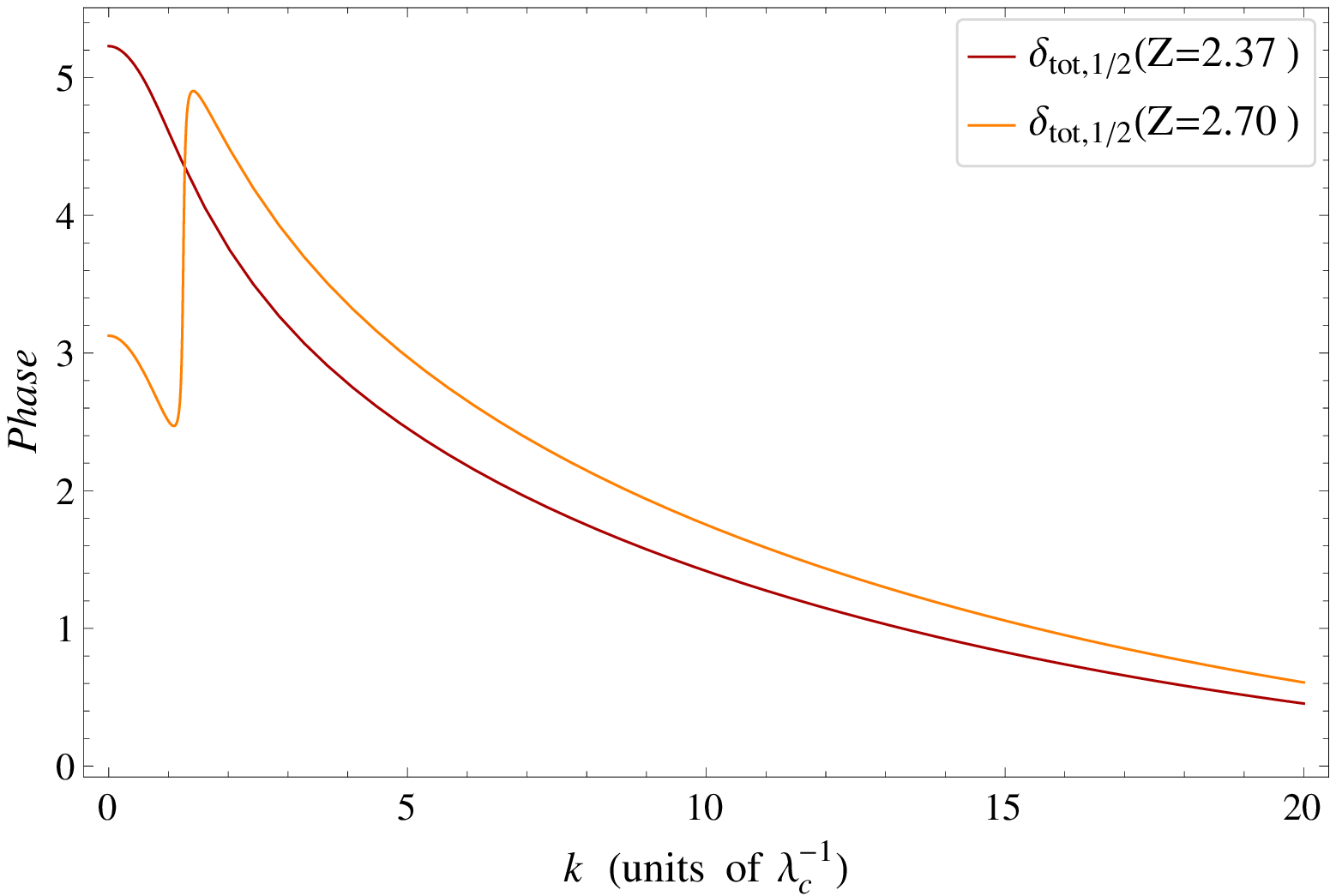}
}
\hfill
\subfigure[]{\label{pic:2}
		\includegraphics[width=\columnwidth]{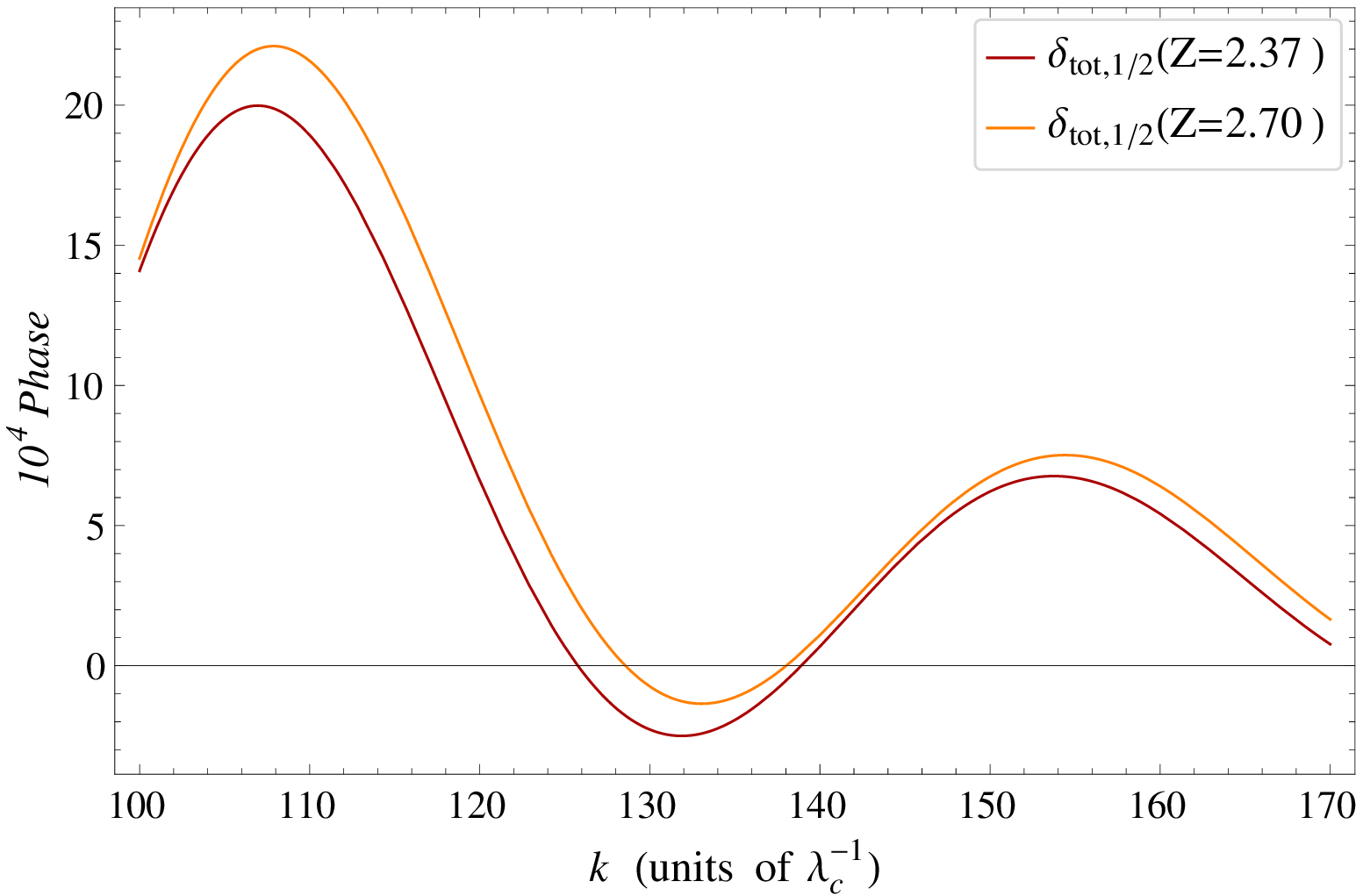}
}
	\caption{(Color online) \small $\d_{tot,|m_j|}(k) $ for $\a=0.4\, , R_0=1/15\, , |m_j|=1/2$  and  $Z=2.37\, ,  Z=2.70 $ on certain intervals of $k$.}\label{pic:1-2}
\end{figure*}

\begin{figure*}[ht!]
\subfigure[]{\label{pic:3}
		\includegraphics[width=\columnwidth]{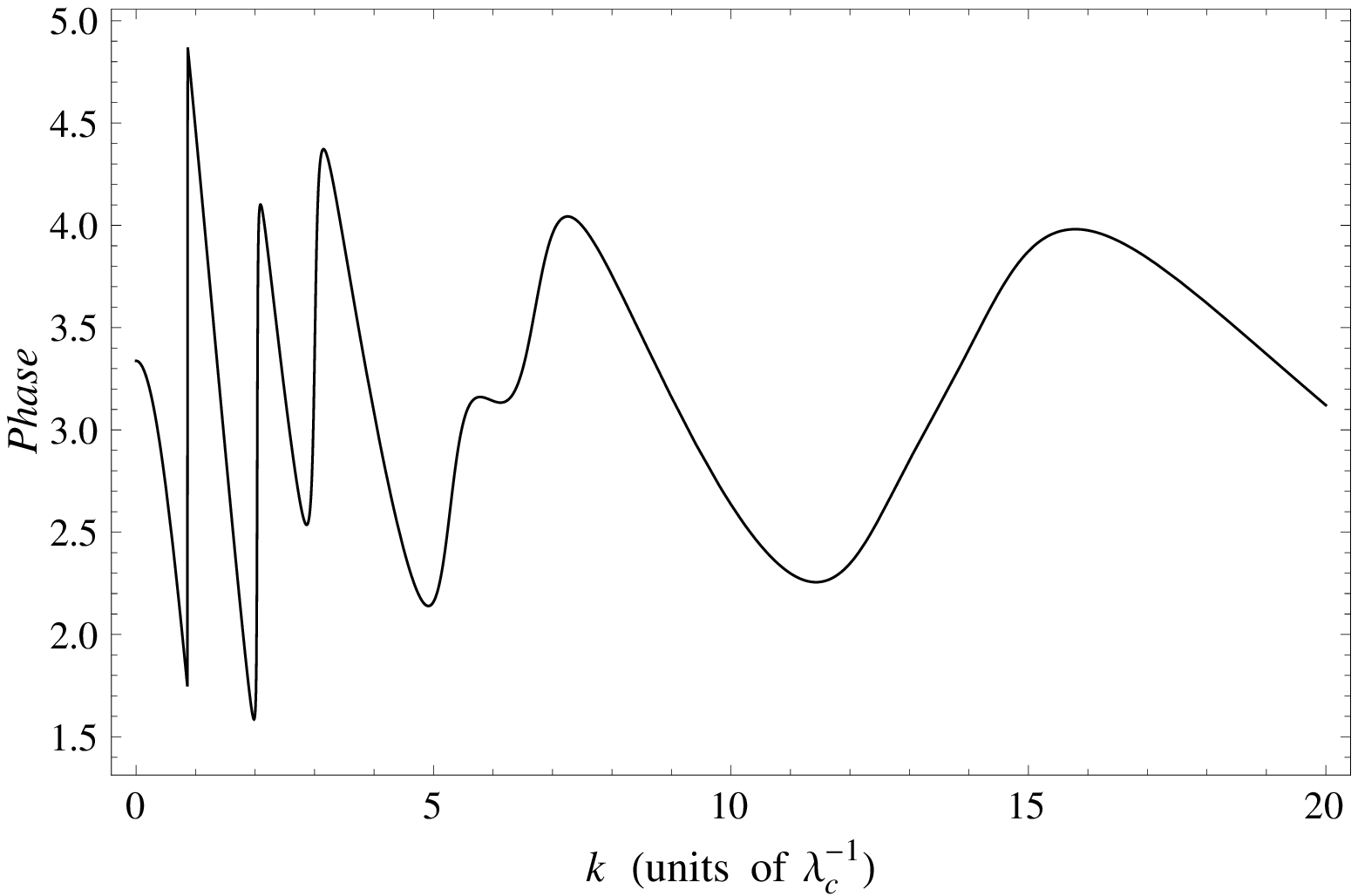}
}	
\hfill
\subfigure[]{\label{pic:4}
		\includegraphics[width=\columnwidth]{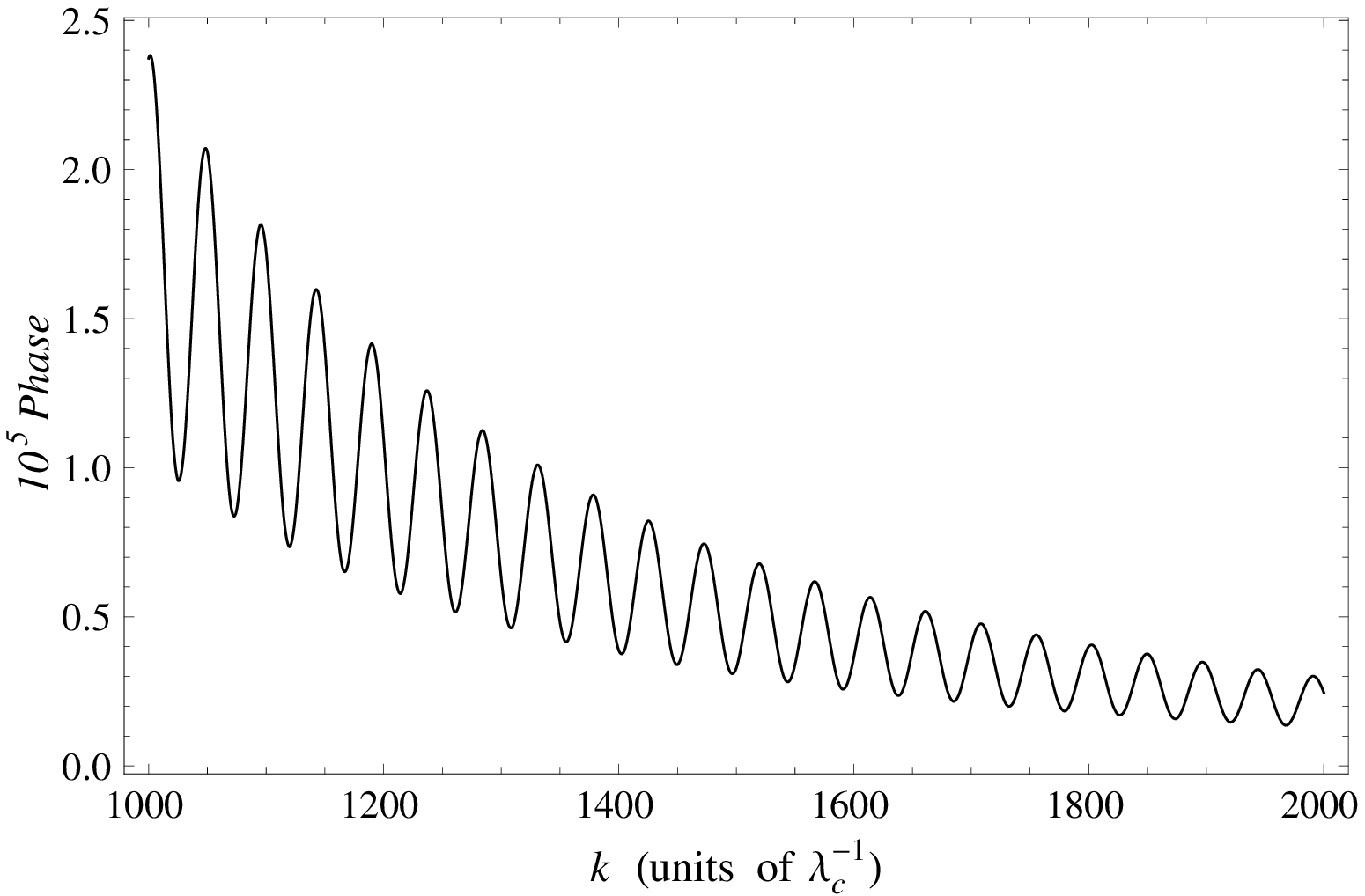}
}
\vfill
\subfigure[]{\label{pic:5}
	\includegraphics[width=\columnwidth]{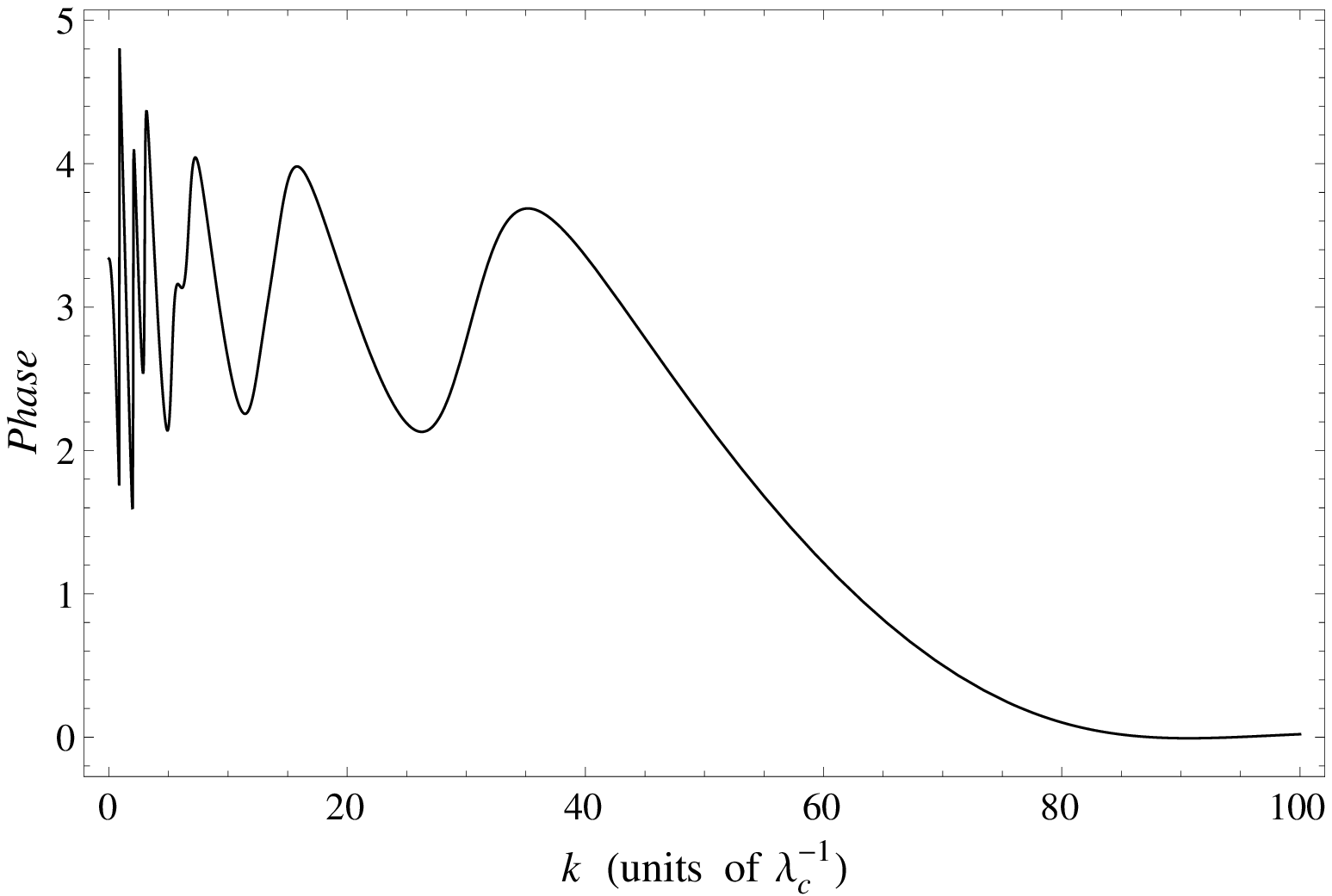}
}
\caption{\small $ \d_{tot,|m_j|}(k) $ for $\a=0.4$, $R_0=1/15$, $|m_j|=1/2$ and $Z=10$. }\label{pic:3-5}
\end{figure*}

The most significant feature in the behavior of $ \d_{tot,|m_j|}(k) $ is the emergence of elastic resonances upon diving of discrete levels into the lower continuum. The typical behavior of the partial total phase with $ |m_j|=1/2 $  as a function of the wavenumber $k$ is shown in Figs.\ref{pic:1-2},\ref{pic:3-5} for $\a=0.4$, $R_0=1/15$, and certain values of $Z$. Figs.\ref{pic:1},\ref{pic:2} represent the total phase $ \d_{tot,1/2}(k) $ for $ Z=2.37 $ and $Z=2.70$ on separate intervals of $k$. For $ Z=2.37 $ none of the discrete levels have reached the lower continuum yet. For $ Z=2.70 $ the first discrete level has already dived into the lower continuum, and hence, there appears in the phase the first and yet sufficiently narrow low-energy elastic resonance. With growing $ Z $ the resonances, which initially show up in the phase as the jumps by $\pi$, undergo  broadening and move towards the higher $k$. Figs.\ref{pic:3},\ref{pic:4} demonstrate the behavior of the total partial phase $ \d_{tot,1/2}(k) $ at small and large values of $ k $ for $Z=10$.  As it follows from Figs.\ref{pic:2},\ref{pic:4},  for large $k$ the total partial phase is a decreasing and oscillating function of $k$.     Fig.\ref{pic:5} represents the  behavior of $ \d_{tot,1/2}(k) $ including the effects from all the nine discrete levels dived into the lower continuum (here the common degeneracy factor 2 is dropped).

\begin{figure*}[ht!]
\subfigure[]{\label{pic:6}
		\includegraphics[width=\columnwidth]{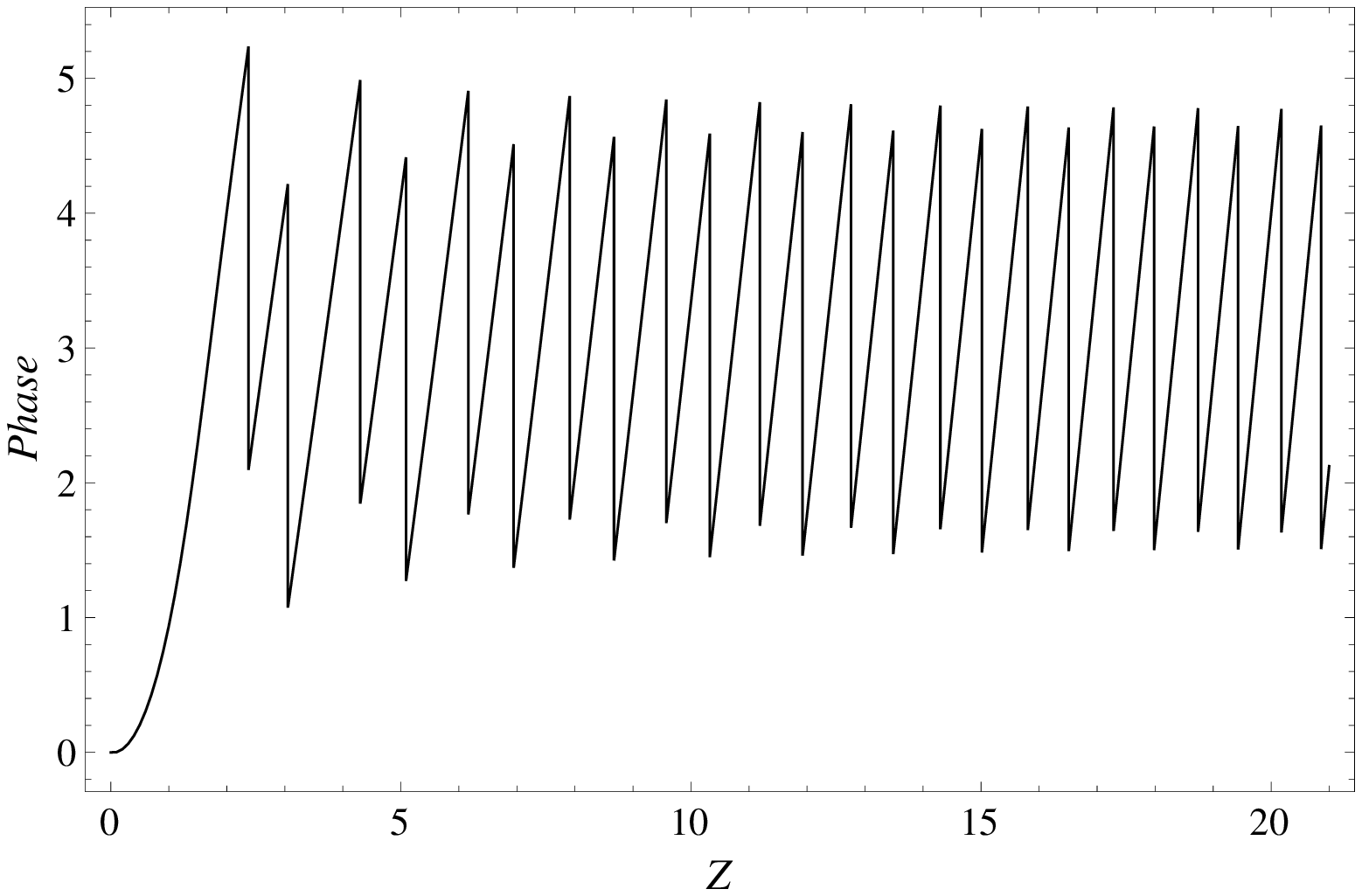}
}
\hfill
\subfigure[]{\label{pic:7}
		\includegraphics[width=\columnwidth]{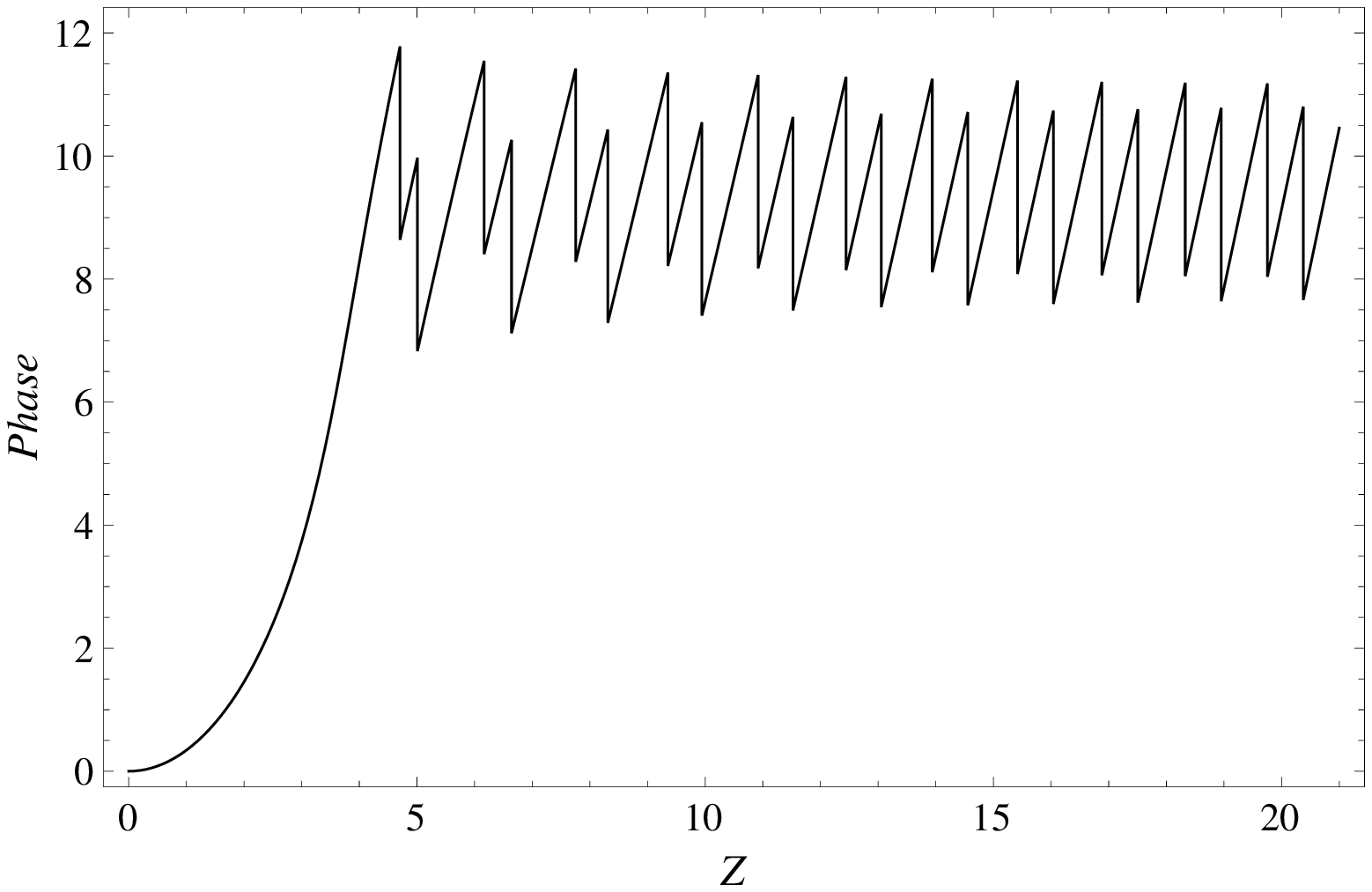}
}
\vfill
\subfigure[]{\label{pic:8}
	\includegraphics[width=\columnwidth]{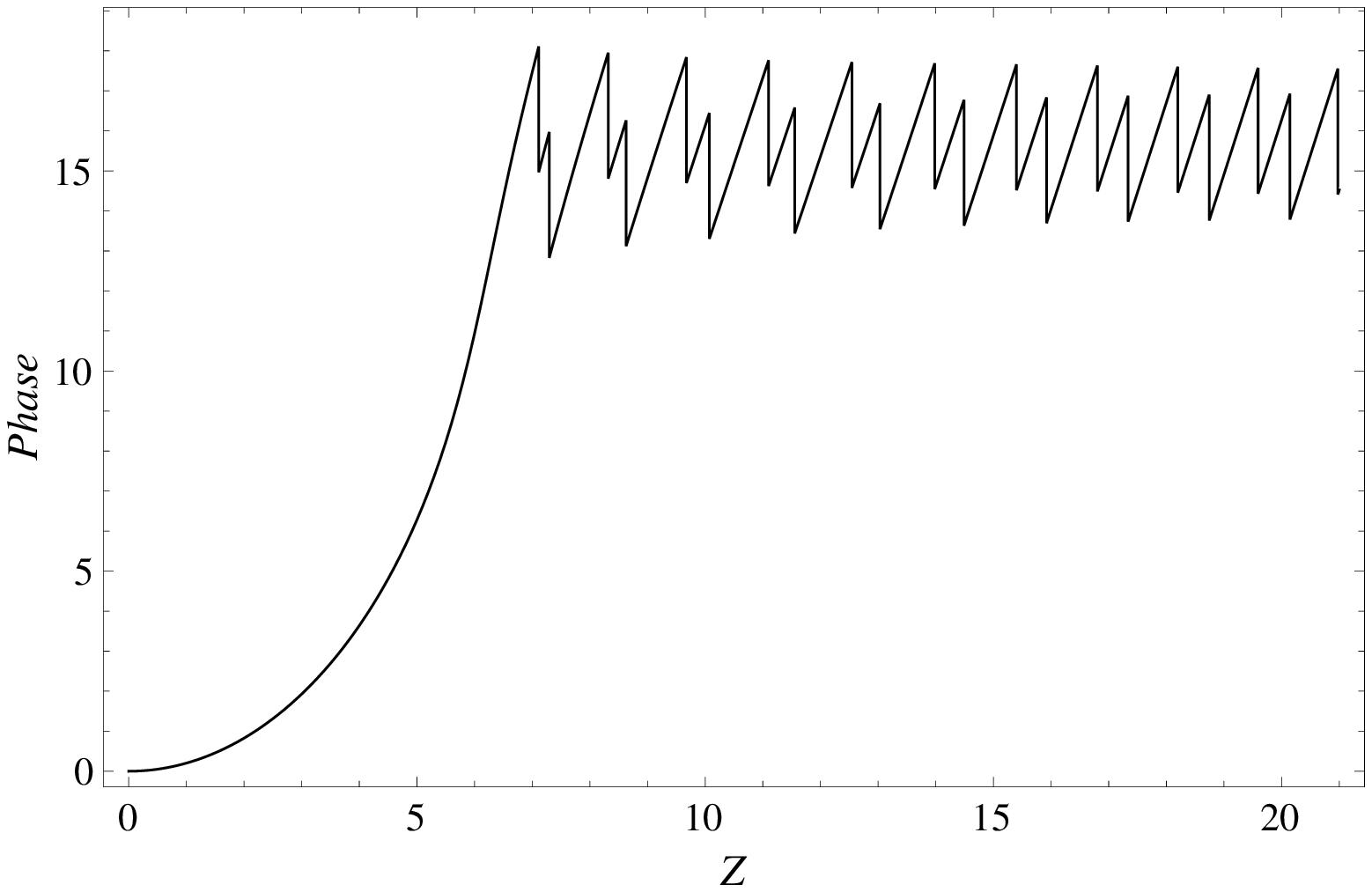}
}
\hfill	
\subfigure[]{\label{pic:9}
	\includegraphics[width=\columnwidth]{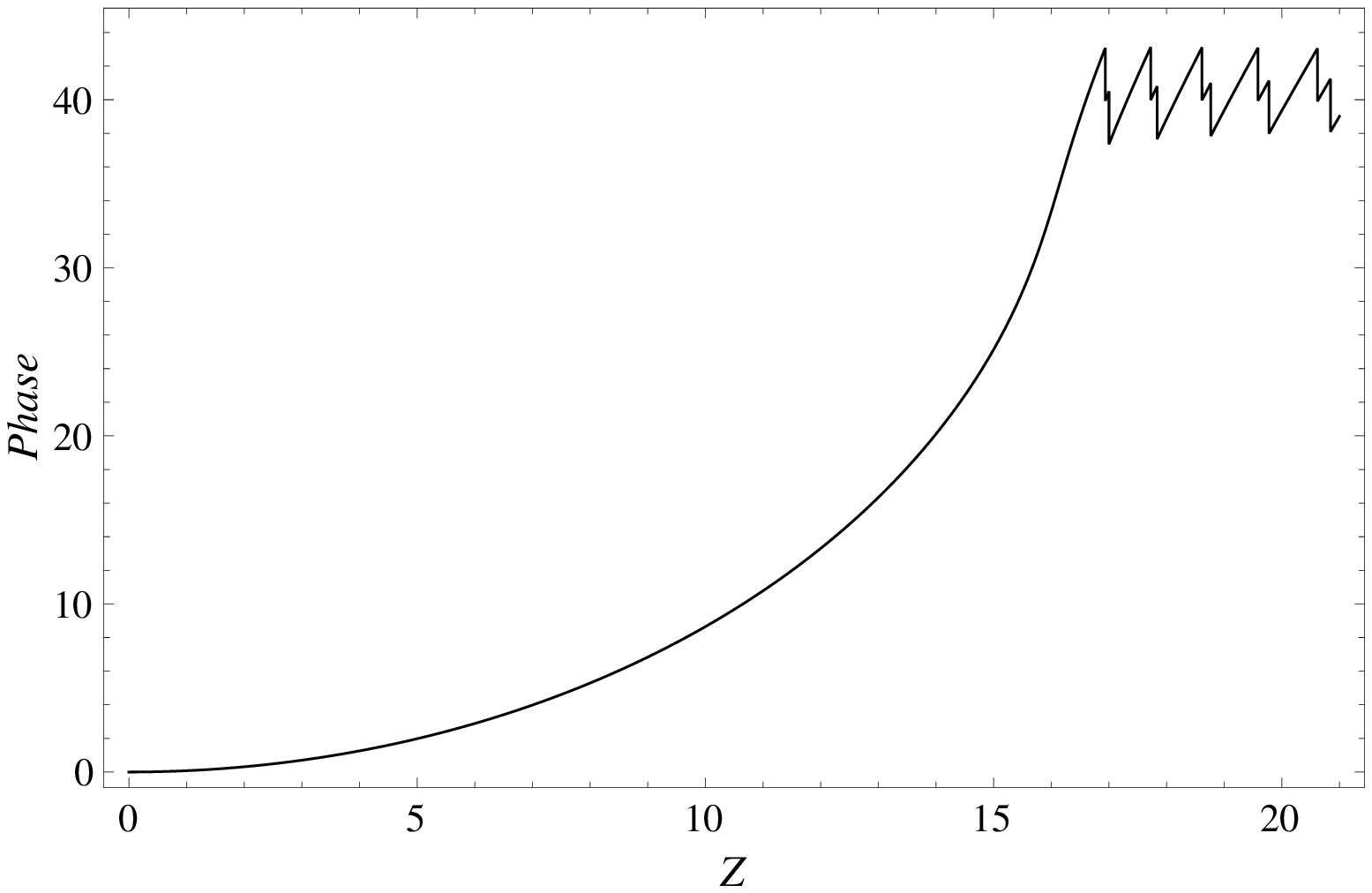}
}
\caption{\small $\delta_{tot,|m_j|}(0)$ for $\alpha=0.4$, $R_0={1\over15}$ and ~\subref{pic:6} $|m_j|={1\over 2}$, ~\subref{pic:7} $|m_j|={3\over 2}$, ~\subref{pic:8} $|m_j|={5\over 2}$, ~\subref{pic:9} $|m_j|={13\over 2}$.}\label{pic:6-9}
\end{figure*}

For other values of $|m_j|$ the partial total phases behave in the similar manner. The formation of  resonances also affects the dependence of $ \d_{tot,|m_j|}(0) $ on $Z$, but keeps it finite according to (\ref{f45-13})-(\ref{f45-17}). Figs.\ref{pic:6-9} show this dependence for  $\a=0.4$, $R_0=1/15$ and certain most representative values of  $|m_j|$. Thus,  $ \d_{tot,|m_j|}(k) $ is regular on the whole $k$-half-axis, while for $k \to \inf$ its rate of decrease  is fast enough to provide the convergence of partial phase integrals in (\ref{f41a}), which therefore can be evaluated via standard numerical recipes.

\begin{figure*}[ht!]
\subfigure[]{\label{pic:10}
		\includegraphics[width=\columnwidth]{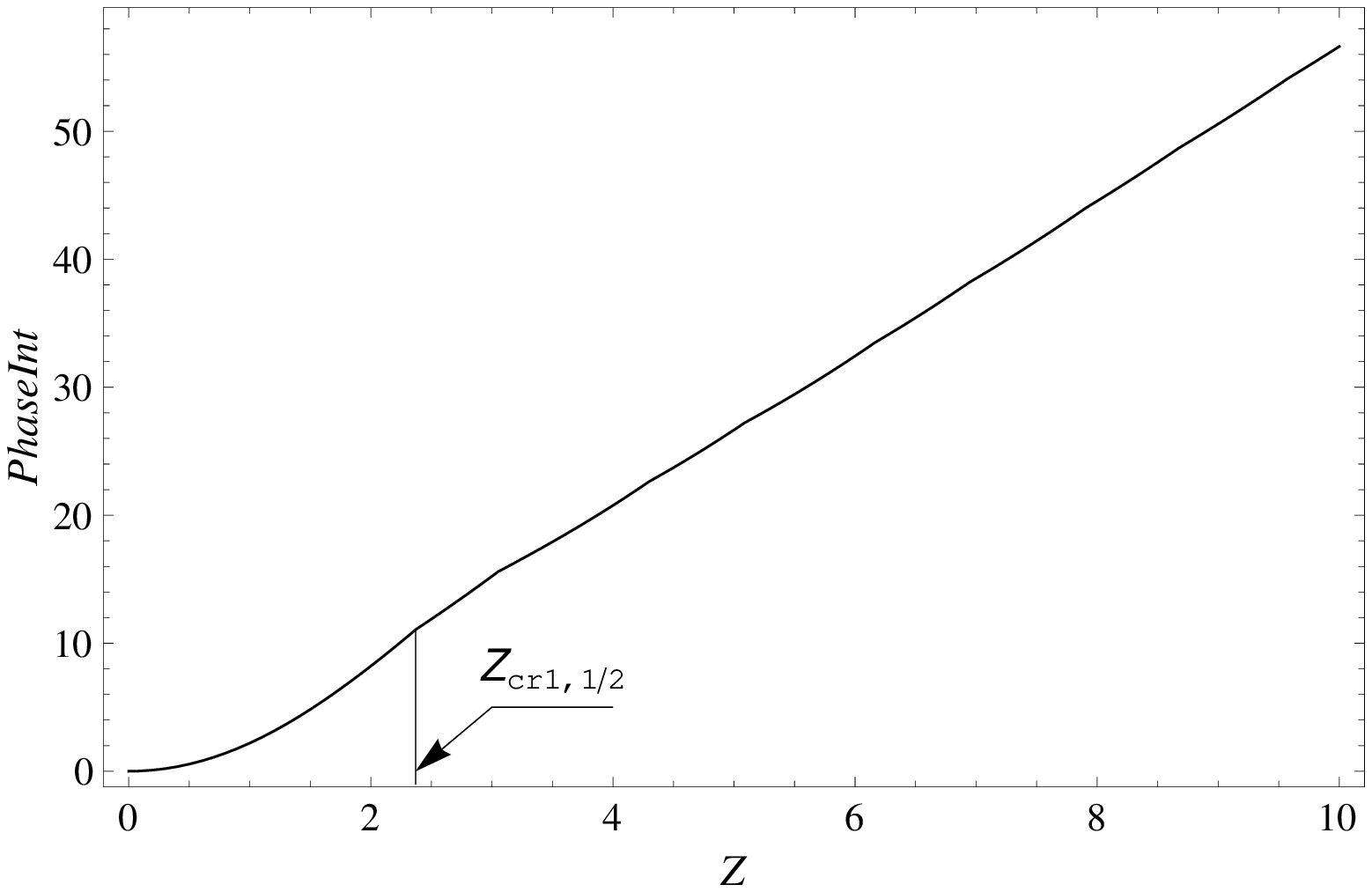}
}
\hfill
\subfigure[]{\label{pic:11}
		\includegraphics[width=\columnwidth]{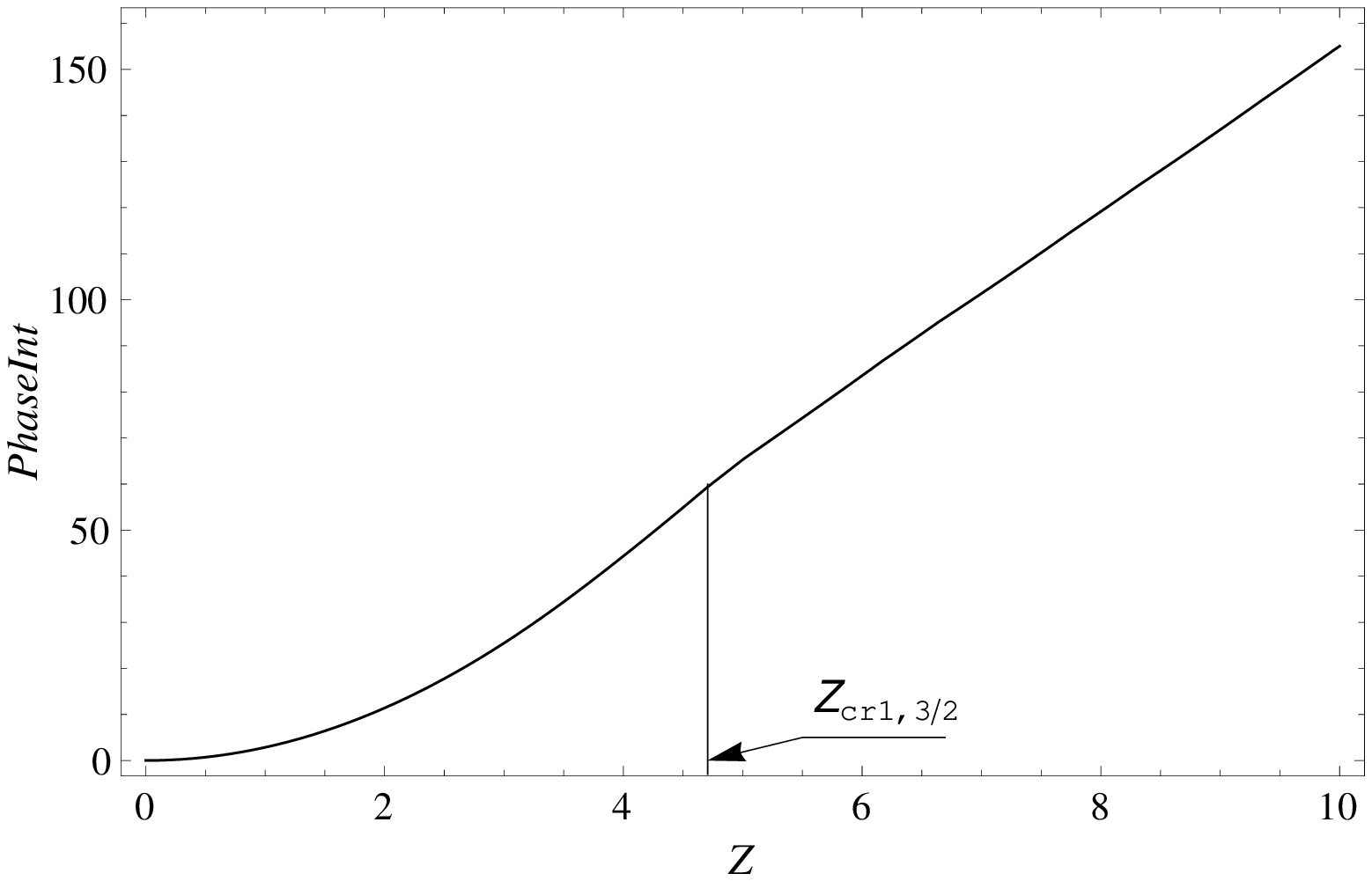}
}
	\caption{\small Phase integral as a function of $Z$ for $\alpha=0.4$, $R_0={1\over 15}$ and  ~\subref{pic:10} $|m_j|={1\over 2}$, ~\subref{pic:11} $|m_j|={3\over 2}$. }\label{pic:10-11}	
\end{figure*}

\begin{figure*}[ht!]
\subfigure[]{\label{pic:12}
		\includegraphics[width=\columnwidth]{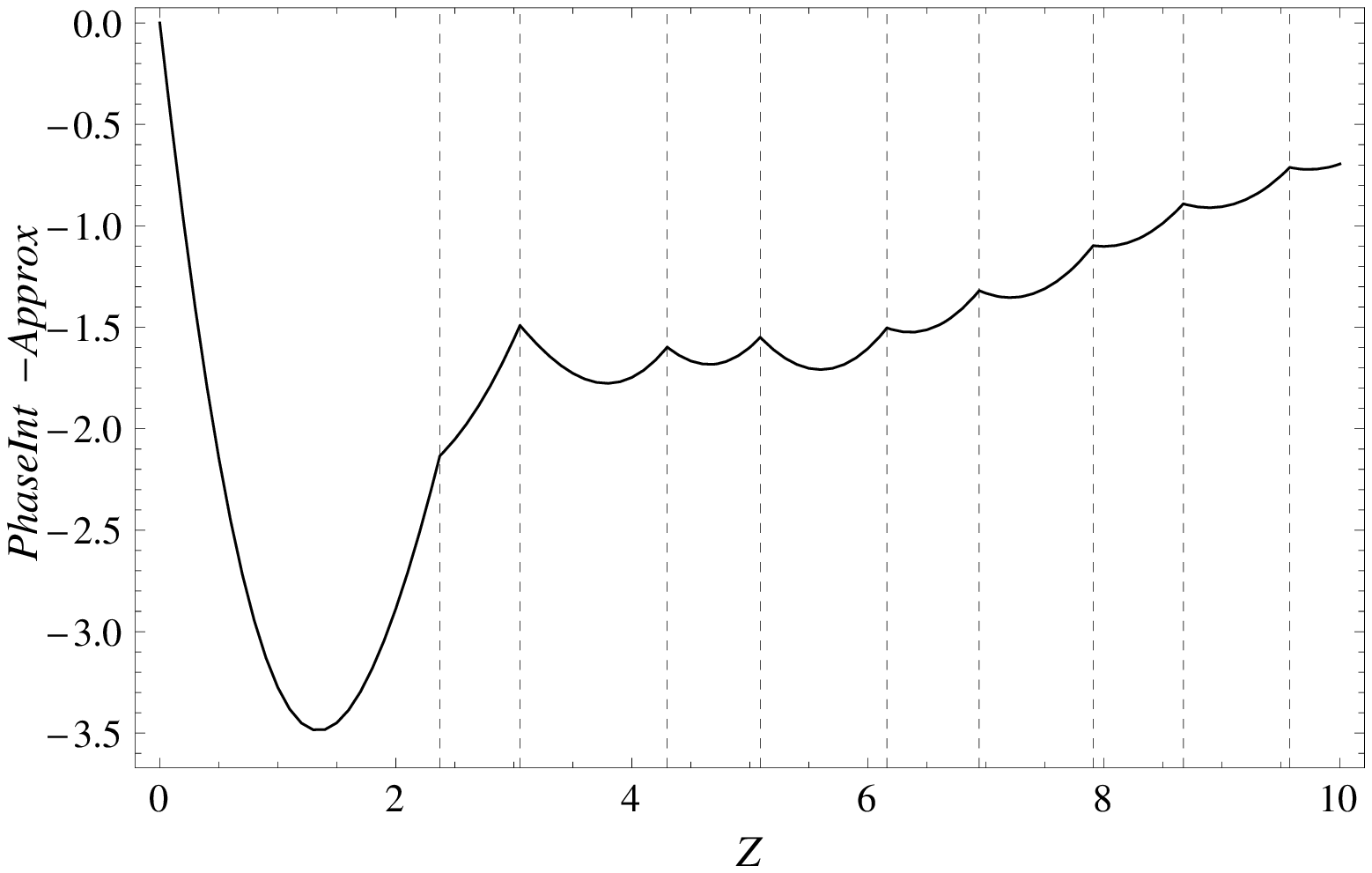}
}
\hfill
\subfigure[]{\label{pic:13}
		\includegraphics[width=\columnwidth]{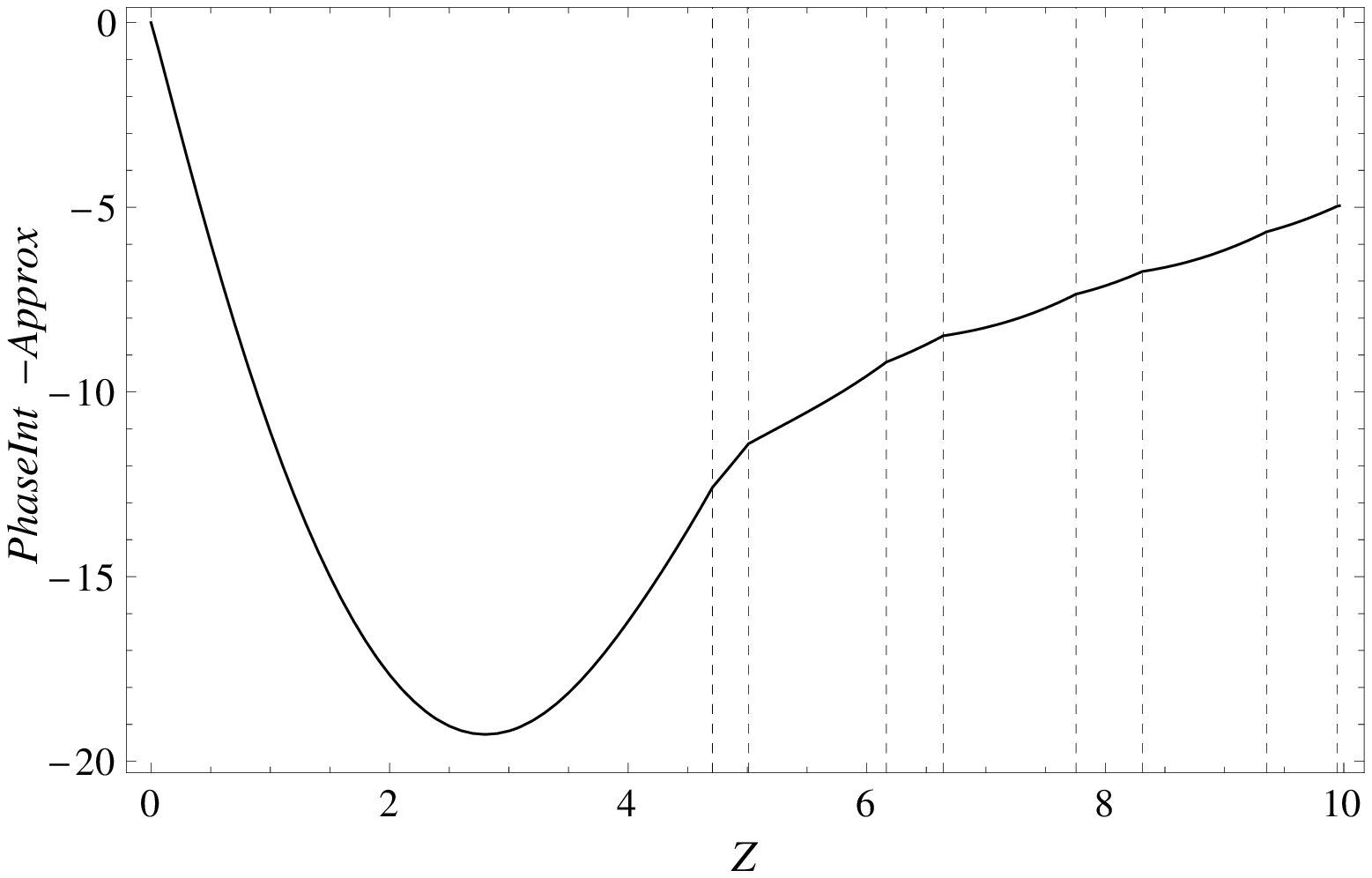}
}
	\caption{\small The difference between phase integral and its power approximation for $\alpha=0.4$, $R_0={1\over15}$ and ~\subref{pic:12} $|m_j|={1\over 2}$, ~\subref{pic:13} $|m_j|={3\over 2}$. The vertical lines show the positions of critical $Z$'s on the $Z$-axis. }\label{pic:12-13}
\end{figure*}

The typical behavior of the partial phase integral as a function of $ Z $ is shown in Figs.\ref{pic:10-11}. Each partial phase integral turns out to be a monotonically increasing and non-negative function of $Z$. In the perturbative region for   $Z \ll Z_{cr}$ it shows up a square growth, which in Fig.\ref{pic:10} is estimated as $\sim 2.219\, Z^2$, in  Fig.\ref{pic:11} as $\sim 2.956\, Z^2$. However, upon the start of  discrete levels diving into the lower continuum the behavior of the phase integral changes remarkably. For large $Z$ the each partial phase integral can be approximated by an almost linear function. In particular, for the cases shown in Figs.\ref{pic:10-11} the behavior of  phase integrals is estimated as $\sim 5.5\, Z^{1.02}$ and $\sim 20.3\, Z^{1.09}$ for $Z\gg Z_{cr1,m_j} $ in the partial channels under consideration. It is the significant difference of the two-dimensional problem from the one-dimensional one, wherein there is no  growing component in the phase integral at all, at least for the considered in ~\cite{Davydov2017, Sveshnikov2017, Voronina2017} values of $Z$. On the other hand, there is a common feature in the behavior of phase integrals for one-dimensional and two-dimensional problems, namely, their derivatives turn out to be discontinuous at each $Z_{cr}$. It is mostly clearly seen in Figs.\ref{pic:12-13}, which display the difference between the phase integrals and their power approximations, given above.

\begin{figure*}[ht!]
\subfigure[]{\label{pic:14}
		\includegraphics[width=\columnwidth]{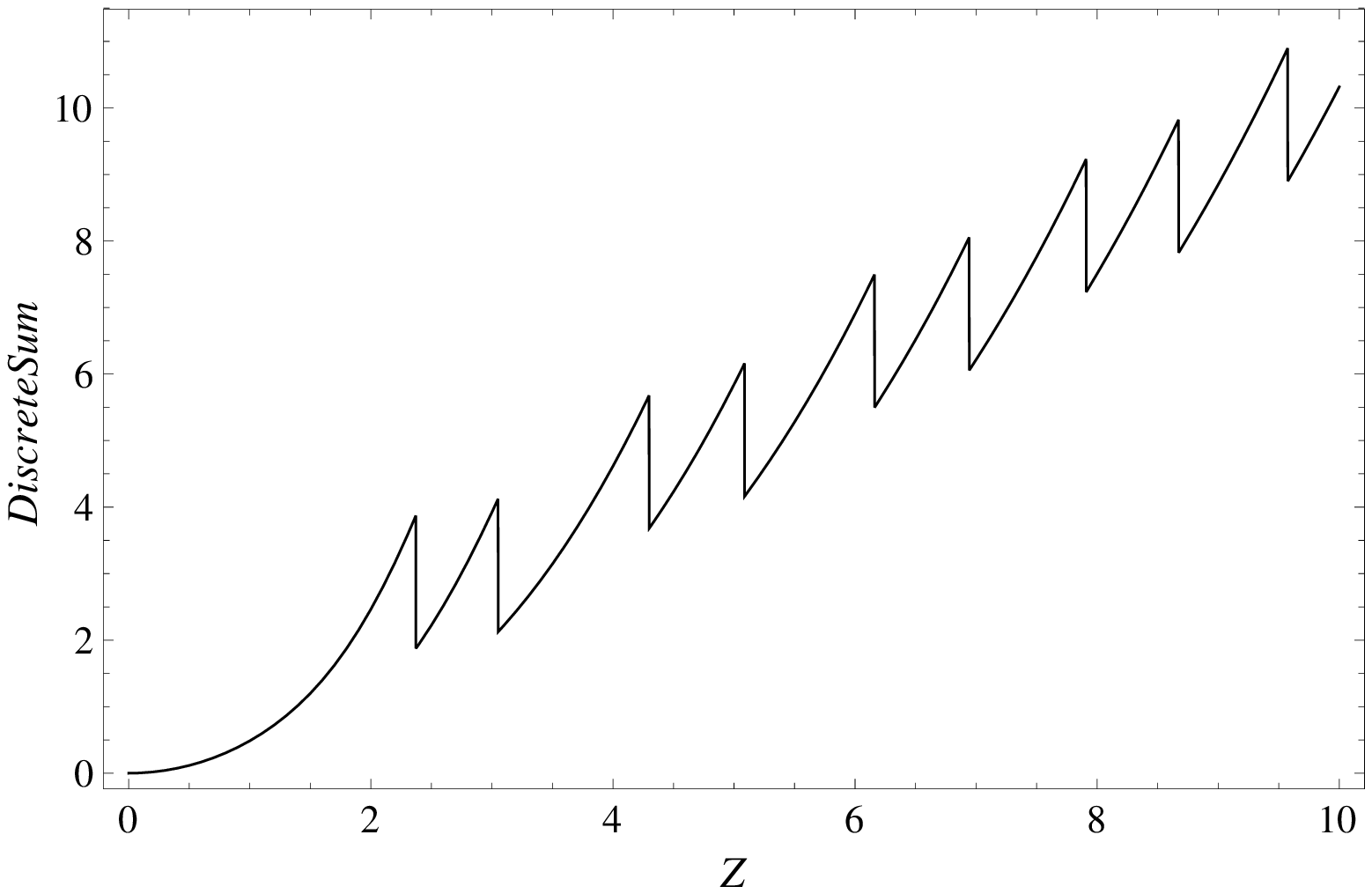}
}
\hfill
\subfigure[]{\label{pic:15}
		\includegraphics[width=\columnwidth]{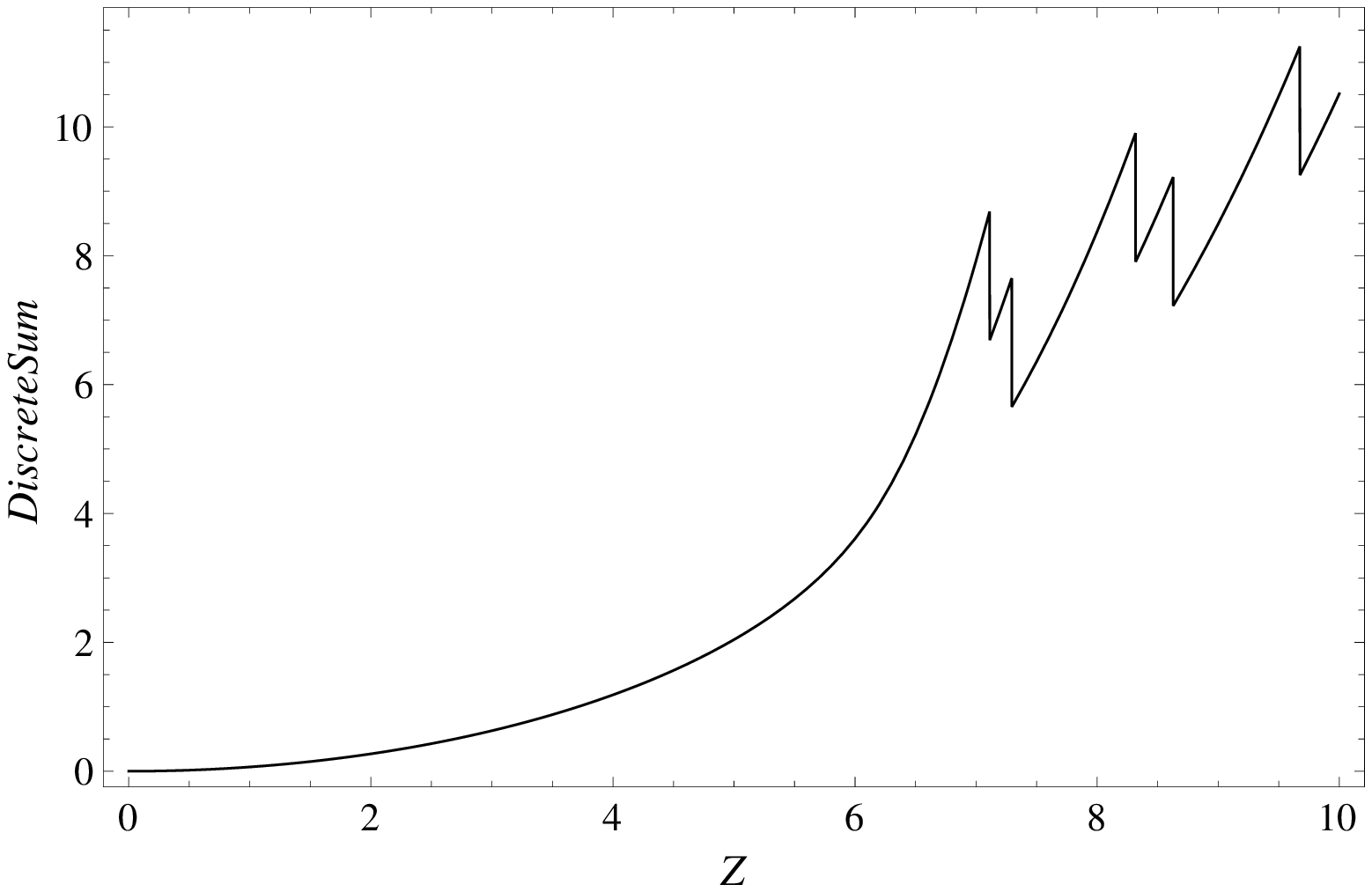}
}
\vfill
\subfigure[]{\label{pic:16}
		\includegraphics[width=\columnwidth]{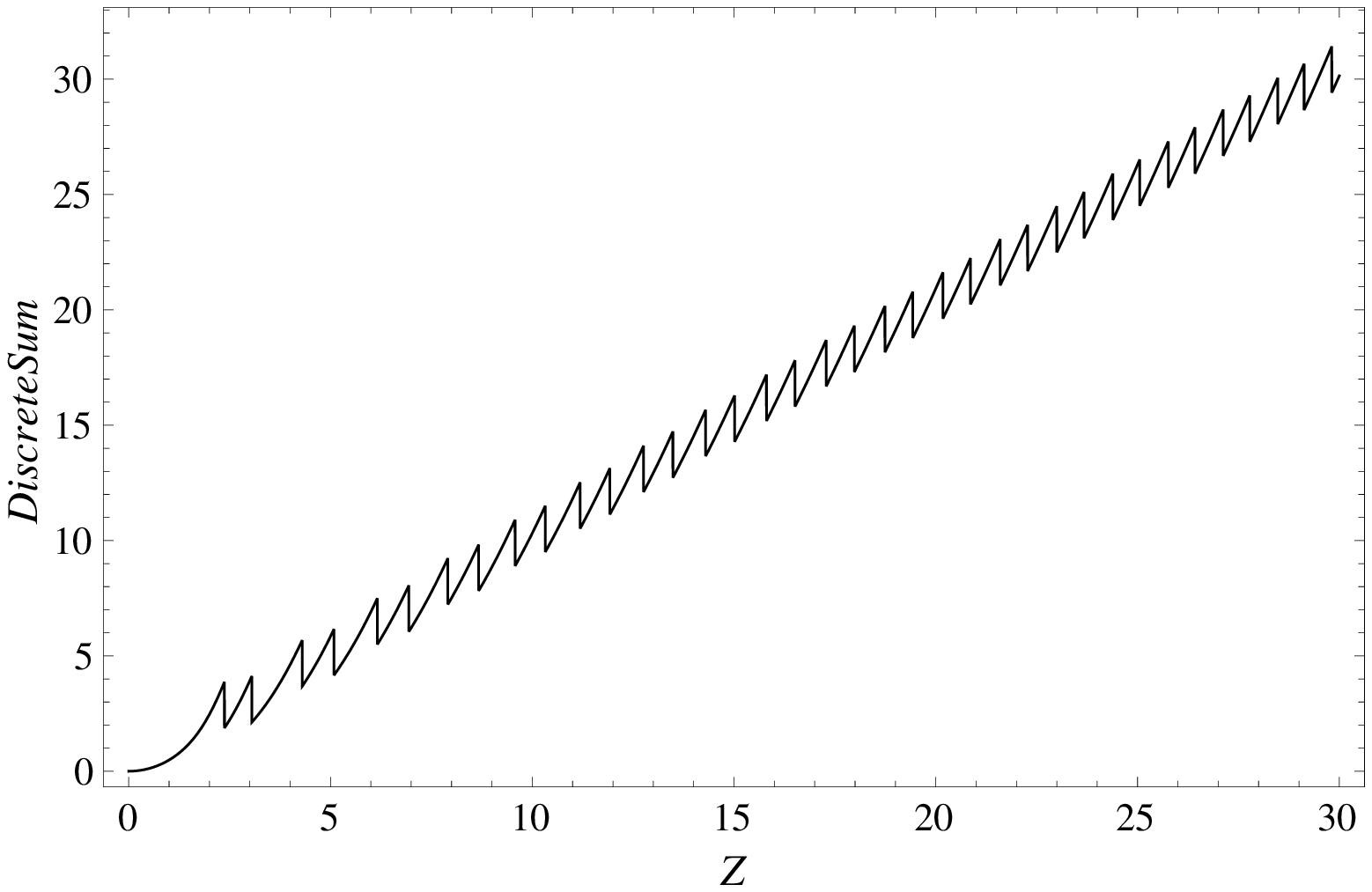}
}
\hfill
\subfigure[]{\label{pic:17}
		\includegraphics[width=\columnwidth]{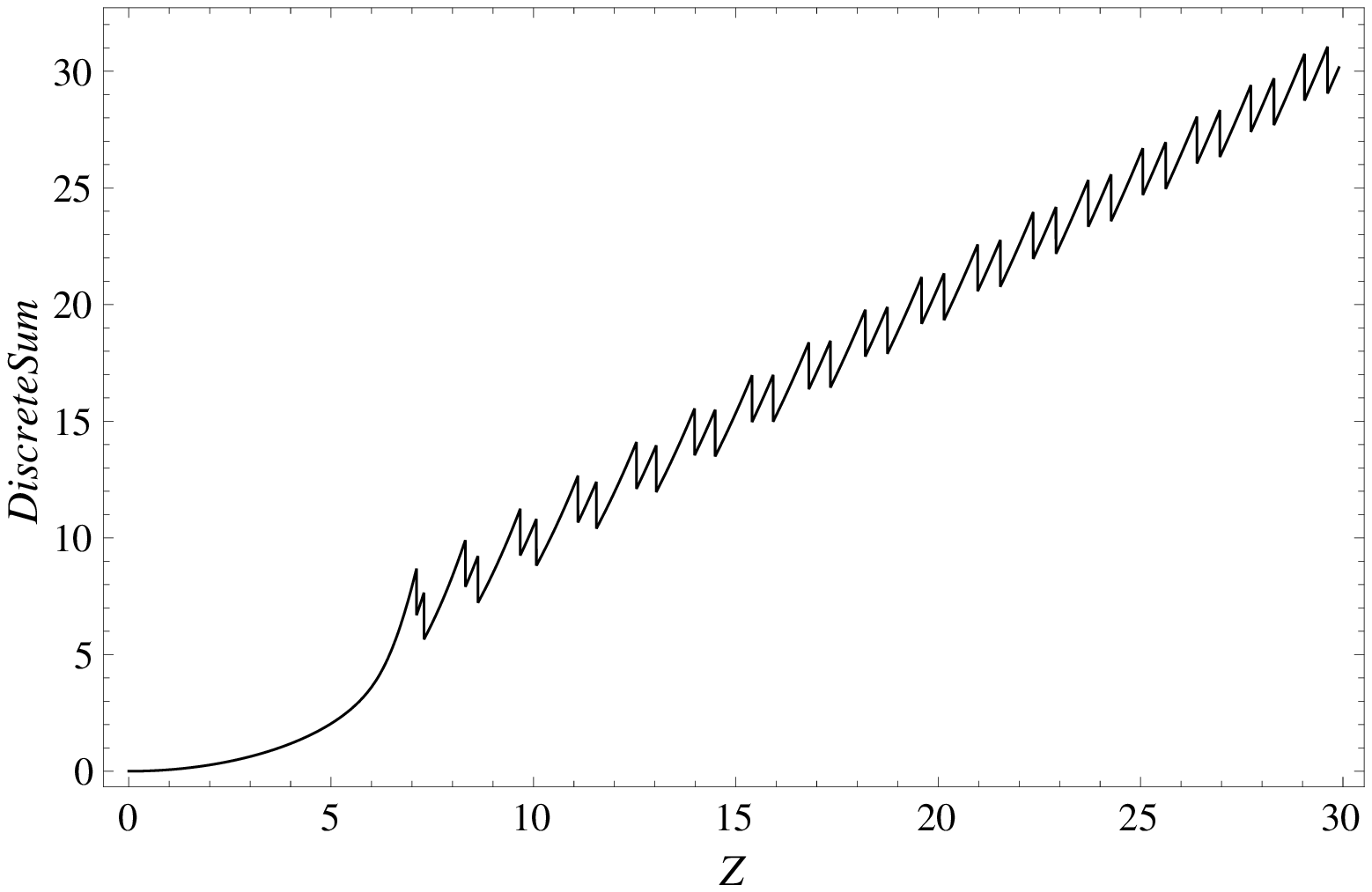}
}
	\caption{\small The total bound energy of discrete levels as a function of $ Z $ for $\a=0.4$, $R_0=1/15$ and ~\subref{pic:14}, ~\subref{pic:16} $ |m_j|=\1/2 $, ~\subref{pic:15}, ~\subref{pic:17} $|m_j|={5 \over 2}$.}\label{pic:14-17}	
\end{figure*}

The typical behavior of the partial total bound energy of discrete levels is shown in Figs.\ref{pic:14-17}. For any given $|m_j|$ it is positive and monotonically increasing function on the intervals between neighboring $ Z_{cr,m_j} $, since there grows with $Z$ the bound energy of each level. At the critical points $ Z_{cr,m_j} $ there take place the jumps of total bound energy equal to $ (-2)\times mv_F^2 $, caused by diving of discrete levels into the lower continuum. At small $Z$ the partial total bound energy shows up a square growth: ~\ref{pic:14} $\sim 0.469\, Z^2$, ~\ref{pic:15} $\sim 0.066\, Z^2$. As in the one-dimensional case, at large $Z$ these functions (without jumps) grow almost linearly, namely for ~\ref{pic:16} $\sim 2.0\, Z^{1.16}$, for ~\ref{pic:17} $\sim 1.2\, Z^{1.3}$. The jumps significantly reduce this rate of growth. Moreover, there follows from Figs. \ref{pic:10-11}-\ref{pic:14-17} that in two-dimensional problem for large $Z$ the phase integral dominates in $ \E_{VP} $, while the total bound energy contributes to $ \E_{VP} $ sufficiently less. On the contrary, in the one-dimensional case the  situation is  opposite  ~\cite{Davydov2017, Sveshnikov2017, Voronina2017}.

\begin{figure*}[ht!]
\subfigure[]{\label{pic:18}
		\includegraphics[width=\columnwidth]{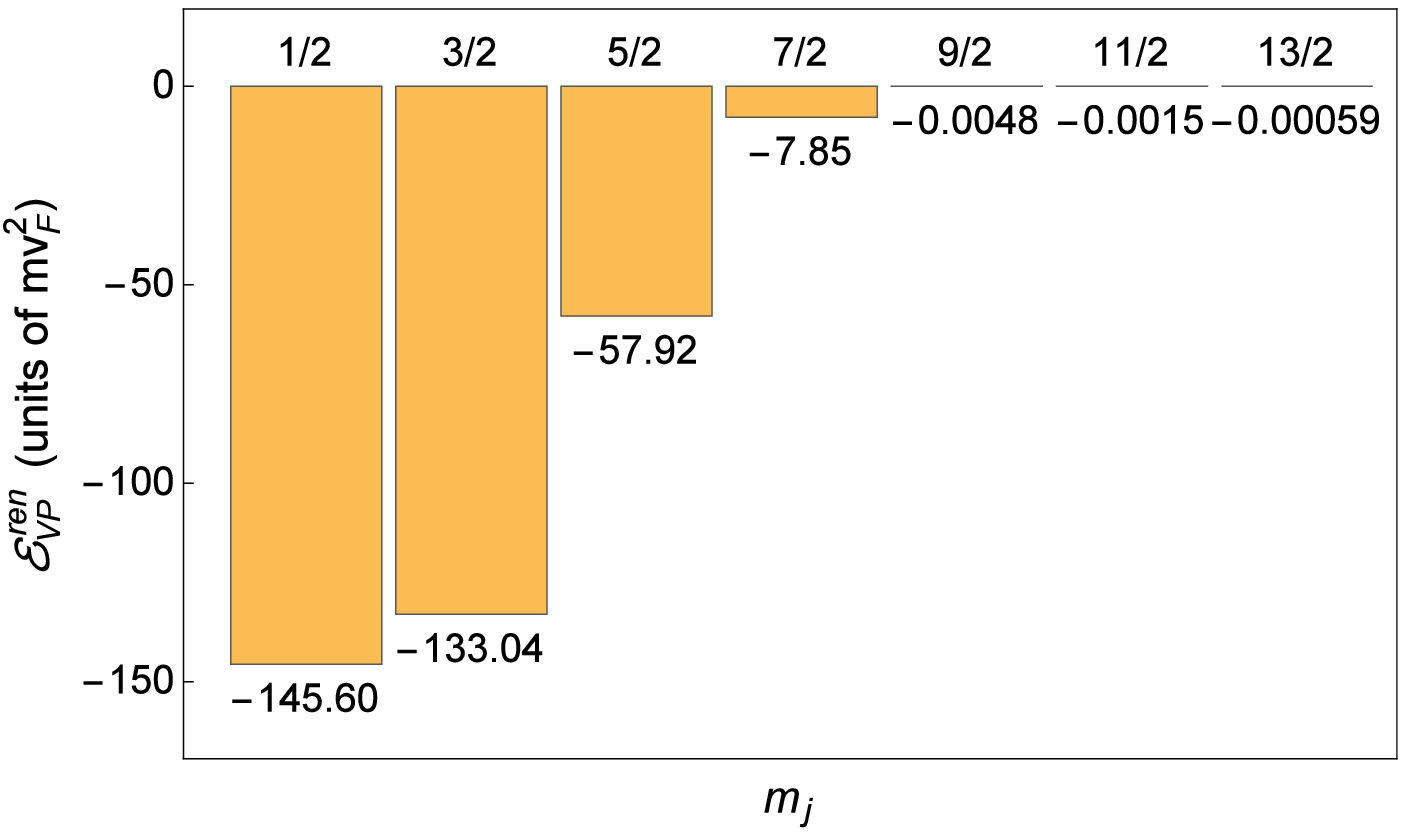}
}
\hfill
\subfigure[]{\label{pic:19}
		\includegraphics[width=\columnwidth]{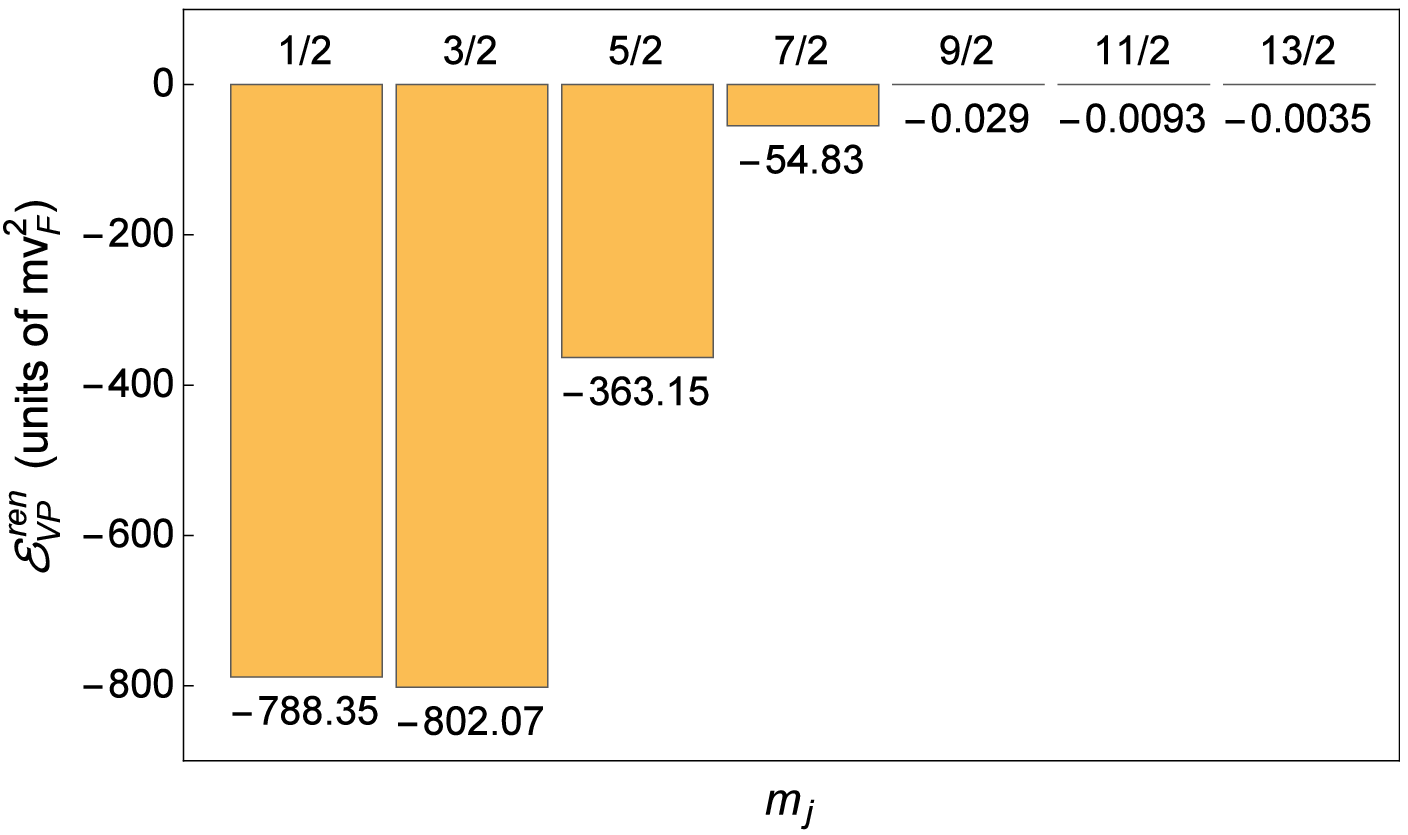}
}	
\vfill
\subfigure[]{\label{pic:20}
		\includegraphics[width=\columnwidth]{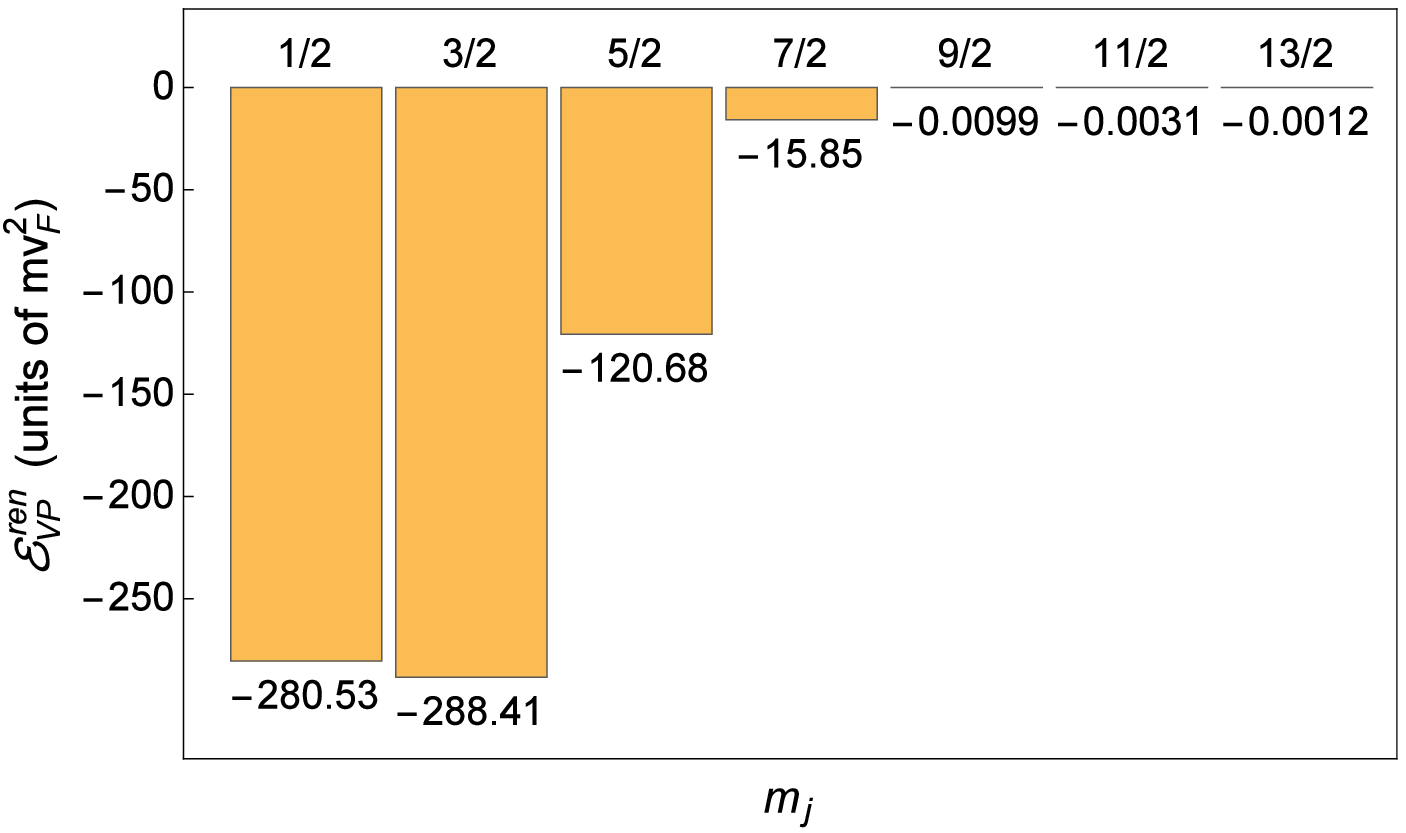}
}
\hfill
\subfigure[]{\label{pic:21}
		\includegraphics[width=\columnwidth]	{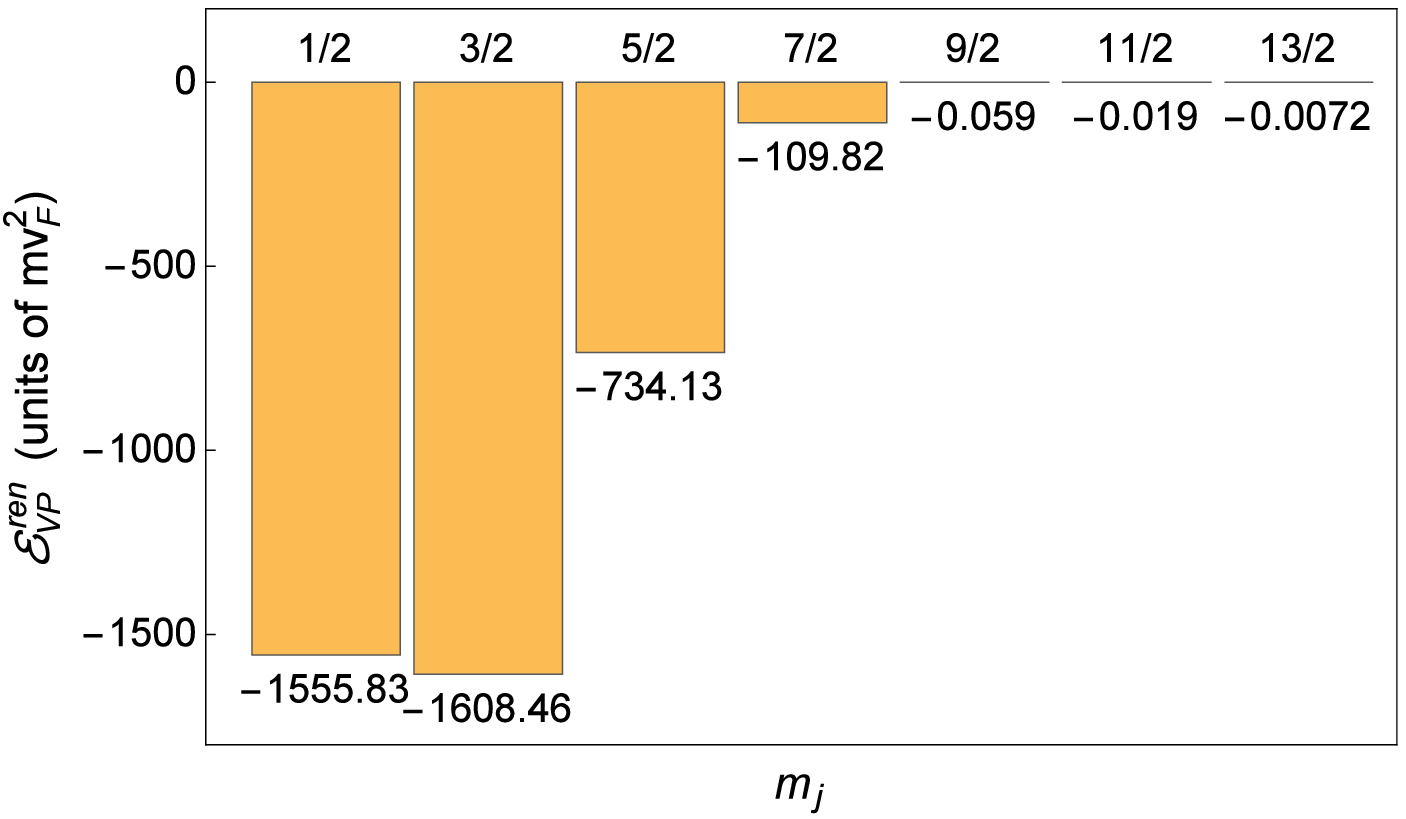}
}
\caption{(Color online) \small Contributions from channels with different $ m_j $ to the vacuum energy for ~\subref{pic:18} $\a=0.4$, $R_0=1/15$, $Z=10$, ~\subref{pic:19} $\a=0.8$, $R_0=2/175$, $Z=5$, ~\subref{pic:20} $\a=0.4$, $R_0=1/30$, $Z=10$, ~\subref{pic:21} $\a=0.8$, $R_0=1/175$, $Z=5$.}\label{pic:18-21}
\end{figure*}

As it was already stated in the end of the preceding Section, each partial term $\E^{ren}_{VP,|m_j|}(Z)$ of the series (\ref{f42}) for the total vacuum energy is in essence  quite similar to $\E_{VP}^{ren}(Z)$ in 1+1 D ~\cite{Davydov2017, Sveshnikov2017, Voronina2017}. The direct consequence of the latter is that the negative contribution from the renormalization term $\eta_{|m_j|} Z^2$ (recall, that all the $\eta_{|m_j|}$'s  are strictly negative)  turns out to be the dominant one in $\E^{ren}_{VP,|m_j|}(Z)$ in the overcritical region, since in this region   the non-renormalized energy in each separate channel, as in 1+1 D, behaves like $\sim Z^\n \ , \ 1<\n<2$. In our case this growth rate is close to linear, as it follows from the estimates for the partial phase integrals and total discrete levels bound energies,  presented above. However, now the total number of the levels, which have sunk into the lower continuum  for the given  $Z$, is determined by  the set of partial channels  with $|m_j|\leq |m_j|_{max}(Z)$, where  $|m_j|_{max}(Z)$ is the last one, in which the number of dived into the lower continuum discrete levels is non-zero. Indeed these levels and channels yield the main contribution to the whole vacuum energy (see histograms in Figs.\ref{pic:18-21}), while  $|m_j|_{max}(Z)$  grows approximately linearly with  Z. And since the total vacuum energy   $\E_{VP}^{ren}(Z)$ is mainly determined by the contributions from  these channels, its rate of decrease  acquires an additional factor of order  $O(Z)$, which in turn leads to the final answer  $\E_{VP}^{ren}(Z) \sim - \eta_{eff}\, Z^3$ in the overcritical region.
Exact numerical calculations confirm this conclusion quite well. Indeed, for $a=0.4\, , R_0=1/15$ the total renormalized vacuum energy can be approximated as $\sim -0.37\, Z^{3.05}$. As it was already stated above, this behavior of $\E_{VP}^{ren}(Z)$ is quite different from the one-dimensional case, when $\E_{VP}^{ren}(Z)$ decreases almost quadratically~\cite{Davydov2017, Sveshnikov2017, Voronina2017}.

\begin{figure*}[ht!]
\subfigure[]{\label{pic:22}
		\includegraphics[width=\columnwidth]	{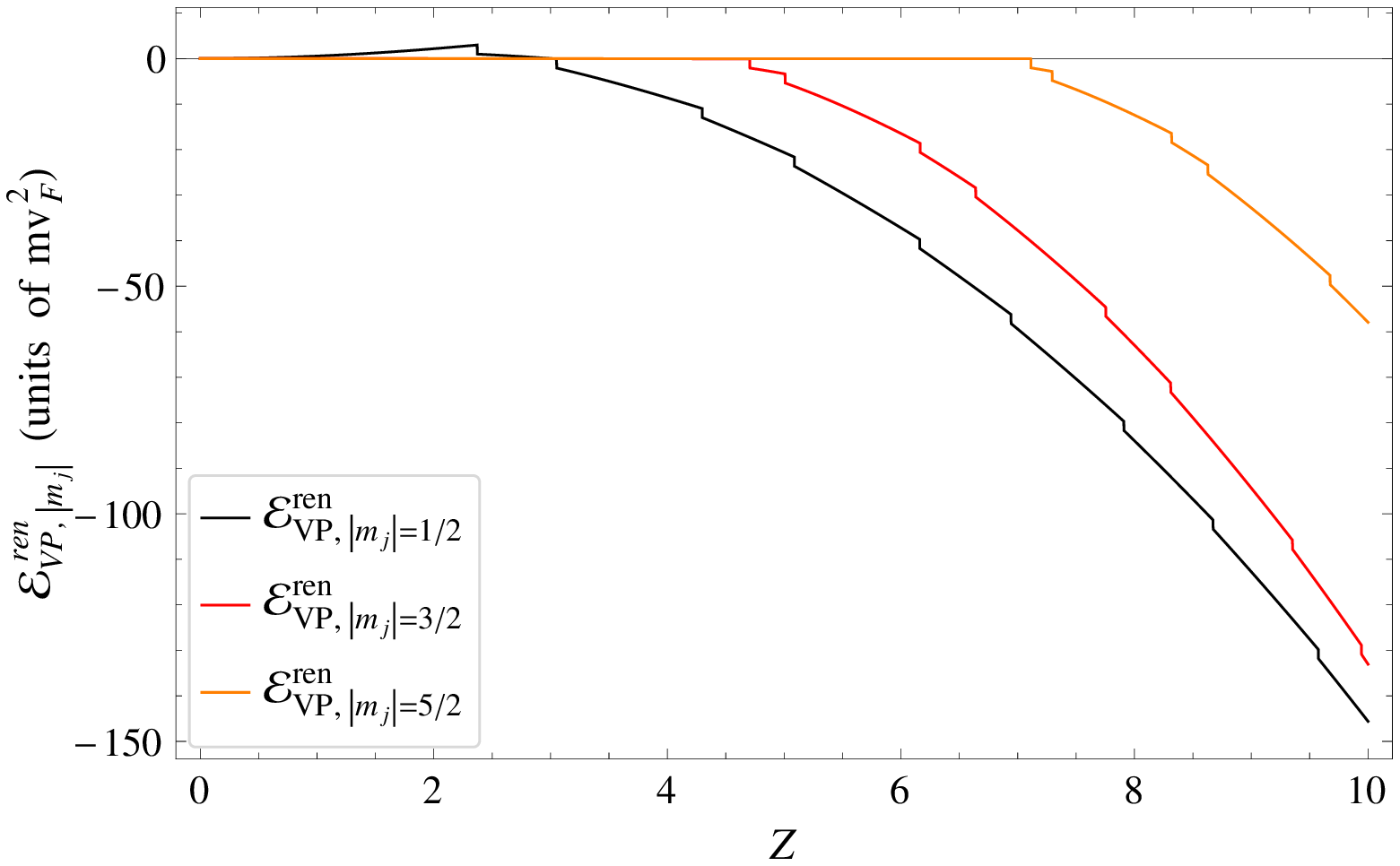}
}
\hfill
\subfigure[]{\label{pic:23}
		\includegraphics[width=\columnwidth]	{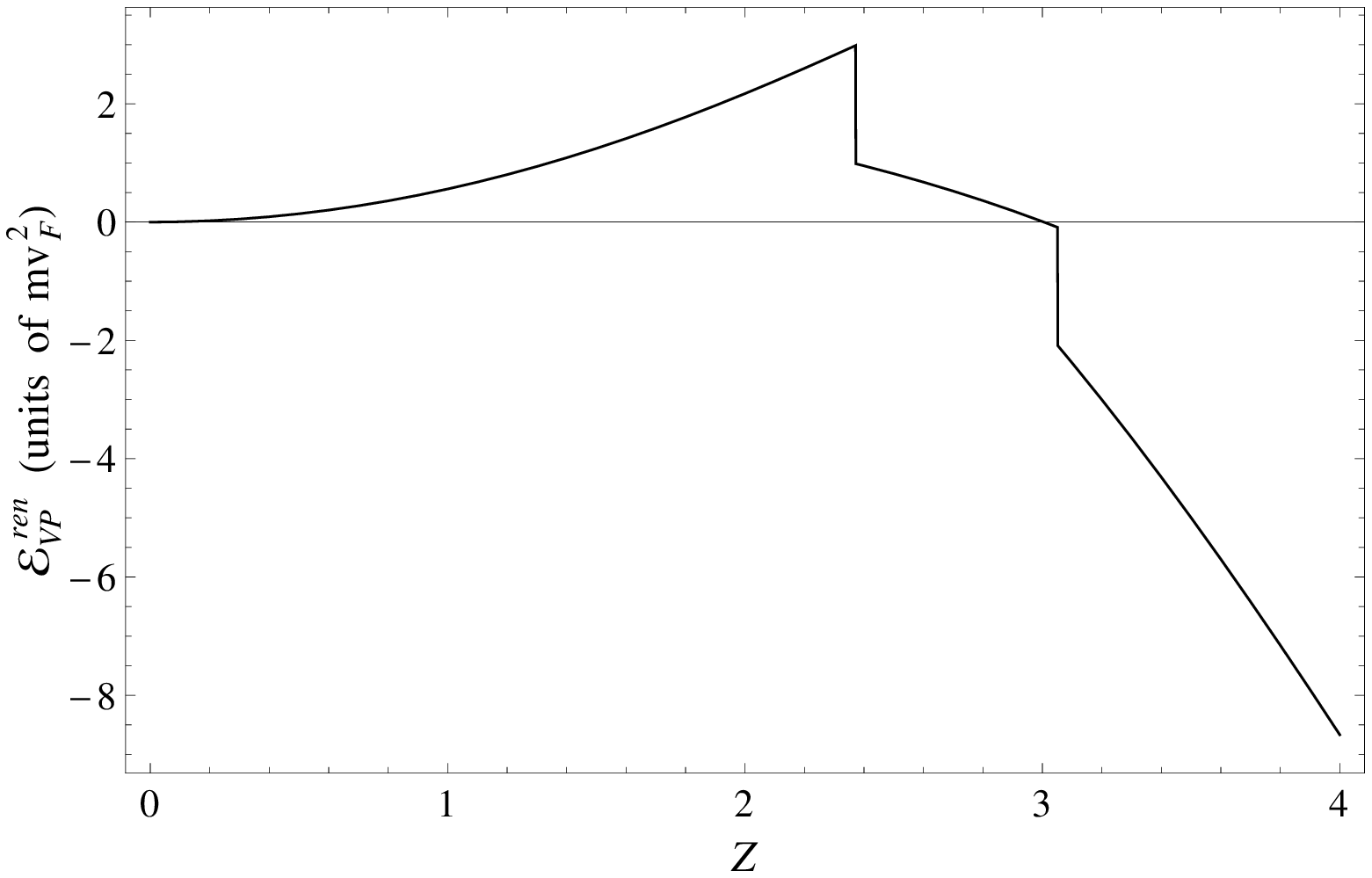}
}
\caption{(Color online) \small \subref{pic:22} $ \E_{VP,|m_j|}^{ren}(Z)$ for $\a=0.4$, $R_0=1/15$ and  $ |m_j|=1/2\, , 3/2\, , 5/2\, $; ~\subref{pic:23} $ \E_{VP}^{ren}(Z) $ for $\a=0.4$, $R_0=1/15$.}\label{pic:22-23}
\end{figure*}

Figs.\ref{pic:22-23} show the behavior of partial $ \E_{VP,|m_j|}^{ren}(Z) $ for certain $|m_j|$ and the total $ \E_{VP}^{ren}(Z) $ for $\a=0.4$, $R_0=1/15$. When considered  as a function of $Z$, $ \E_{VP,|m_j|}^{ren}(Z) $  behaves differently for $Z < Z_{cr1,|m_j|}$ and for $Z > Z_{cr1,|m_j|}$. Most clearly it is seen on the behavior of $ \E_{VP}^{ren}(Z) $. In the subcritical region the dominant contribution in $\E_{VP}^{ren}(Z)$ comes from the term $\E_{VP}^{(1)}$ and therefore the total vacuum energy shows up a square growth, but already the first level diving transforms the behavior of $\E_{VP}^{ren}(Z)$ into the decreasing one.

\begin{figure*}[ht!]
\subfigure[]{\label{pic:24}
		\includegraphics[width=\columnwidth]	{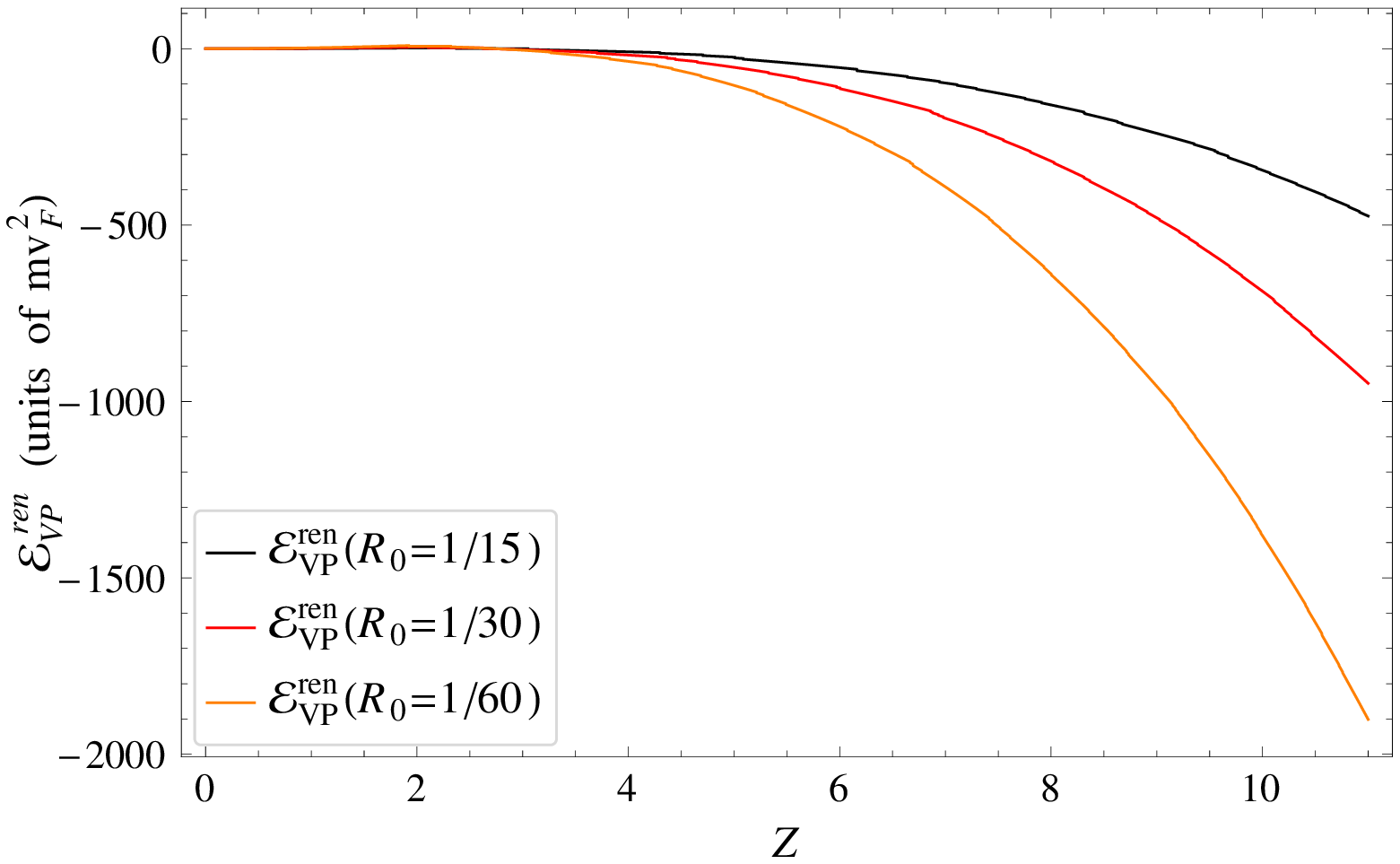}
}
\hfill
\subfigure[]{\label{pic:25}
		\includegraphics[width=\columnwidth]	{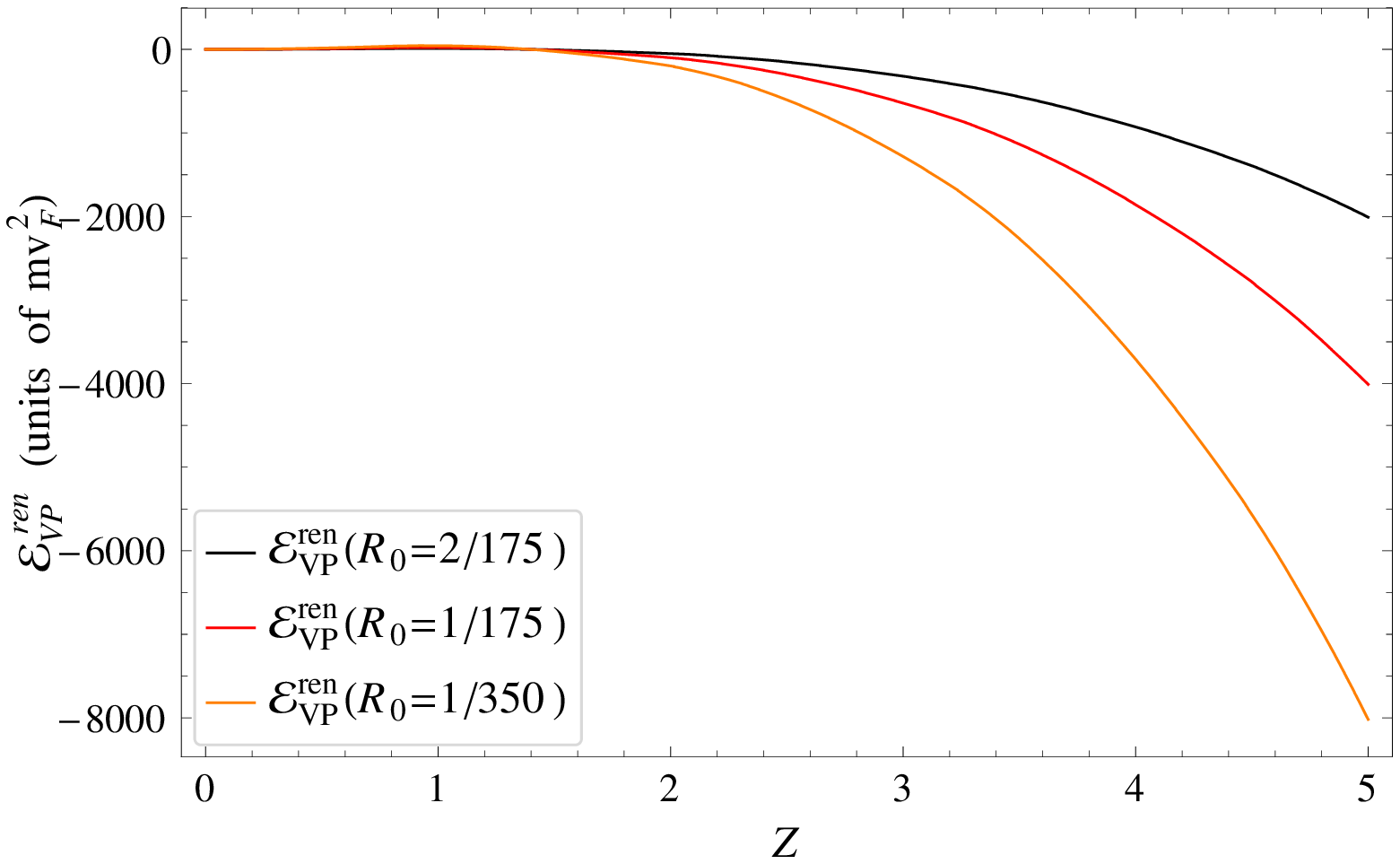}
}
\caption{(Color online) \small $ \E_{VP}^{ren}(Z)$ for ~\subref{pic:24} $\a=0.4$, $R_0=1/15$,  $R_0=1/30$, $R_0=1/60$, ~\subref{pic:25} $\a=0.8$, $R_0=2/175$,  $R_0=1/175$,  $R_0=1/350$.}\label{pic:24-25}
\end{figure*}

In Figs.\ref{pic:24-25} the behavior of $ \E_{VP}^{ren}(Z) $ is shown for other values of $\a$ and $R_0$. For $\a=0.4$ the total renormalized vacuum energy is estimated as $-0.37\, Z^{3.05}$, $-0.74\, Z^{3.05}$ and $-1.47\, Z^{3.05}$ for $R_0=1/15\, , 1/30\, , 1/60$, correspondingly, while for $\a=0.8$ the estimates are $-18.1\, Z^3$, $-36.2\, Z^3$ and $-72.4\, Z^3$ for $R_0=2/175\, , 1/175\, , 1/350$.

\begin{figure*}[ht!]
\subfigure[]{\label{pic:26}
		\includegraphics[width=\columnwidth]	{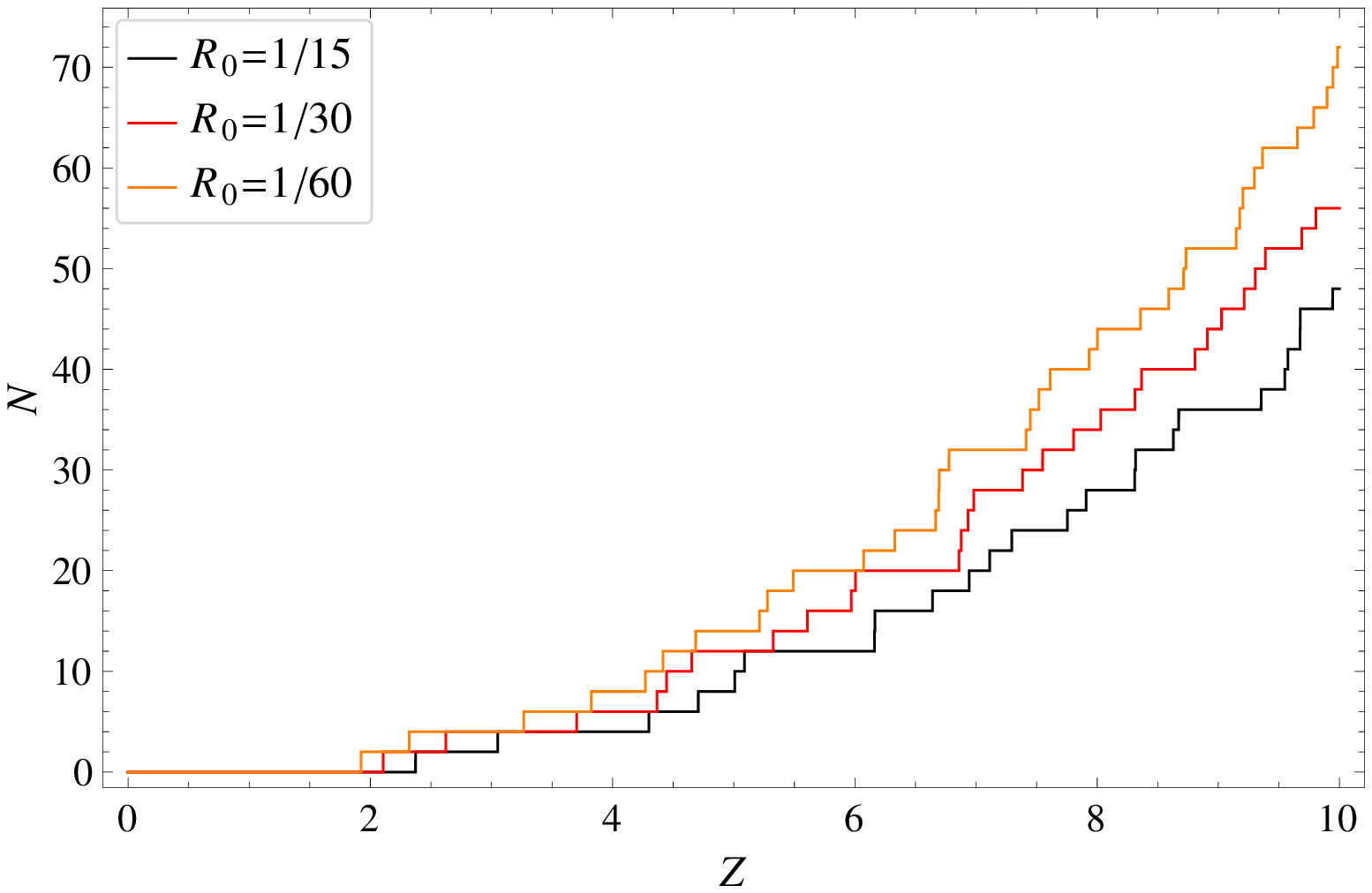}
}
\hfill
\subfigure[]{\label{pic:27}
		\includegraphics[width=\columnwidth]{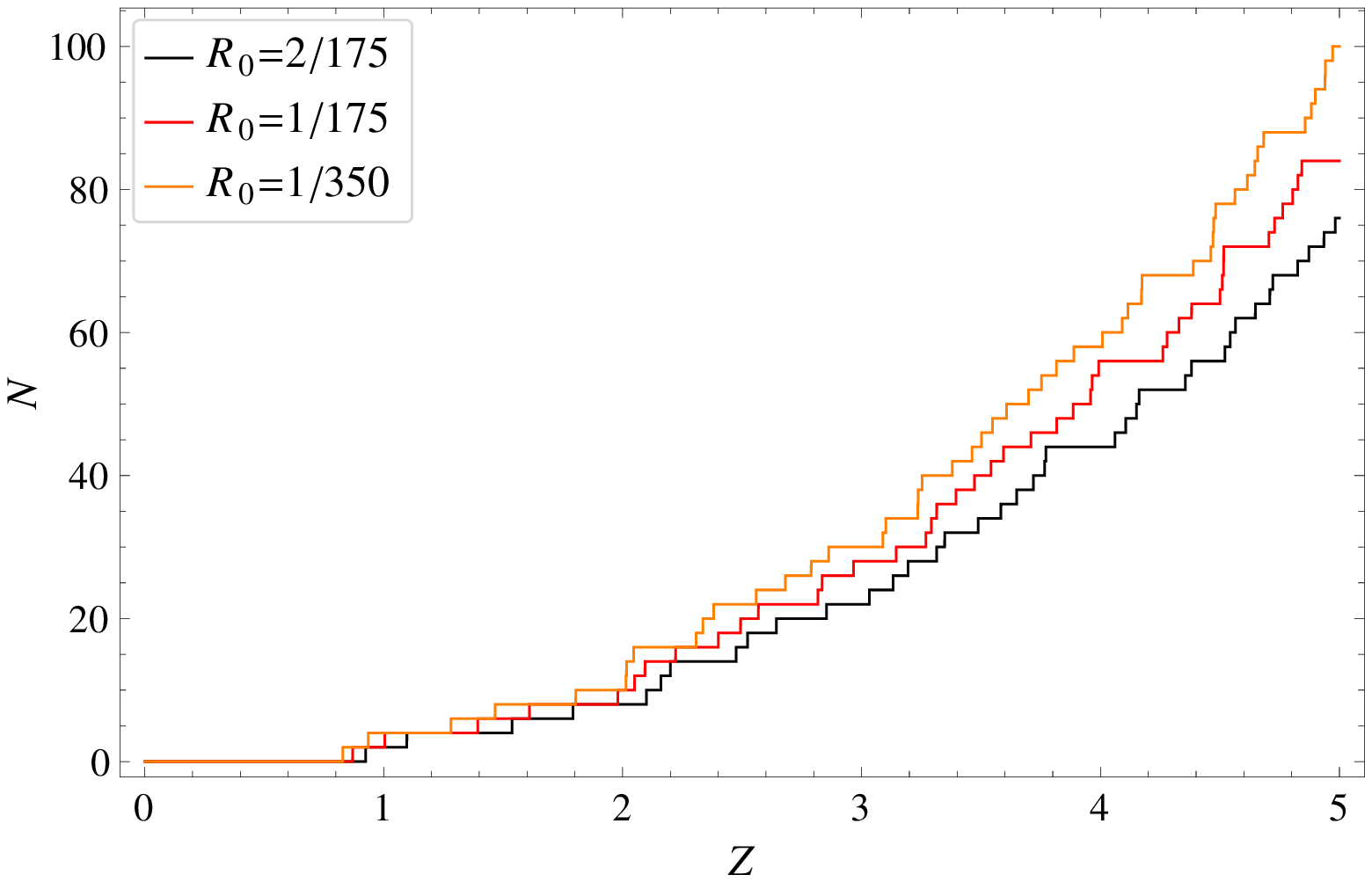}
}
\vfill
\subfigure[]{\label{pic:28}
		\includegraphics[width=\columnwidth]{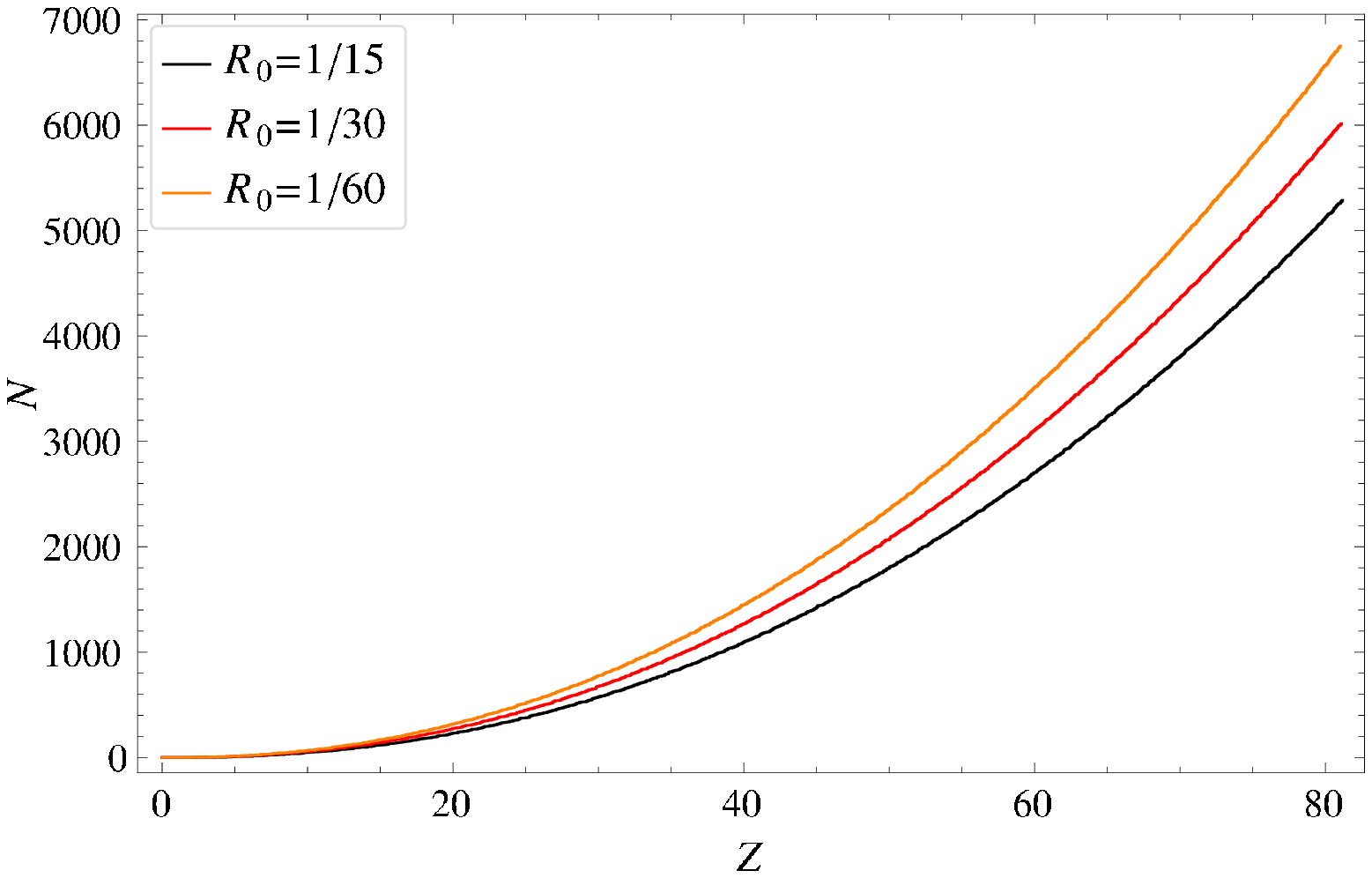}
}
\hfill
\subfigure[]{\label{pic:29}
		\includegraphics[width=\columnwidth]{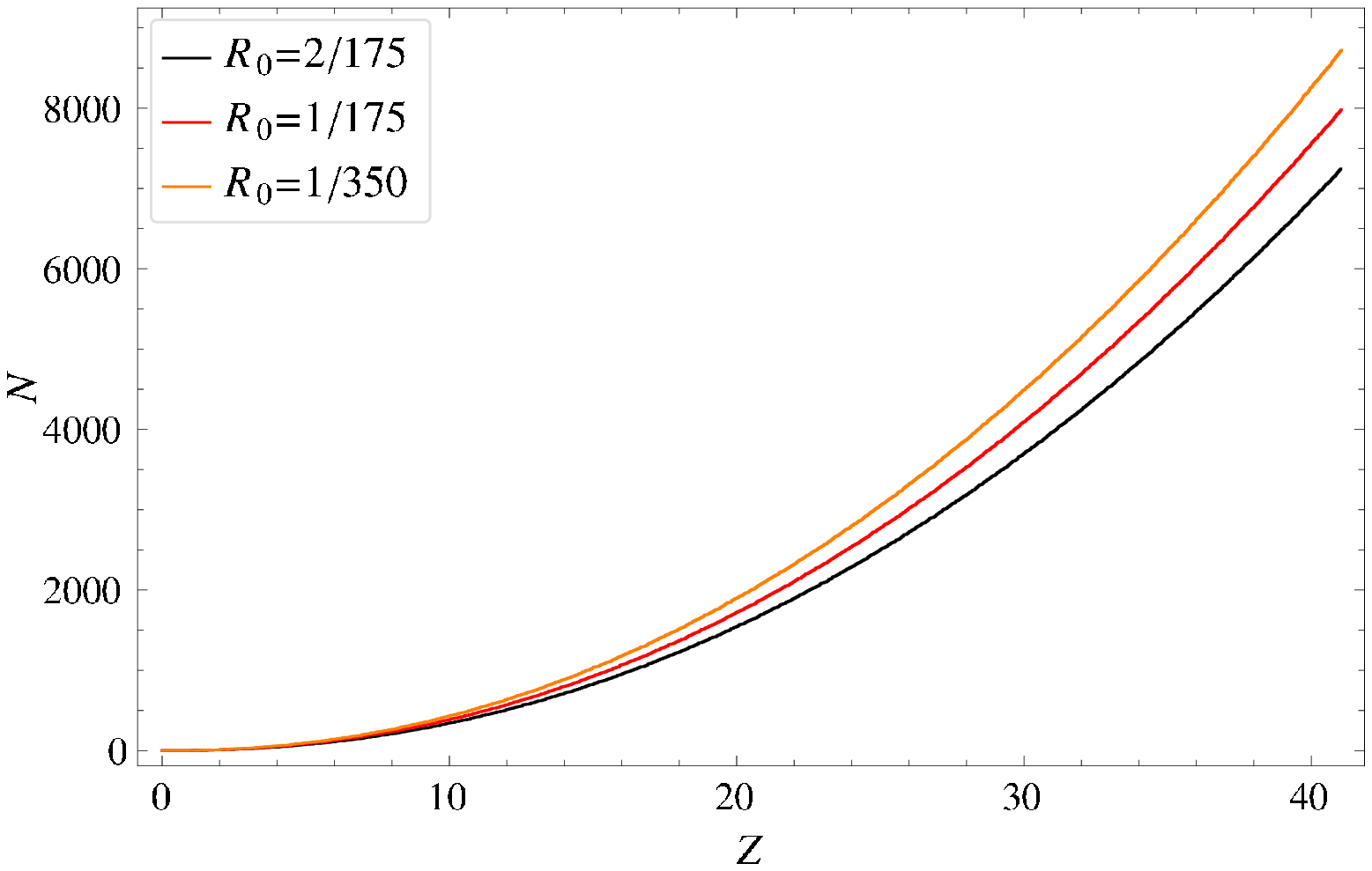}
}
\caption{(Color online) \small The vacuum shells total numbers and their approximations  as functions of $Z$ for ~\subref{pic:26},~\subref{pic:27} $\a=0.4$, ~\subref{pic:28},~\subref{pic:29} $\a=0.8$. }\label{pic:26-29}
\end{figure*}

Thus, in 2+1 D $\E_{VP}(Z)$ decreases faster as in the  one-dimensional case due to the higher rate of growth of the vacuum shells total number. Figs.\ref{pic:26-29} show the shells total  number as a function of $Z$. The  partial shells number for the given $|m_j|$ is an almost linear function for $Z\gg Z_{cr1,|m_j|}$ in each separate partial channel, as in the one-dimensional problem. In particular, the number of energy levels dived into the lower continuum $N_{|m_j|}$ for $\a=0.4\, , R_0=1/15$ is estimated as $1.3\, Z^{1.19}\, , 1.2\, Z^{1.21}\, , 1.1\, Z^{1.23}$ for $|m_j|=1/2\, , 3/2\, , 5/2$, correspondingly. At the same time, the sum of $N_{|m_j|}$ over  $|m_j|\leq |m_j|_{max}(Z) $ shows up an almost square growth. Depending on $R_0$,  the total number of shells is estimated as $0.32 \, Z^{2.21}\, , 0.41\, Z^{2.18}\, , 0.51\, Z^{2.16}$ for $\a=0.4$ and $2.58 \, Z^{2.14}\, ,
2.95\, Z^{2.13}\, , 3.34\, Z^{2.12}$ for $\a=0.8$. As for the total $\E_{VP}^{ren}$, an additional factor of order  $O(Z)$ in the total shells number compared to the one-dimensional case is caused by the linear growth of $|m_j|_{max}(Z)$ as a function of $Z$.

\begin{figure*}[ht!]
\subfigure[]{\label{pic:30}
		\includegraphics[width=\columnwidth]		{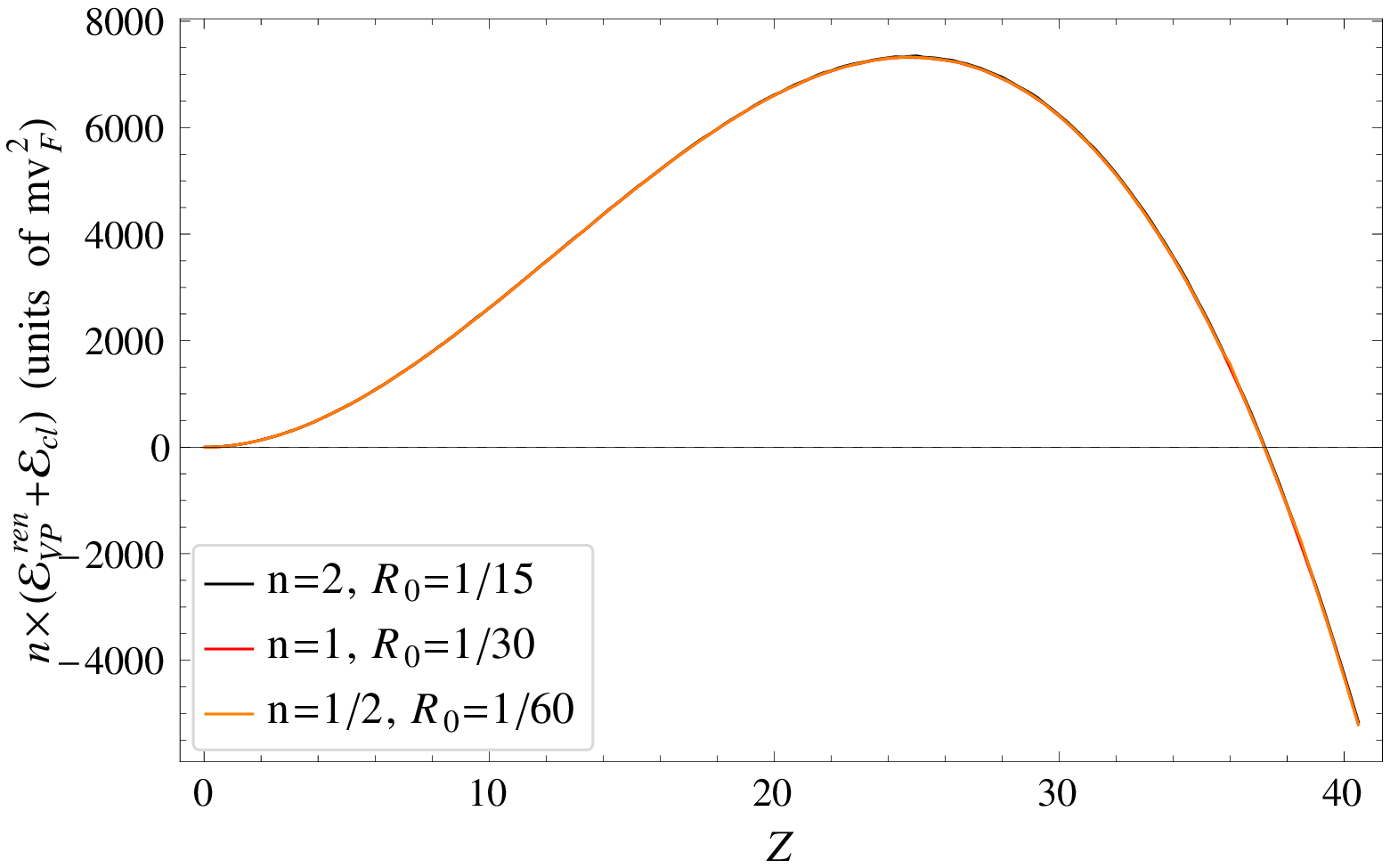}
}
\hfill
\subfigure[]{\label{pic:31}
		\includegraphics[width=\columnwidth]	{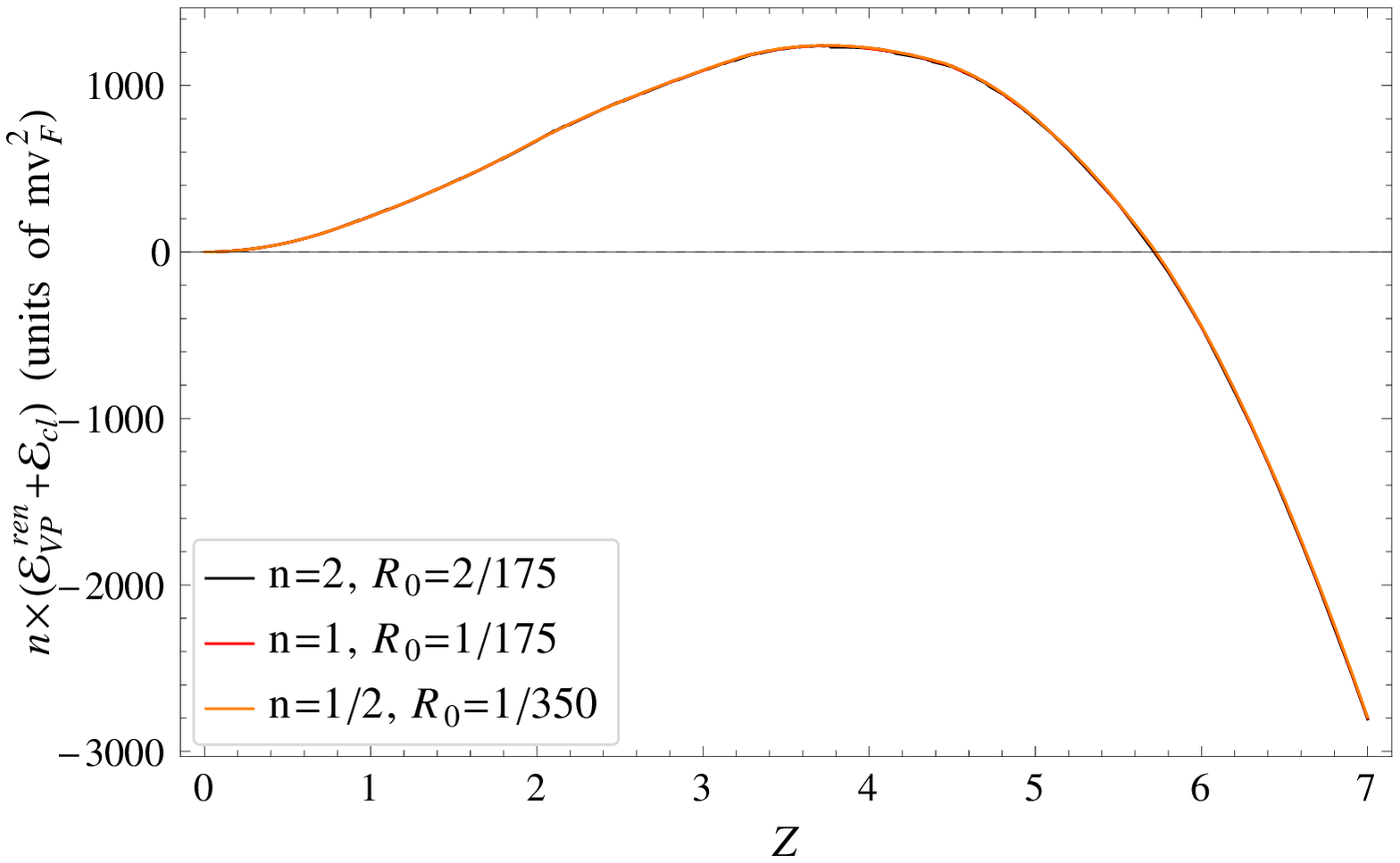}
}
\caption{(Color online) \small  The sum $ \E_{VP}^{ren} + \E_{cl}$ as a function of $Z$ and of the scaling parameter $n$: ~\subref{pic:30} for $\a=0.4$ and $n=1/2\, , R_0=1/60$; $n=1\, , R_0=1/30$; $n=2\, , R_0=1/15$, ~\subref{pic:31} for $\a=0.8$ and $n=1/2\, , R_0=1/350$; $n=1\, , R_0=1/175$; $n=2\, , R_0=2/175$.}\label{pic:30-31}
\end{figure*}

As a result, for such a planar graphene-like Dirac-Coulomb system with strong coupling the rate of decrease  of the renormalized Casimir energy turns out to be such that it  becomes competitive with the electrostatic repulsive self-energy of the external Coulomb source for quite reasonable impurity charges. In our case the latter coincides with the classical electrostatic energy of a charged sphere $\E_{cl}(Z)=Z^2 \a_0 /2R_0$, which in the dimensionless form contains the ``bare'' fine-structure constant of graphene $\a_0= e^2/\h v_F \simeq 2.2$, because $\(Z^2 e^2/2R_0\)/m v_F^2=Z^2/(2 R_0/\l_c) \times (e^2/\h v_F)$. The performed calculations show that  $ \E_{VP}^{ren}(Z) $ surpasses $\E_{cl}(Z)$ for $\a=0.4$ at  $Z^{\ast} \simeq 37$ and for $\a=0.8$ at  $Z^{\ast} \simeq 6$ (see Figs.\ref{pic:30-31}).
The curves, shown in Fig.\ref{pic:30-31}, demonstrate that the nonperturbative vacuum effects are able to compensate completely the classical repulsion energy and so could significantly affect the basic properties of the  system. In addition, Figs.\ref{pic:30-31}  confirm another property of $\E_{VP}^{ren}$, namely,  the inverse proportionality with respect to the source size $R_0$. If $R_0$ is multiplied by the scaling factor $n$, the sum $\E_{VP}^{ren} + \E_{cl}$ scales as $1/n$.
So to the leading order the behavior of $\E_{VP}^{ren}$ as a function of $Z\, ,R_0$ should be estimated as
  \beq
\label{f46}
\E_{VP}^{ren} \simeq - \eta_{eff} Z^3/R_0 \
\eeq
with $\eta_{eff}>0$. More details concerning the corrections to this relation are given in~\cite{Davydov2018b}.

 In turn, (\ref{f46})  means that in the considered range of the source sizes the value of the charge $Z^{\ast}$, which provides the exact compensation between $ \E_{VP}^{ren}$ and $\E_{cl}$
  \beq
\label{f47}
\E_{VP}^{ren}(Z^{\ast})+ \E_{cl}(Z^{\ast}) \simeq 0  \ ,
\eeq
is almost independent of $R_0$.

\section{ Casimir energy for the screened external potential (\ref{1.0})}\label{sec:screening}

For finite $R_1$ the main relations  (\ref{f40}),(\ref{f41a}) and  (\ref{f44a}), which define the renormalized vacuum energy, remain the same, but all the ingredients of $\E_{VP}$ undergo significant changes. The first one takes place in the  lowest-order perturbative  energy $\E_{VP}^{(1)}$. For the screened potential with discontinuity at $r=R_1$, considered in \cite{partI2018}, the expression for $\E_{VP}^{(1)}$ would be
\begin{multline}
\label{4.4}
\E^{(1)}_{VP}=\frac{Q^2}{32}\int\limits_{0}^{\infty}dq\,\[\frac{2 }{q} + \(1-\frac{4}{q^2}\) \arctan \( \frac{q}{2}\)\] \times  \\  \times \(2\[J_1(q R_0)+q R_1 J_0(q R_1)-q R_0 J_0(q R_0)\] + \right. \\ \left. + \pi q R_0\[ J_0(q R_0) \mathbf{H}_1(q R_0)- J_1(q R_0) \mathbf{H}_0(q R_0)\] - \right. \\ \left. - \pi q R_1\[ J_0(q R_1) \mathbf{H}_1(q R_1)- J_1(q R_1) \mathbf{H}_0(q R_1)\] \)^2 \ .
\end{multline}
It is easy to see that in (\ref{4.4}) the integrand  behaves for $q \to \inf$ as $\sim 1/q$ and so leads to the logarithmic divergence in $\E_{VP}^{(1)}$. The origin of the latter is quite similar to the singularity in the induced density at $r=R_1$, described in \cite{partI2018, Sveshnikov2018a}, and lyes in the slow decrease of the Fouriet-transform $\widetilde{A}_{0}(q)$ in the momentum space due to vertical jump in $A^{ext}_{0}(r)$  at $r=R_1$ in this case. So for the correct evaluation of $\E_{VP}^{ren}$ in the screened  case some kind of smoothing of this jump is required, the most simple version of which is achieved by means of the continuous external potential (\ref{1.0}). For the latter one finds
\begin{multline}
\label{4.5}
\E^{(1)}_{VP}=\frac{Q_1^2}{32} \int\limits_{0}^{\infty}dq\,\[\frac{2 }{q} + \(1-\frac{4}{q^2}\) \arctan \( \frac{q}{2}\)\] \times  \\  \times \(2\[J_1(q R_0)-J_1(q R_1)+q R_1 J_0(q R_1)-q R_0 J_0(q R_0)\] +\right. \\ \left. + \pi q R_0\[ J_0(q R_0) \mathbf{H}_1(q R_0)- J_1(q R_0) \mathbf{H}_0(q R_0)\] - \right. \\ \left. - \pi q R_1\[ J_0(q R_1) \mathbf{H}_1(q R_1)- J_1(q R_1) \mathbf{H}_0(q R_1)\] \)^2 \ ,
\end{multline}
where
 \beq\label{4.51}
Q_1=Z \a/(1-R_0/R_1) \ .
\eeq
Besides the latter relation, the only difference between the expressions (\ref{4.4}) and (\ref{4.5}) is  the structure of the second line. However, this difference  is crucial, since now the integrand in  (\ref{4.5}) behaves for $q \to \inf$ as $\sim 1/q^3$ and so  $\E_{VP}^{(1)}$ becomes  well-defined again. Moreover, screening in the form  (\ref{1.0}) allows to perform the most part of calculations required for evaluation of $\E_{VP}^{ren}$ in the analytical form.

Now let us consider the changes in the solutions of the DC problem (\ref{3.7}). For $0< r \leqslant R_0$ the solutions remain the same as in the unscreened case (\ref{f450}), while for $R_0<r<R_1$ their convenient form is given now in terms of the Kummer  $\F(b,c,z)$ and modified Kummer $\tilde \F(b,c,z)=z^{1-c} \F(b-c+1,2-c,z)$ functions. The formulae, presented below, work equally well both for the continuous spectra, where $\e = \pm\sqrt{k^2 + 1}$ for the upper and lower ones, respectively, and for the discrete one $-1 \leqslant \e <1$.

For $|m_j| > Q_1$ one obtains
\beq \label{4.6}\bal
	&\p_{1,\, m_j}^{mid}(r,\e)= \sqrt{|\e -V_1 + 1|}\,\, r^{\vk-1/2}\, \times \\& \times \(F_{1,r}+B_{m_j}(\e)\tilde{F}_{1,r} \) \ ,
	\\
	&\p_{2,\, m_j}^{mid}(r,\e)= \sign(1+V_1-\e)\sqrt{|\e -V_1 - 1|}\,\, r^{\vk-1/2}\, \times \\& \times  \( F_{2,r}+B_{m_j}(\e)\, \tilde{F}_{2,r} \) \ ,
\eal\eeq
where
\beq \label{4.6a}
 V_1=Q_1/R_1 \ ,
\eeq
\beq\label{4.7}
F_{1,r}= \left\lbrace \bal
&\mathrm{e}^{-\g_1 r}\,\((m_j+ Q_1/\g_1)\F_{r} \ +  \right.\\ & \left. + \ b\F_{r}(b+)\) \ , \quad &|\e-V_1|\leq 1 \ , \\
&\mathrm{Re}\[\mathrm{e}^{i\f_+}\mathrm{e}^{-\g_1 r} \F_r\] \ , \quad &|\e-V_1|>1 \ ,
\eal \right.
\eeq
\beq\label{4.8}
F_{2,r}= \left\lbrace \bal
&\mathrm{e}^{-\g_1 r}\,\(-(m_j+ Q_1/\g_1)\F_{r} \ +  \right.\\ & \left. + \ b\F_{r}(b+)\) \ , \quad &|\e-V_1|\leq 1 \ , \\
&\mathrm{Im}\[\mathrm{e}^{i\f_+}\mathrm{e}^{-\g_1 r} \F_r\] \ , \quad &|\e-V_1|>1 \ ,
\eal \right.
\eeq
\beq\label{4.9}
\tilde{F}_{1,r}= \left\lbrace \bal
&\mathrm{e}^{-\g_1 r}\,\( (m_j+ Q_1/\g_1)\tilde \F_r \ + \right.\\ & \left.  + \ (1+b-c)\tilde \F_r(b+)\) \ , \quad &|\e-V_1|\leq 1 \ , \\
&\mathrm{Re}\[i \mathrm{e}^{-i\pi\vk}\mathrm{e}^{i\f_-}\mathrm{e}^{-\g_1 r}  \tilde \F_{r}\], \quad &|\e-V_1|>1 \ ,
\eal \right. \eeq
\beq\label{4.10}
\tilde{F}_{2,r}= \left\lbrace \bal
&\mathrm{e}^{-\g_1 r}\,\(-(m_j+ Q_1/\g_1)\tilde \F_r \ + \right.\\ & \left.  + \ (1+b-c)\tilde \F_r(b+)\) \ , \quad &|\e-V_1|\leq 1 \ , \\
&\mathrm{Im}\[i \mathrm{e}^{-i\pi\vk}\mathrm{e}^{i\f_-}\mathrm{e}^{-\g_1 r}  \tilde \F_{r}\], \quad &|\e-V_1|>1 \ .
\eal \right.
\eeq
In eqs. (\ref{4.7})-(\ref{4.10}) the following denotations are used
\beq\label{4.11}\bal
&\vk=\sqrt{m_j^2-Q_1^2} \ , \quad b=\vk - (\e-V_1) Q_1 / \g_1 \ , \quad c=1+2\vk \ , \\
& \f_+=\1/2 \mathrm{Arg}\[{m_j+Q_1/\g_1 \over b}\], \quad \f_-=\1/2 \mathrm{Arg}\[{b \over m_j-Q_1/\g_1}\] \ ,
\eal\eeq
with $\g_1$ being defined as
\beq\label{4.11a}
\g_1= \left\lbrace \bal
&\sqrt{1-\(\e-V_1\)^2} \ , \quad &|\e-V_1|\leq 1 \ , \\
&-i\sqrt{\(\e-V_1\)^2-1} \ , \quad &|\e-V_1|>1 \ ,
\eal \right.
\eeq
while
\beq\bal\label{4.11b}
&\F_r=\F\(b,c,2 \g_1 r\) \ , \quad \F_r(b+)=\F\(b+1,c,2 \g_1 r\) \\& \tilde \F_r=\tilde \F \(b,c, 2 \g_1 r\) \ , \quad \tilde \F_r(b+)=\tilde \F \(b+1,c, 2 \g_1 r\) \ .
\eal\eeq
The coefficients $B_{m_j}(\e)$ are determined via matching relations for $\psi^{int}_{m_j}(r,\e)$ and $\psi^{mid}_{m_j}(r,\e)$ at $r=R_0$, what gives
 \beq\label{4.12}
 B_{m_j}(\e)=-{C_{1,\, m_j}(\e)\,F_{2,R_0} - C_{2,\, m_j}(\e)\, F_{1,R_0} \over C_{1,\, m_j}(\e)\, \tilde{F}_{2,R_0} - C_{2,\, m_j}(\e)\,   \tilde{F}_{1,R_0}} \ ,
 \eeq
where
\beq
\bal
&C_{1,\,m_j}(\e)=\sign(1+V_1-\e)\,\sqrt{|\e-V_1 - 1|}\,\, \psi^{int}_{1, m_j}(R_0,\e ) \ , \\
&C_{2,\,m_j}(\e)=\sqrt{|\e-V_1 + 1|}\,\, \psi^{int}_{2, m_j}(R_0,\e ) \ .\\
\eal\eeq

For $|m_j| < Q_1$ the solutions should be written as follows
\beq\label{4.13}\bal
&\p_{1,\, m_j}^{mid}(r,\e)=\sqrt{|\e -V_1 + 1|} \times \\& \mathrm{Re}\Big[\mathrm{e}^{i\l_{m_j}(\e)} \mathrm{e}^{-\g_1 r}(2  \g_1 r)^{i|\vk|-\1/2}
\(b \F_r(b+)+(m_j+Q_1/\g_1)\F_r\)\Big] \ ,
\\
&\p_{2,\, m_j}^{mid}(r,\e)=\sign(1+V_1-\e)\,\sqrt{|\e-V_1 - 1|}\, \times \\  & \times \mathrm{Re}\Big[i^{\theta{((\e-V_1)^2-1)}} \,\mathrm{e}^{i\l_{m_j}(\e)} \mathrm{e}^{-\g_1 r} (2 \g_1 r)^{i|\vk|-\1/2} \times  \\&  \times \(b \F_r(b+)-(m_j+Q_1/\g_1)\, \F_r\)\Big] \ ,
\eal\eeq
where
\beq\bal\label{4.14}
& |\vk|=\sqrt{Q_1^2-m_j^2} \ , \\ & b=i|\vk| - {\(\e-V_1\) Q_1/ \g_1} \ , \quad  c=1+2i|\vk| \ ,
\eal\eeq
while $\lambda_{m_j}(\e)$ is determined via matching at $r=R_0$
\begin{widetext}
\begin{multline}\label{4.15}
\lambda_{m_j}(\e)= -\mathrm{Arg}\left[i \mathrm{e}^{-\g_1 R_0}(2 \g_1 R_0)^{i |\vk|} \times \right.\\ \left. \times \(-\(C_{2,\, m_j}+i^{\theta{((\e-V_1)^2-1)}}\, C_{1,\, m_j}\)  (m_j+Q_1/\g_1)\F_{R_0} +   (-C_{2,\, m_j}+i^{\theta{((\e-V_1)^2-1)}} C_{1,\, m_j})\, b\F_{R_0}(b+) \)\right] \ .
\end{multline}

In the region   $r\geqslant R_1$ the  solutions   for both continua with $\e = \pm\sqrt{k^2 + 1}$ are expressed by means of $J_{\nu}(z)$ and $N_{\nu}(z)$, namely
\beq\label{4.16}\bal
& \p_{1,\, m_j}^{out}(r,\e)=\sqrt{|\e+1|}\,  \(J_{m_j-1/2}(kr) + D_{m_j}(\e)N_{m_j-1/2}(kr)\) \ , \\
& \p_{2,\, m_j}^{out}(r,\e)=-\sign(\e)\sqrt{|\e - 1|}\,  \(J_{m_j+1/2}(kr) +  D_{m_j}(\e)N_{m_j+1/2}(kr)\) \ , \\
\end{aligned}
\eeq
 \beq
\label{4.17}
D_{m_j}(\e)=-{\sqrt{|\e+1|}\,J_{m_j-1/2}(k R_1) \p_{2,\, m_j}^{mid}(R_1,\e) + \sign(\e)\sqrt{|\e - 1|}\,J_{m_j+1/2}(k R_1) \p_{1,\, m_j}^{mid}(R_1,\e)  \over \sqrt{|\e+1|}\,N_{m_j-1/2}(k R_1)\p_{2,\, m_j}^{mid}(R_1,\e) + \sign(\e)\sqrt{|\e - 1|}\,N_{m_j+1/2}(k R_1)\p_{1,\, m_j}^{mid}(R_1,\e)} \ ,
\eeq
whereas  for the discrete levels with $-1 \leqslant \e <1$ the  corresponding solutions in this case should be written as
\beq\label{4.21}
\p_{1,\, m_j}^{ext}(r,\e) = \sqrt{1+\e} \, K_{m_j-1/2}(\g r) \ , \quad \p_{2,\, m_j}^{ext}(r,\e) = -\sqrt{1-\e} \, K_{m_j+1/2}(\g r) \ , \quad \g=\sqrt{1-\e^2} \ .
\eeq
The equation for discrete spectrum is obtained by matching the corresponding solutions at $r=R_0$ and $r=R_1$. For $|m_j| > Q_1$ it reads
\begin{multline}\label{4.22}
\Big(C_{1,m_j}(\e)\, \tilde{F}_{2,R_0}-C_{2,m_j}(\e)\,\tilde{F}_{1,R_0}\Big)\times \\ \times \Big(\sign(1+V_1-\e)\,\sqrt{|(\e-V_1 - 1)(\e + 1)|}\,K_{m_j-1/2}(\g R_1)\,F_{2,R_1} +\sqrt{|(\e-V_1 + 1)(\e - 1)|}\, K_{m_j+1/2}(\g R_1)\, F_{1,R_1}\Big) - \\ -\Big(C_{1,m_j}(\e)\,F_{2,R_0}-C_{2,m_j}(\e)\,F_{1,R_0}\Big) \times \\ \times
\Big(\sign(1+V_1-\e)\,\sqrt{|(\e-V_1 - 1)(\e + 1)|}\,K_{m_j-1/2}(\g R_1)\,\tilde{F}_{2,R_1}+\sqrt{|(\e-V_1 + 1)(\e - 1)|}\,K_{m_j+1/2}(\g R_1)\,\tilde{F}_{1,R_1}\Big)=0 \ ,
\end{multline}
while for $|m_j| < Q_1$ it takes the form
\begin{multline}
	\label{4.23}
	\mathrm{Im}\Big[\mathrm{e}^{-\g_1 R_1-\g_1^* R_0}(2\g_1 R_1)^{i |\vk|-1/2}(2\g_1^* R_0)^{-i |\vk|-1/2}\Big(\sqrt{|(\e-V_1 + 1)(\e - 1)|}\, K_{m_j+1/2}(\g R_1)\,\Big((m_j+ Q_1/\g_1)\F_{R_1} +\\ b \F_{R_1}(b+)\Big) + \sign(1 + V_1 - \e)\,\sqrt{|(\e-V_1 - 1)(\e + 1)|}\,K_{m_j-1/2}(\g R_1)\,i^{\theta{((\e-V_1)^2-1)}} \(-(m_j+ Q_1/\g_1)\,\F_{R_1} + b \F_{R_1}(b+)\)\Big) \times \\
	\times\Big(C_{1,m_j}(\e)\,i^{\theta{((\e-V_1)^2-1)}}\(-(Q_1/\g_1+m_j)\,\F_{R_0}+b\, \F_{R_0}(b+)\)
	-C_{2,m_j}(\e)\,\((Q_1/\g_1+m_j)\,\F_{R_0}+b\, \F_{R_0}(b+)\)\Big)^*\Big]=0 \ .
\end{multline}
\end{widetext}

Now let us turn to the critical charges. In the screened case the notion of  critical charges turns out to be more diverse, since the condensation point for levels with $\e \to 1$ disappears and the total number of discrete levels becomes finite. So in the screened case there remain the lower  critical charges, which as before imply the diving of levels into the lower continuum, and in addition there appear the upper critical charges, when the virtual levels transform into the real ones (and vice versa) at the upper threshold. The equations for both types of critical charges can be easily deduced from the equations (\ref{4.22}) and (\ref{4.23})  in the limit $\e \to \pm 1$ by taking account of  the well-known  limiting relations, which replace the MacDonald  functions by the power-like ones~\cite{Bateman1953}. Moreover, in what follows we intentionally will consider mainly the case $|m_j|<Q_1$, since only by fulfilment of the latter condition the levels attain the lower threshold, what is the most important condition for emergence of essentially non-perturbative polarization effects under consideration. It is convenient to represent the corresponding equations in the form
  \beq
\label{4.23a}
X^{\mp}_{\pm|m_j|}=0 \ ,
\eeq
where
\beq\label{4.23c} \bal
& X^{-}_{|m_j|}=\p_{1,\, |m_j|}^{mid}(R_1,-1) \ , \\ & X^-_{-|m_j|}=\p_{1,\, -|m_j|}^{mid}(R_1,-1)  + {|m_j| - 1/2 \over R_1}\p_{2,\, -|m_j|}^{mid}(R_1,-1)
\eal
\eeq
are responsible for the lower critical charges, when the levels with $\pm |m_j|$ attain the lower continuum, while
\beq \bal\label{4.23b}
&	X^+_{|m_j|}=\p_{2,\, |m_j|}^{mid}(R_1,1)  +  \ {|m_j| -  1/2 \over R_1}\p_{1,\, |m_j|}^{mid}(R_1,1)\ , \\ &X^+_{-|m_j|}=\p_{2,\, -|m_j|}^{mid}(R_1,1), \ ,
\eal\eeq
define the upper ones, when the levels with $\pm |m_j|$ appear at the upper threshold.
Moreover, the eqs.(\ref{4.23a})-(\ref{4.23b}) cover all the cases including the peculiar ones with $m_j=\pm 1/2\,  ,\pm 3/2$, when the emerging solutions with $\e = \pm 1$ don't refer neither to discrete spectrum nor to the scattering states, what has been discussed in detail in \cite{partI2018}, Section 5.

The total partial phases are still defined via (\ref{f45-10}), while the separate  shifts are determined now from the asymptotics of solutions (\ref{4.16})
\beq\label{4.24}\bal
	\d_{m_j}(\e)=\mathrm{Arg}\[ 1- i D_{m_j}(\e) \] \ .
\eal\eeq
Screening of the type (\ref{1.0}) of the external potential doesn't significantly affect the asymptotics of the total partial phases $\d_{tot,|m_j|}(k)$ for $k\to\infty$, since  it proceeds without discontinuities and so the partial phases correspond still merely to the Coulomb one, rather than  to the scattering on the potential well of finite depth and size. As a result, quite similar to the unscreened case,  the total partial phases for $k\to\infty$ decrease as $\sim 1/k^3$, namely
\begin{widetext}\begin{multline}
\label{4.25}
\d_{tot,|m_j|}(k) \to {Q_1\over k^3}\Bigg[{2 Q_1\over 3}\left\{ \left({1\over R_1}-{1\over R_0}\right) \left(6+m^2_j \left({1\over R^2_1}+{1\over R_0 R_1}-{2\over R^2_0}\right)\right) - {6\over R_1}\ln\left({R_0\over R_1}\right)\right\}+\\
+|m_j|(-1)^{|m_j|-1/2}\left\{ {\sin(2Q) \sin(2 k R_0) \over R^3_0} - {\sin\left(2Q_1\,\ln\left(R_1/R_0\right)\right) \sin(2 k R_1) \over R^3_1}\right\}
\Bigg]+O(1/k^4) \ .
\end{multline}\end{widetext}
For $R_1 \to \infty$ this result exactly reproduces the answer for unscreened case (\ref{f45-18}). It should be noted that the derivation of the asymptotics (\ref{4.25}) itself and especially with account for next-to-leading orders of expansion in $1/k$ for a reasonable time is possible only be means of symbolic computer algebra tools.

Screening affects the IR-behavior of the total partial phases as well. Namely, now the limiting value for $\d_{tot,|m_j|}(k)$ for $k\to 0$ is determined via
  \beq
\label{4.26}
\delta_{tot,\, |m_j|}(k\to 0)=\mathrm{Arg}\[\prod \limits_{\pm} \! X^-_{\pm |m_j|}\,X^+_{\pm|m_j|}\] \ ,
\eeq
what gives rise to a jump-like behavior of $\d_{tot,|m_j|}(0)$ as a function of $Z$, since by passing through each upper or lower critical value of $Z$ the corresponding $X^{\mp}_{\pm|m_j|}$ changes its sign, and so there appears a jump by $\pm \pi$ in $\d_{tot,|m_j|}(0)$.
It should be noted that the expression (\ref{4.26})   defines the limiting value of the total partial phase  up to  $2\pi$ only. Removing this uncertainty requires to keep the imaginary part of the function under the sign of Arg. In any case, however, this relation correctly reproduces the jumps in  $\d_{tot,|m_j|}(k \to 0)$, accompanying diving or emergence of levels at both thresholds.

\begin{figure*}[ht!]
\subfigure[]{\label{pic:32}
		\includegraphics[width=\columnwidth]{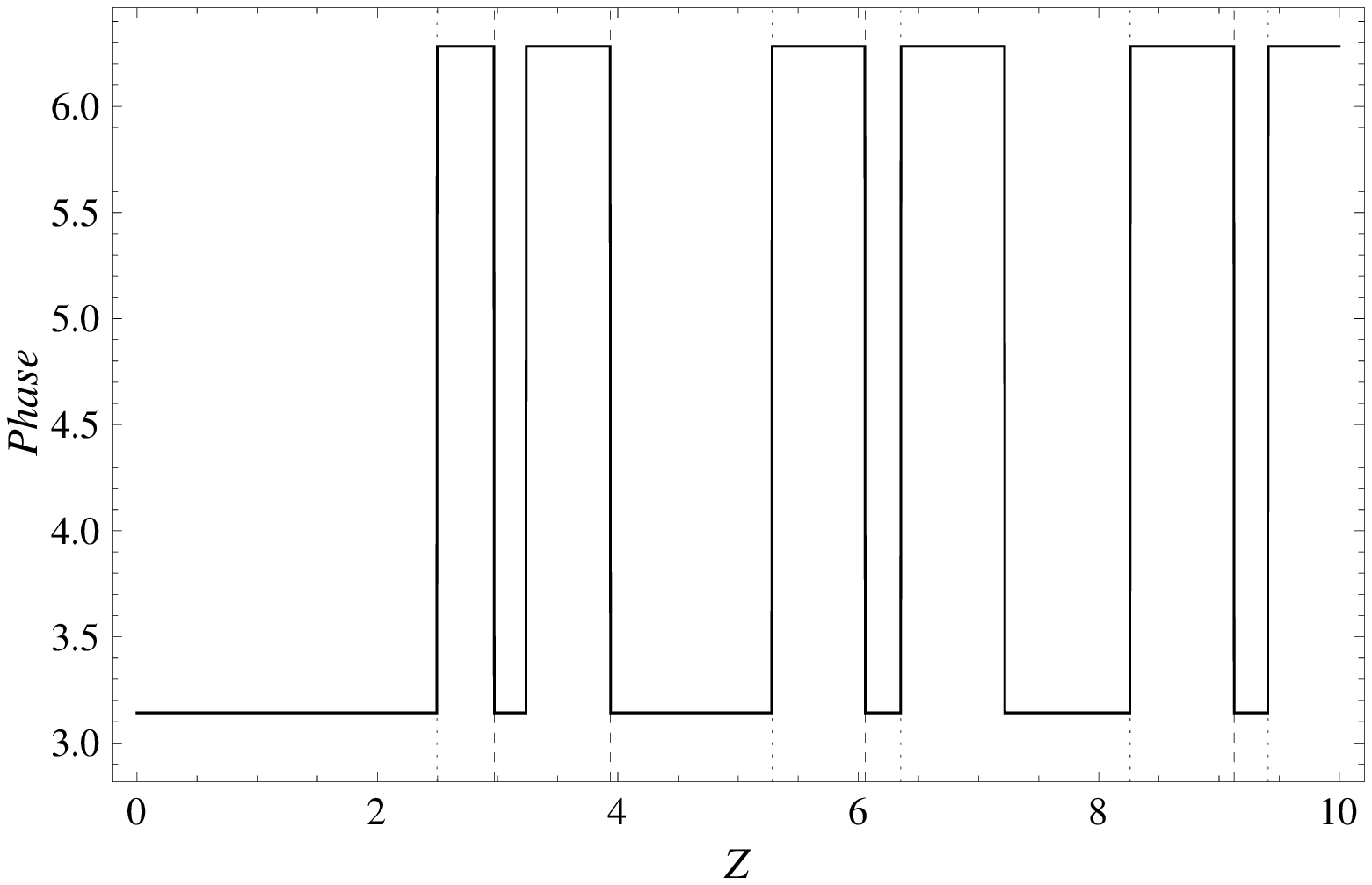}
}
\hfill
\subfigure[]{\label{pic:33}
		\includegraphics[width=\columnwidth]{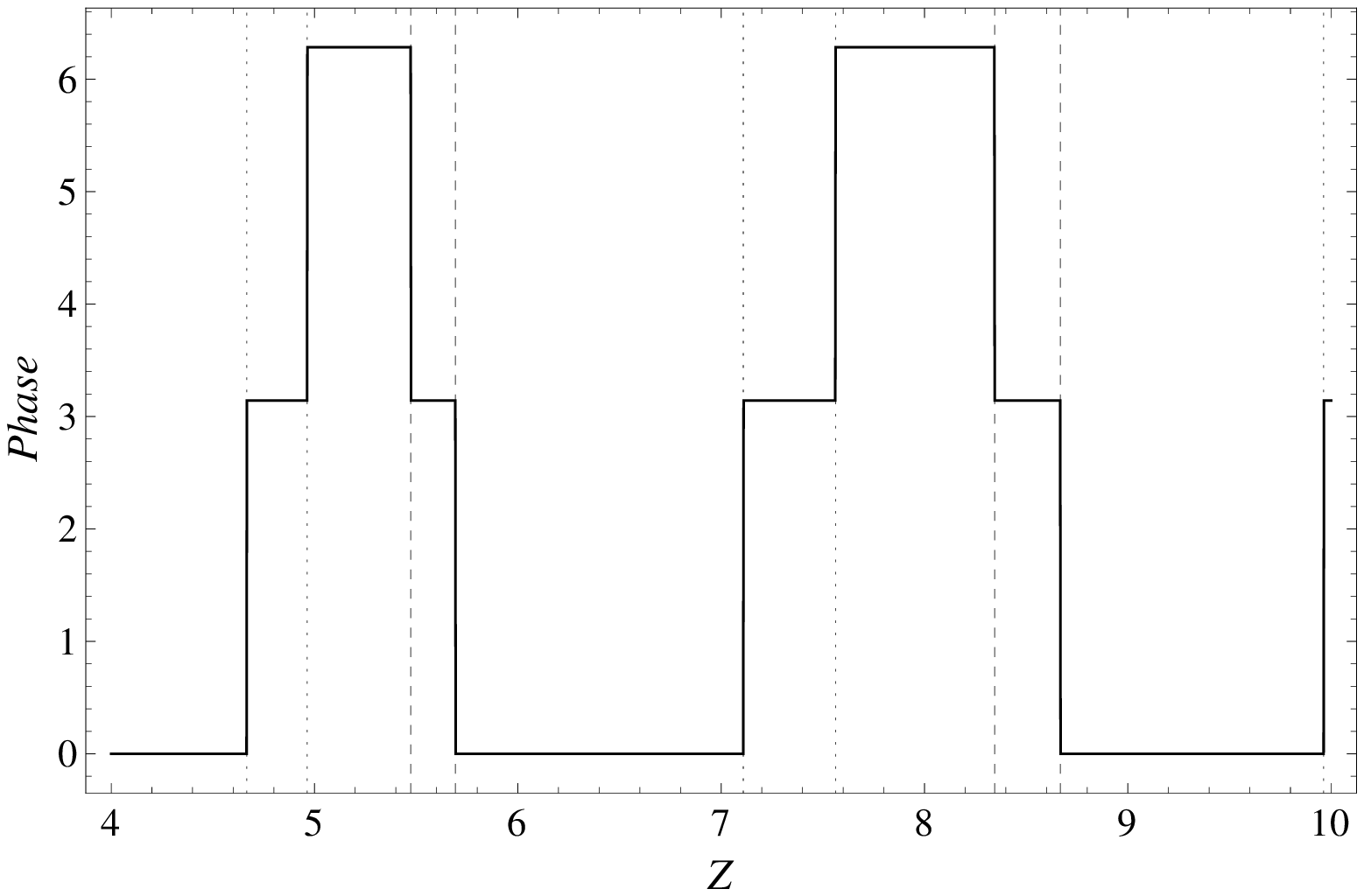}
}
\caption{\small $  \d_{tot,|m_j|}(0) $ for $\a=0.4$, $R_0=1/15$, $R_1=10 R_0$ and ~\subref{pic:32} $|m_j|=1/2$, ~\subref{pic:33} $|m_j|=3/2$ on the interval $0< Z < 10$. The vertical dashed lines denote the positions of lower critical charges, while the dotted ones indicate the positions of the  upper ones.}\label{pic:32-33}	
\end{figure*}

In Figs.\ref{pic:32-33} the behavior of $\d_{tot,|m_j|}(0)$ as a function of $Z$ is shown for $\a=0.4$, $R_0=1/15$, $R_1=10 R_0$ and  $|m_j|=1/2\, ,3/2$  on the interval $0< Z < 10$.
In contrast to the unscreened case $R_1\to\inf$, now for $k =0$ the total partial phase $\d_{tot,|m_j|}(0)$ on this interval of $Z$  takes only 3 values $0$, $\pi $ and $2 \pi$, which replace each other in a jump-like fashion by passing through  $Z_{cr}$ of both types.

\section{Peculiar effects in  the screened case for the channel with $|m_j|=1/2$ }\label{sec:peculiareffects}

\begin{figure*}[ht!]
\subfigure[]{\label{pic:39}
		\includegraphics[width=\columnwidth]{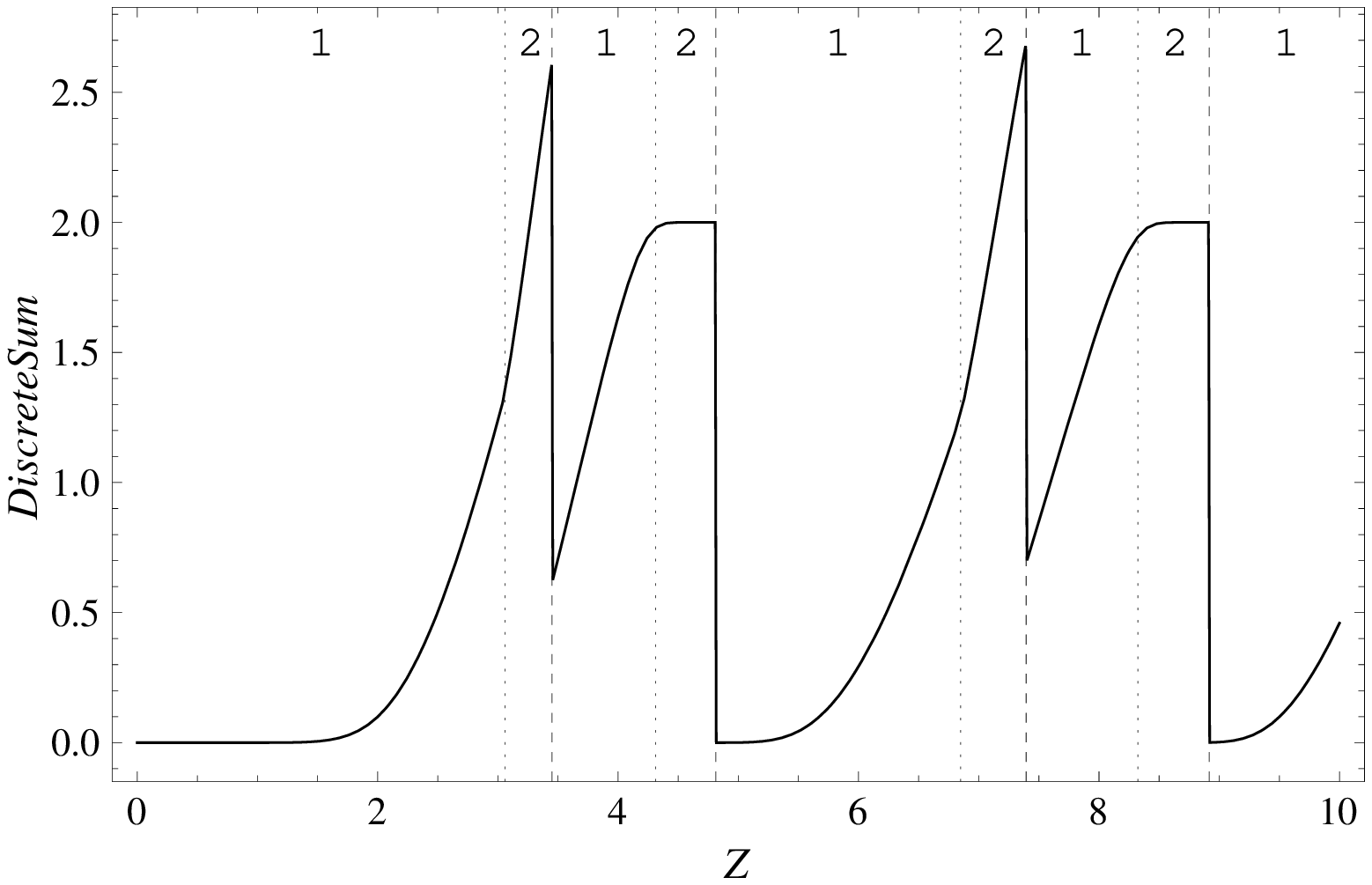}
}
\hfill
\subfigure[]{\label{pic:40}
		\includegraphics[width=\columnwidth]{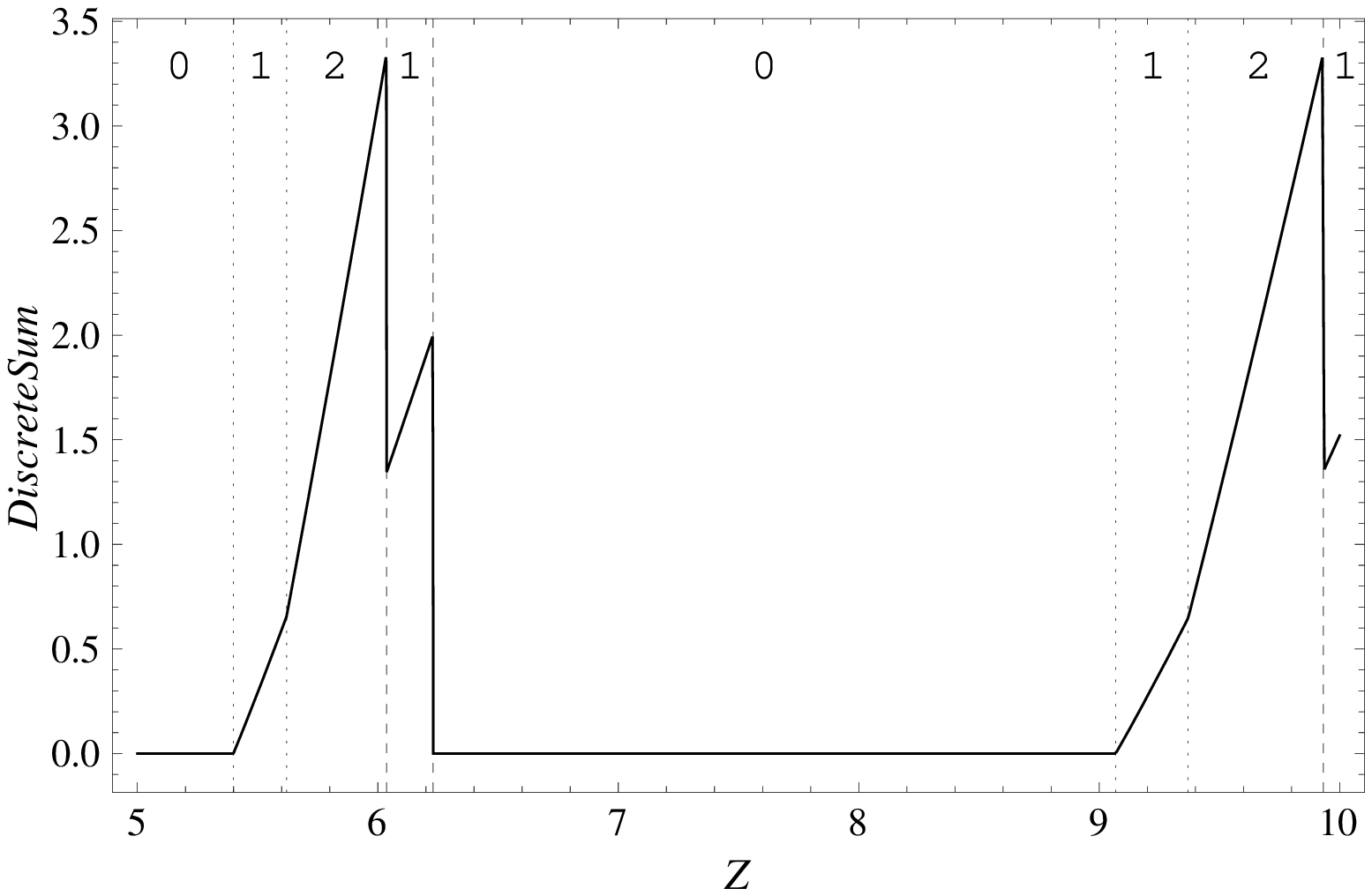}
}
\caption{\small The dependence of the total bound energy of discrete levels on $Z$ for  $\a=0.4$, $R_0=1/15$,$R_1=5 R_0$  and ~\subref{pic:39} in the channel with $|m_j|=1/2$, ~\subref{pic:40} in the channel with $|m_j|=3/2$. The vertical dashed lines denote the positions of lower critical charges, while the dotted ones indicate the upper ones. The number on top denotes the total number of existing discrete levels in this channel. }\label{pic:39-40}	
\end{figure*}

The peculiar effects for the screened planar DC problem in the channel with $|m_j|=1/2$, discussed already in \cite{partI2018} in terms of the principally different evolution of discrete levels by growing $Z$, show up in the behavior of the ingredients of the Casimir energy in this channel   as well. In Fig.\ref{pic:39} for $\a=0.4$, $R_0=1/15$, $R_1=5 R_0$ the dependence of the total bound energy of discrete levels in the channel $|m_j|=1/2$ on  $Z$ is given. The vertical dashed lines denote the positions of lower critical charges, while the dotted ones indicate the upper ones. The number on top denotes the total number of existing discrete levels in this channel between vertical lines. The first and third divings of levels into the lower continuum correspond to $m_j=1/2$, while the second and forth ones to $m_j=-1/2$, and vice versa, the first and third creations of levels at the upper threshold correspond to $m_j=-1/2$, while the second and forth ones to $m_j=1/2$. To underline this difference of the channel with $|m_j|=1/2$ from the others, in Fig.\ref{pic:40} the dependence on  $Z$ of the total bound energy of discrete levels in the channel $|m_j|=3/2$  is presented.

\begin{figure*}[ht!]
\subfigure[]{\label{pic:41}
		\includegraphics[width=\columnwidth]{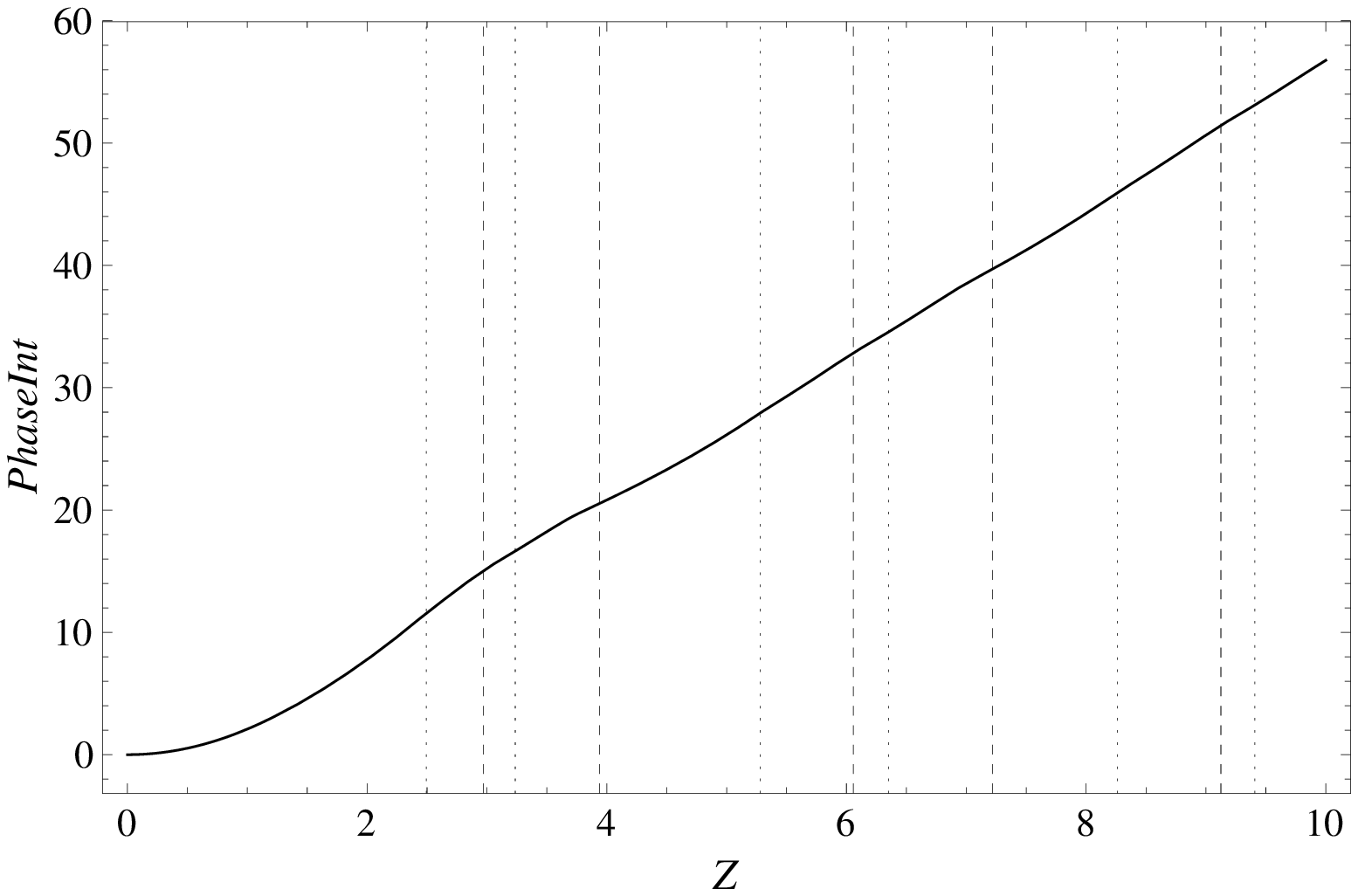}
}
\hfill
\subfigure[]{\label{pic:42}
		\includegraphics[width=\columnwidth]{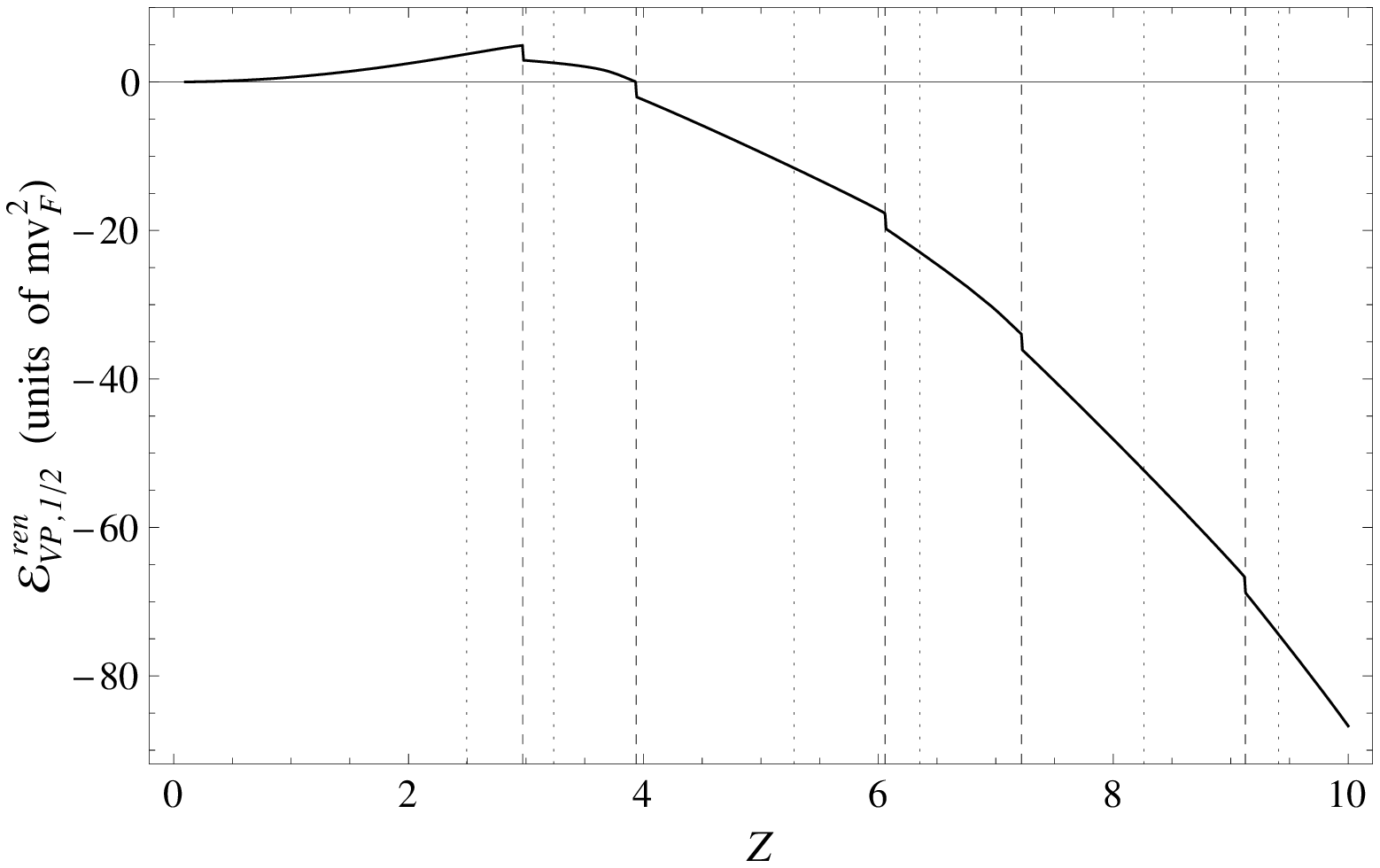}
}
\caption{\small  The dependence on $Z$ for  $\a=0.4$, $R_0=1/15$, $R_1=10 R_0$ in the channel with $|m_j|=1/2$ of: ~\subref{pic:41} the phase integral and ~\subref{pic:42}  $\E^{ren}_{VP,1/2}(Z)$. The vertical dashed lines denote the positions of lower critical charges, while the dotted ones indicate the upper ones. }\label{pic:41-42}	
\end{figure*}

\begin{figure*}[ht!]
\subfigure[]{\label{pic:43}
		\includegraphics[width=\columnwidth]{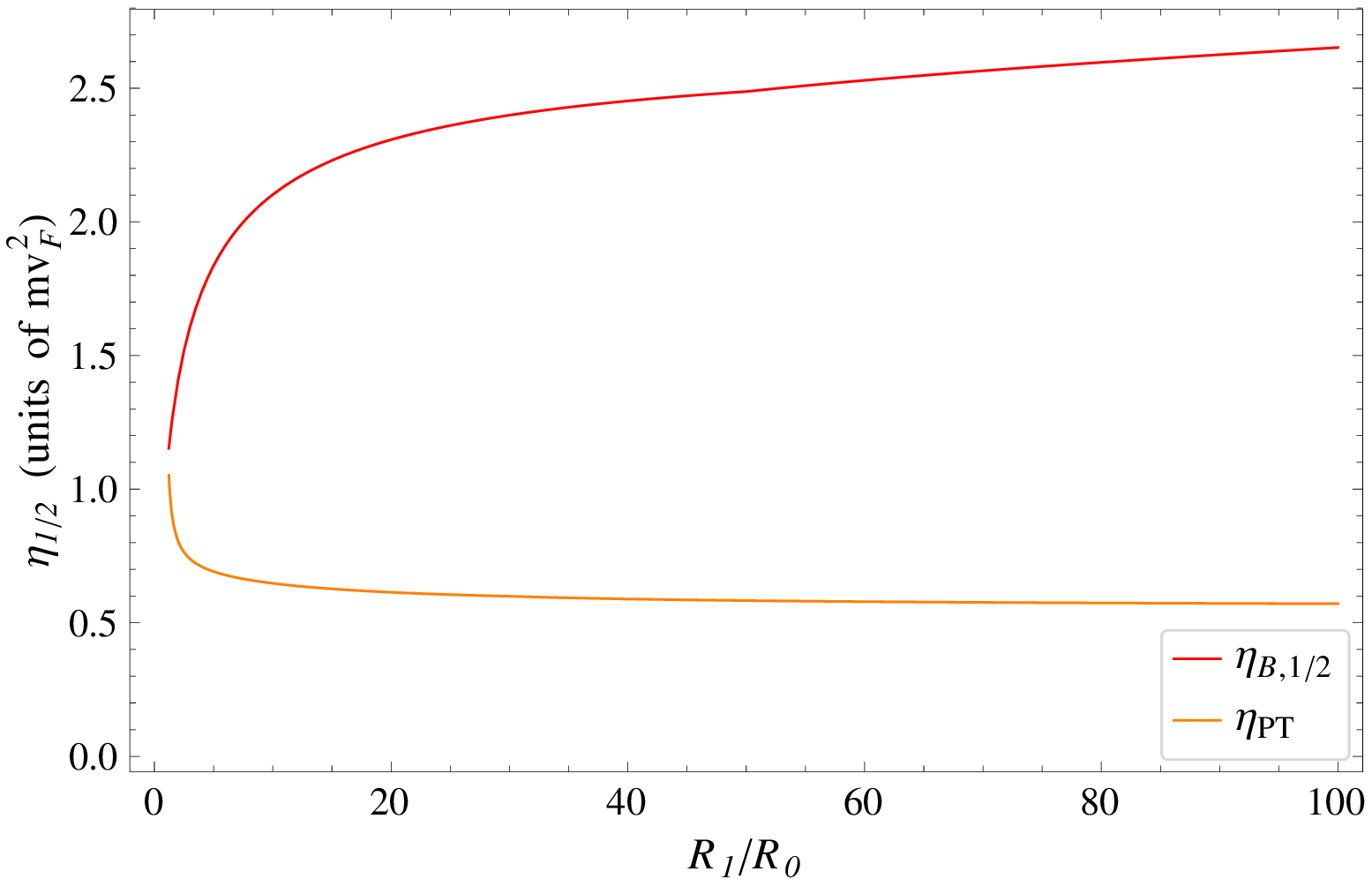}
}
\hfill
\subfigure[]{\label{pic:44}
		\includegraphics[width=\columnwidth]{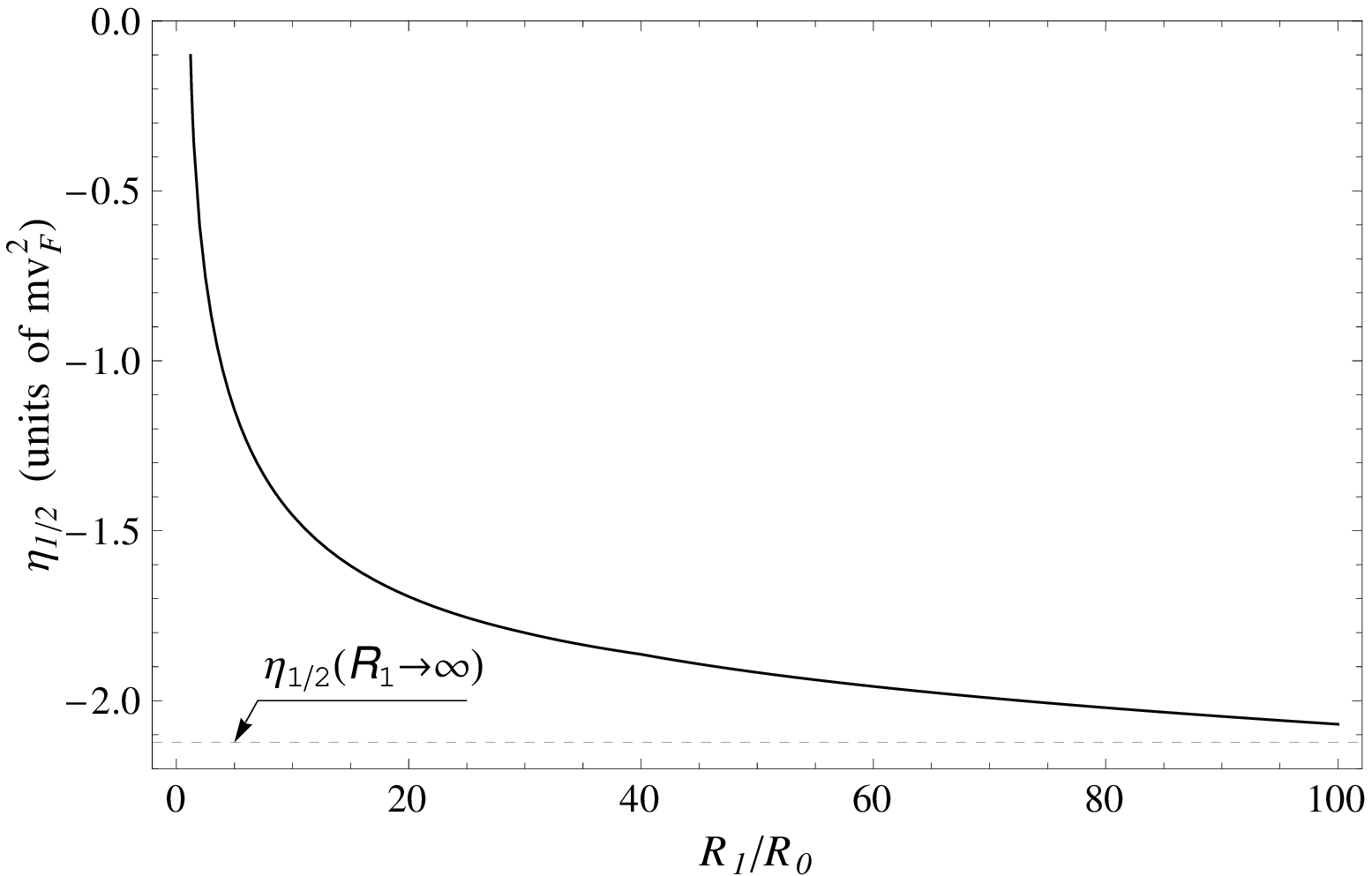}
}
\caption{(Color online) \small The dependence on $R_1$ for  $\a=0.4$, $R_0=1/15$ of: ~\subref{pic:43} $\et_{B,1/2}$ and $\et_{PT}$; ~\subref{pic:44} $\et_{1/2}=\et_{PT} -\et_{B,1/2}$.  }\label{pic:43-44}	
\end{figure*}

Actually the same specificity in the channel $|m_j|=1/2$ shows up in the behavior of the phase integral and the renormalized Casimir energy. In Fig.\ref{pic:41} the dependence of the phase integral on $Z$ for this channel is shown, while in Fig.\ref{pic:42} ---  $\E^{ren}_{VP,1/2}$ as a function of $Z$. The behavior of $\E^{ren}_{VP,1/2}(Z)$ turns out to be an increasing one up to the first level diving, since the renormalization coefficient $\et_{1/2}$, defined in (\ref{f43}), in this case turns out to be also strictly negative for all $R_1 > R_0$, and so the influence of the growing perturbative component (\ref{4.5}) turns out to be pronounced only up to $Z_{cr,1}$. It should be noted that the latter monotonically increases with decreasing $R_1$. In particular,  for $\alpha=0.4\, , \ R_0=1/15$ and $R_1=\infty\, , 20 R_0\, , 10 R_0\, ,  5 R_0\, , 2 R_0$ it takes the values $2.373\, , 2.685\, , 2.971\, , 3.451\, , 4.592$, respectively,  while for $\alpha=0.8\, , \ R_0=1/175$ and the same $R_1$ one gets instead $0.870\, , 1.200\, , 1.373\, , 1.633\, , 2.216$.
The dependence of the renormalization coefficient $\et_{1/2}=\et_{PT} -\et_{B,1/2}$ on the screening parameter $R_1$ for $\a=0.4$ and $R_0=1/15$ is shown in Figs.\ref{pic:43-44}. For all $R_1> R_0$ it is strictly negative and for $R_1\to\inf$  tends to the unscreened value. All the other $\eta_{|m_j|}$'s with  $|m_j| \geq 3/2$ are negative by construction, since in this case $\et_{|m_j|}=-\et_{B,|m_j|}$ and so behave in the same fashion.

\section{Conclusion}

\begin{figure*}[ht!]
\subfigure[]{\label{pic:45}
		\includegraphics[width=\columnwidth]{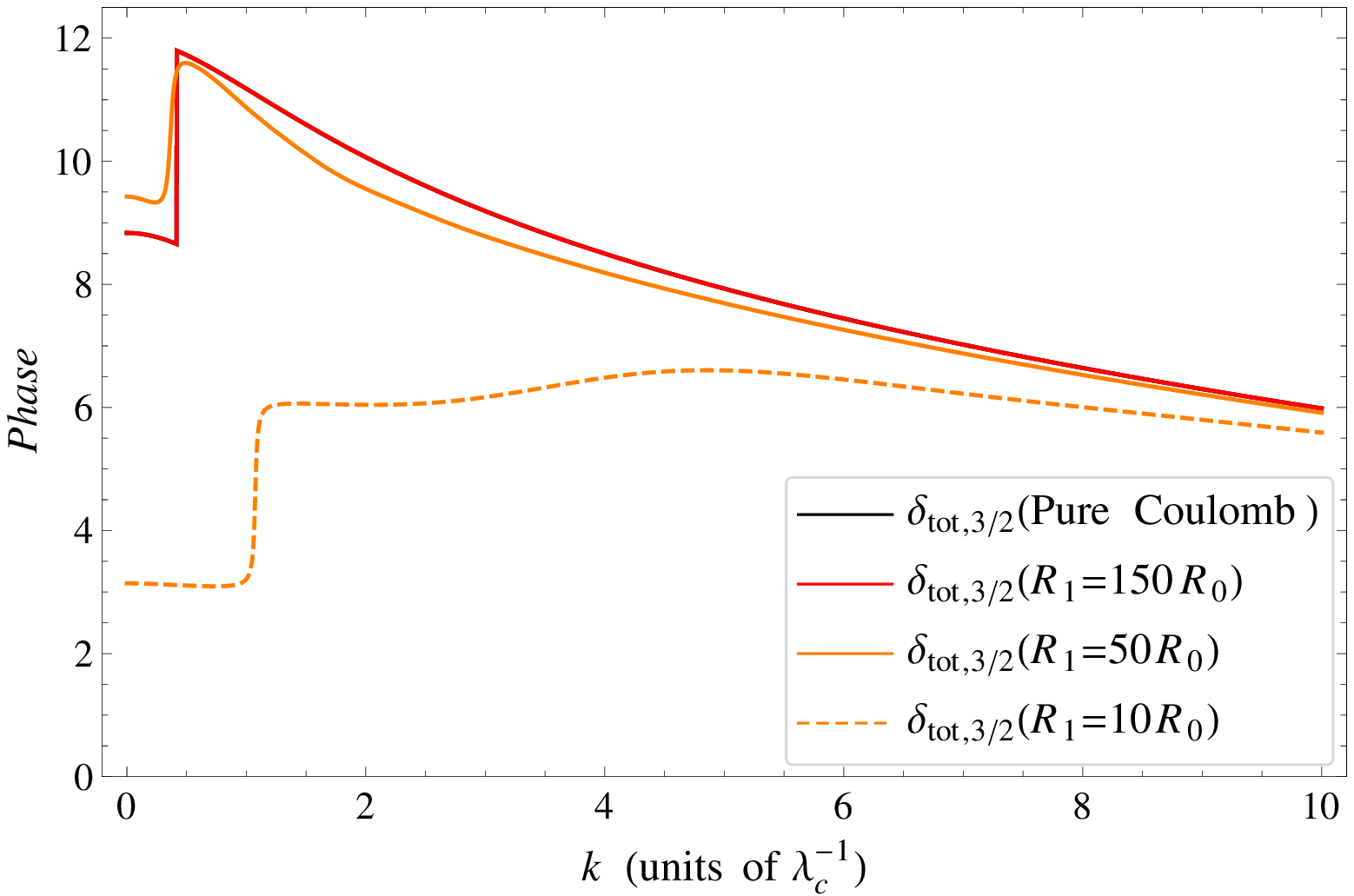}
}
\hfill
\subfigure[]{\label{pic:46}
		\includegraphics[width=\columnwidth]	{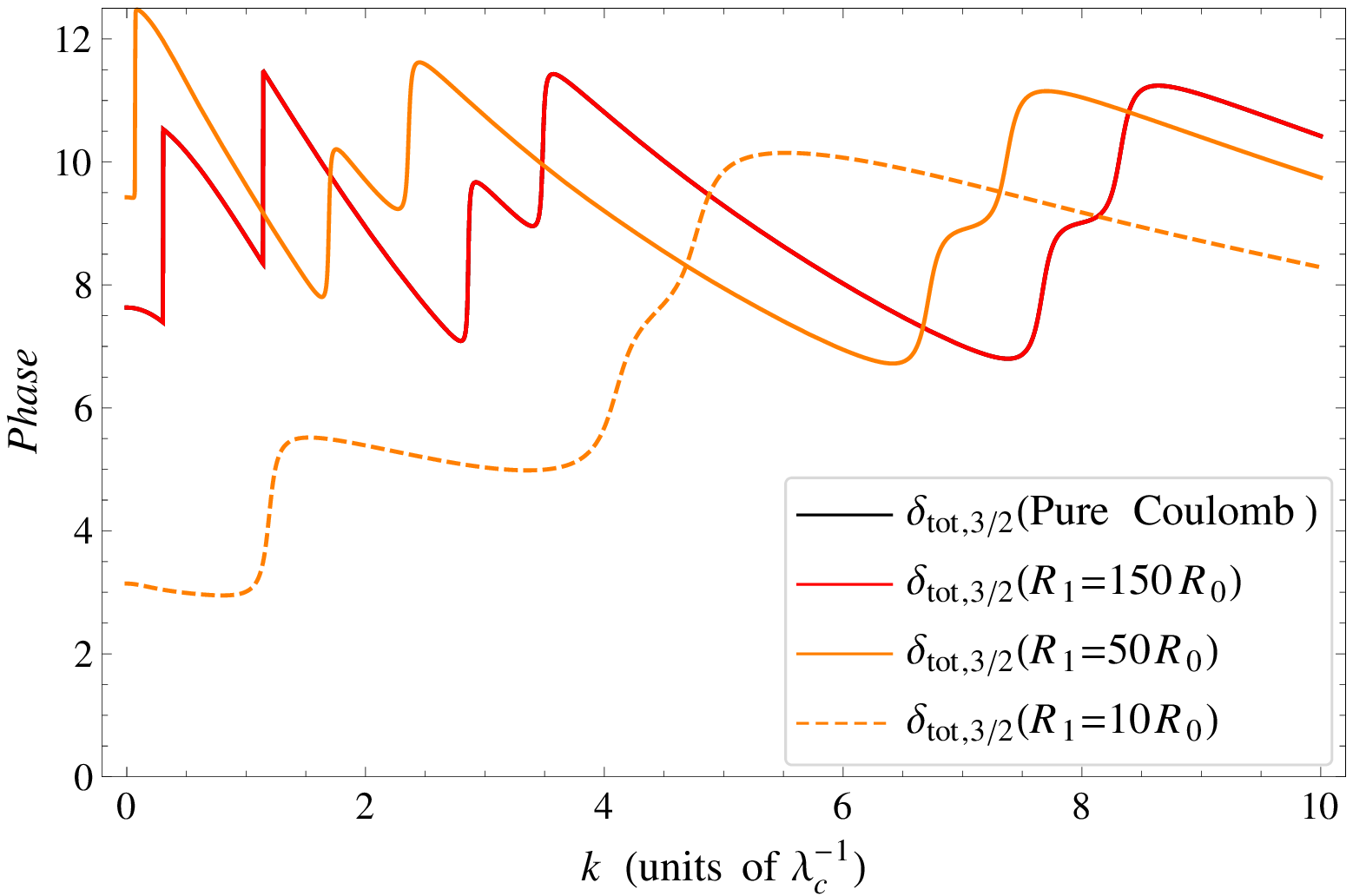}
}
\caption{(Color online) \small $ \d_{tot,3/2}(k) $ for $\a=0.4$, $R_0=1/15$ and ~\subref{pic:45} $Z=4.75$, ~\subref{pic:46} $Z=10$ with and without screening of the Coulomb  asymptotics.}\label{pic:45-46}	
\end{figure*}

However, besides the peculiarities in the channel $|m_j|=1/2$, the behavior of other ingredients of $\E^{ren}_{VP}$ is much more regular, providing a smooth transition into the unscreened case for $R_1 \to \inf$.
Namely, in  Figs.\ref{pic:45-46} there are shown the curves of  $\delta_{tot,|m_j|}(k)$ for $a=0.4$, $R_0=1/15$,  $|m_j|=3/2$ without screening and for  $R_1=10\, , 50\, , 150 R_0$ and charges $Z=4.75$, i.e. just after diving the first discrete level in the unscreened case,   and  $Z=10$. Actually for  $R_1=50 R_0$ the difference between the cases with and without screening is quite moderate, while for  $R_1=150 R_0$ it becomes negligibly small for the considered values of the external charge $Z$. The evaluation of the total Casimir energy also confirms this result.

\begin{figure*}[ht!]
	\subfigure[]{
		\includegraphics[width=\columnwidth]{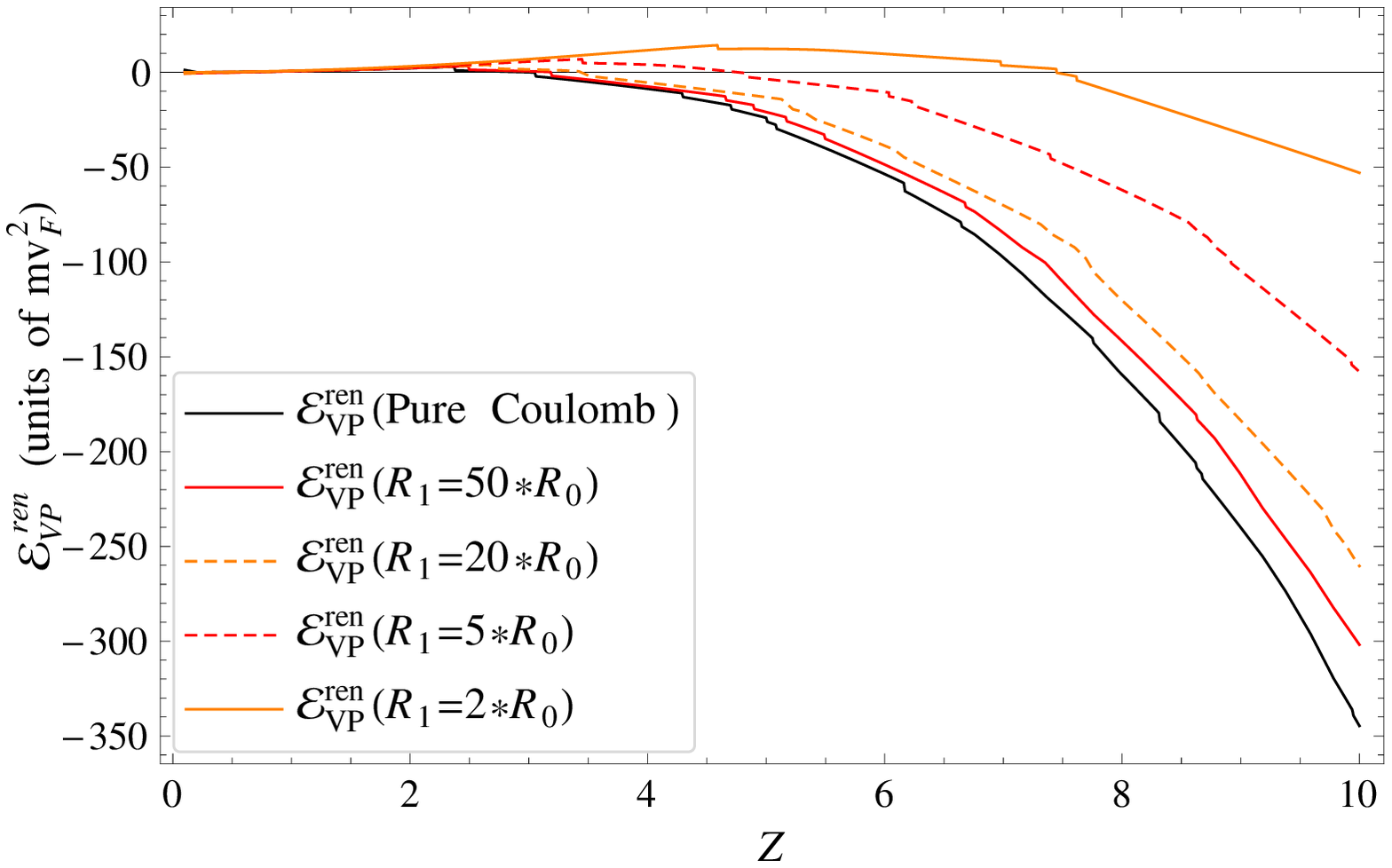}
}
	\caption{(Color online) \small $ \E_{VP}^{ren}(Z) $ for $\a=0.4$, $R_0=1/15$ and different values of the screening parameter $R_1$.}\label{pic:59}
\end{figure*}

The dependence of the Casimir energy $\E_{VP}^{ren}(Z)$ on screening effects is shown in Fig.\ref{pic:59} for $\a=0.4$, $R_0=1/15$ and $R_1=2 R_0\, , 5 R_0\, , 20 R_0\, , 50 R_0\, , \inf$. For decreasing $R_1$ the values of critical charges   increase, and hence, the growth rate of vacuum shells number decreases.  On the curves $\E_{VP}^{ren}(Z)$ the moments of discrete levels diving are clearly seen as jumps. In particular, for $Z=10$ without screening 46 discrete levels reach the lower continuum, for $R_1=50 R_0$ their  number equals to 40, for $R_1=20 R_0$ --- to 30, for $R_1=5 R_0$ ---  to 18, while for  $R1=2R0$ --- just to 8.  In the same way the decrease of vacuum energy into the negative range also slows down.

Thus, our calculation indicates  that in the graphene-like two-dimensional QED-systems with strong coupling  the decline of the Casimir energy could surpass the repulsive self-energy of the external Coulomb source for such impurity charges, which might seem  large enough in view of present experiments, especially for $\a=0.4$, but at the same time are not unattainable. Moreover, to a certain degree this effect turns out to be insensitive to the impurity size and to the screening of Coulomb asymptotics of the external potential, since the decline of the Casimir energy disappears only for screening of the external potential at the scales close to $R_0$. To some extent the latter circumstance  justifies our choice of screening in the form of the simplest continuous or discontinuous shielding for $r > R_1$. Here it should be noted that in the most of works cited above ~\cite{Katsnelson2006b, Biswas2007, Pereira2007, Shytov2007, Fogler2007,   Kotov2008, Terekhov2008, Milstein2010,  Nishida2014, Khalilov2017} the impurity potentials are considered without any kind  of screening at large distances from the Coulomb source. However, in fact the screening in such systems should definitely take place and is quite complicated, since it turns out to be a composite reaction of the medium combined with vacuum shells. But the consistent study of this question lies beyond the scope of our work.

It should be also mentioned  that  actually in the planar Dirac-Coulomb system of the considered type   the calculation of the Casimir energy by means of  UV-renormalization via fermionic loop could be implemented solely on the basis of relations  (\ref{f40}),(\ref{f41a}) and  (\ref{f44a}) without applying to the shell effects in the induced density. The  essential point here is that by such renormalization we simultaneously ensure the convergence of the whole partial series for $\E_{VP}^{ren}$, since according to  (\ref{term_m}) the divergent terms in the series (\ref{f41a}) are proportional to $(Z\a)^2$. So the renormalization via fermionic loop turns out to be the universal tool, that removes the divergence of the theory both in the purely perturbative and in the essentially non-perturbative regimes of vacuum polarization by the external Coulomb source.

However, in essence the decrease of $\E_{VP}^{ren}$ in the overcritical region is governed first of all by the non-perturbative changes in the induced density for $Z > Z_{cr,1}$ due to discrete levels, reaching the threshold of the lower continuum (``the shell effect''). In 1+1 D the growth rate of  vacuum shells is $\sim Z^s \ , \ 1<s<2$, at least in the considered in Refs.~\cite{Davydov2017},\cite{Voronina2017} range of external parameters. Therefore in the overcritical region the growth rate of the non-renormalized energy $\E_{VP}$ does not exceed $O( Z^\n)\, ,  1<\n<2$, and so the dominant contribution to $\E_{VP}^{ren}$ comes from the renormalization term $\eta Z^2$. In 2+1 D (and especially in 3+1 D) the shell effect is much more pronounced and the growth rate  of the total number of vacuum shells $N(Z)$ exceeds definitely  $O(Z^2)$. As a result, for considered planar QED-systems   $\E_{VP}^{ren}(Z)$ decreases in the overcritical region at least by one order of magnitude faster.

The decline of  the cumulative energy $\E_{VP}^{ren} + \E_{cl}$ into the negative values for $Z >Z^{\ast}$  means the emergence of a kind of attraction, and hence, the possibility of bound states  in the composite system formed from the source and the graphene plane.  Thus, the non-perturbative vacuum polarization effects could play an important role in the  properties of such graphene-like planar systems upon doping by charged impurities with $Z >Z^{\ast}$, leading to  a special type of affinity between the impurities and the graphene plane.

\twocolumngrid

\bibliography{biblio/VP2DG}
\end{document}